\documentclass[fleqn,usenatbib]{mnras}

\usepackage{newtxtext,newtxmath}

\usepackage[T1]{fontenc}

\DeclareRobustCommand{\VAN}[3]{#2}
\let\VANthebibliography\thebibliography
\def\thebibliography{\DeclareRobustCommand{\VAN}[3]{##3}\VANthebibliography}

\usepackage{graphicx}	
\usepackage{amsmath}	
\usepackage{tikz}
\usepackage{float}
\usepackage{booktabs}
\usepackage{longtable}
\usepackage{layouts}
\usepackage{soul}

\newcommand{\gaussian}{
  \begin{tikzpicture}[scale=0.15]
    \draw[thick, domain=-2:2,smooth,variable=\x] plot ({\x},{exp(-(\x*\x)/(0.5)});
  \end{tikzpicture}
}

\newcommand{\logarithm}{
  \begin{tikzpicture}[scale=0.15]
    \draw[thick,domain=0.1875:2.25,smooth,variable=\x] plot ({\x-1.2},({-log10(\x)+0.75});
    \draw[thick] (-2,0) -- (-0.905,0);
    \draw[thick] (-1,0) -- (-1,{-log10(0.2)+0.74});
    \draw[thick] (1,0) -- (1,{-log10(2.2)+0.83});
    \draw[thick] (0.905,0) -- (2.,0);
  \end{tikzpicture}
}

\newcommand{\tophat}{
  \begin{tikzpicture}[scale=0.15]
        \draw[thick] (-2,0) -- (-1,0) -- (-1,1) -- (1,1) -- (1,0) -- (2,0);
  \end{tikzpicture}
}

\title[Massive quiescent galaxies at $2\;<\;z\;<\;5$]{PRIMER \& JADES reveal an abundance of massive quiescent galaxies at $\mathbf{2\;<\;z\;<\;5}$}

\author[S. Stevenson et al.]{Struan D. Stevenson,$^{1}$\thanks{E-mail: struan.stevenson@ed.ac.uk}
Adam C. Carnall,$^{1}$
Ho-Hin Leung,$^{1}$
Elizabeth Taylor,$^{1}$
Fergus Cullen,$^{1}$\and
James S. Dunlop,$^{1}$
Derek J. McLeod,$^{1}$
Ross J. McLure,$^{1}$
Ryan Begley,$^{1}$
Karla Z. Arellano-Córdova,$^{1}$\and
Laia Barrufet,$^{1}$
Cecilia Bondestam,$^{1}$
Callum T. Donnan,$^{2}$
Richard S. Ellis,$^{3}$
Norman A. Grogin,$^{4}$\and
Anton M. Koekemoer,$^{4}$
Feng-Yuan Liu,$^{1}$
Pablo G. P\'erez-Gonz\'alez,$^{5}$
Kate Rowlands, $^{4,6}$
Ryan L. Sanders,$^{7}$\and
Dirk Scholte,$^{1}$
Alice E. Shapley,$^{8}$
Maya Skarbinski,$^{6}$
Thomas M. Stanton,$^{1}$
Vivienne Wild$^{9}$
\\
\\
$^{1}$Institute for Astronomy, University of Edinburgh, Royal Observatory, Edinburgh EH9 3HJ, UK\\
$^{2}$NSF’s National Optical-Infrared Astronomy Research Laboratory, 950 N. Cherry Ave., Tucson, AZ 85719, USA\\
$^{3}$University College London, Department of Physics \& Astronomy, Gower Street, London WC1E 6BT, UK\\
$^{4}$Space Telescope Science Institute, 3700 San Martin Drive, Baltimore, MD 21218, USA\\
$^{5}$Centro de Astrobiolog\'{\i}a (CAB), CSIC-INTA, Ctra. de Ajalvir km 4, Torrej\'on de Ardoz, E-28850, Madrid, Spain\\
$^{6}$William H. Miller III Department of Physics and Astronomy, Johns Hopkins University, Baltimore, MD 21218, USA\\
$^{7}$Department of Physics and Astronomy, University of Kentucky, 505 Rose Street, Lexington, KY 40506, USA\\
$^{8}$Department of Physics \& Astronomy, University of California, Los Angeles, 430 Portola Plaza, Los Angeles, CA 90095, USA\\
$^{9}$School of Physics \& Astronomy, University of St Andrews, North Haugh, St Andrews, KY16 9SS, UK}
\date{Accepted XXX. Received YYY; in original form ZZZ}
\pubyear{\the\year{}}
\begin{document}
\label{firstpage}
\pagerange{\pageref{firstpage}--\pageref{lastpage}}
\maketitle
\begin{abstract}
\noindent We select a mass-complete sample of 225 quiescent galaxies at $z>2$ with $M_*>10^{10}\,\mathrm{M}_\odot$ from PRIMER and JADES photometry spanning a total area of $\simeq320$ sq. arcmin. Our analysis is restricted to only area with optical coverage in three \textit{HST} ACS filters, which we show is important for selecting the most complete and clean samples. We investigate the contamination in our sample via \textit{JWST} NIRSpec spectroscopy, $Chandra$ X-ray imaging, and ALMA interferometry, calculating a modest contamination fraction of $12.9_{-3.1}^{+4.0}$ per cent. The removal of \textit{HST} data increases star-forming galaxy contamination by $\simeq10$ per cent and results in a $\simeq20$ per cent loss of candidates recovered from \textit{HST}+\textit{JWST} data combined. We calculate massive quiescent galaxy number densities at $2<z<5$, finding values three times larger than pre-\textit{JWST} estimates, but generally in agreement with more-recent and larger-area \textit{JWST} studies. In comparison with simulations, we find that most can now reproduce the observed number density at $2<z<3$, however they still increasingly fall short at $z>3$, up to $\simeq1$ dex. We place 14 of our $z>3$  massive quiescent galaxies on the BPT and WHaN diagrams using medium-resolution spectroscopic data from the EXCELS survey, finding a very high incidence of weak AGN ($\simeq50$ per cent), consistent with recent results at cosmic noon. This is interesting in the context of `maintenance-mode’ feedback, which is invoked in many simulations to prevent the re-ignition of quenched galaxies. To properly characterise the evolution of early massive quiescent galaxies, greater coverage in optical filters and significantly larger spectroscopic samples will be required.

\end{abstract}




\section{Introduction}\label{section:introduction}

Since delivering its first data in 2022, the \textit{James Webb Space Telescope} (\textit{JWST}) has investigated the early Universe in unprecedented detail (see \citealt{Adamo2024} for a recent review). A key theme emerging is the abundance of bright galaxies at high redshift, with large numbers of UV-luminous objects discovered beyond $z=10$ (e.g., \citealt{Castellano2022,Curtis-Lake2023, McLeod2024, Castellano2024,Carniani2024, Naidu2025}). Galaxies have also been observed to exhibit significant mass growth and quenching on timescales that challenge current theories of galaxy formation (e.g., \citealt{Hartley2023, Gould2023}), with significant numbers of evolved, passive galaxies in place by $z\simeq5$ \citep[e.g.,][]{Carnall2023a, Valentino2023}. Previously too faint and/or red to be detected, \textit{JWST} has been able to characterise these high-redshift objects in great detail, revealing significant dust production and chemical enrichment \citep{dEugenio2023, Shapley2024, Setton2024, Beverage2025}. 


Observations of high-redshift massive quiescent galaxies have been a key constraint on galaxy formation theory for decades \citep{Dunlop1996, Cimatti2004, Fontana2009}. Such observations have progressively revealed inadequacies in contemporary models and simulations, which have historically struggled to reproduce observed quiescent galaxy number densities; unable to both grow enough massive galaxies in the early Universe and subsequently rapidly quench them \citep[e.g.,][]{Dave2017, Schreiber2018,Cecchi2019,Girelli2019, Merlin2019}. The probability that massive high-redshift quiescent galaxies go on to form the cores of the most massive ellipticals today is also furthering interest into this field \citep[e.g.,][]{Rennehan2024,Beverage2024,Baggen2024}.

The first years of \textit{JWST} have produced a number of publicly available deep photometric surveys using NIRCam. In the first months of \textit{JWST}, \cite{Carnall2023a} calculated high-redshift passive number densities over $\simeq30$ sq. arcmin of Cosmic Evolution Early Release Science (CEERS; \citealt{Finkelstein2025}) imaging data, finding values $3-5$ times higher than previously reported by \cite{Schreiber2018} at $3 < z < 4$. Subsequently, \cite{Valentino2023} explored the passive number density in multiple publicly available surveys, totalling an overall sky area of $\simeq145$ sq. arcmin (with the largest contiguous fields each covering $\simeq30$ sq. arcmin). Their results showed significant field-to-field variations, but followed a general trend of overabundance compared with simulations and pre-\textit{JWST} observations at $z>3$. Several subsequent papers supported these results \citep[e.g.,][]{Long2023,Alberts2024a, Russel2024, Baker2025a, Baker2025b}, implying both a rapid assembly of mass and rapid quenching are not uncommon well before cosmic noon. 

Further fuelling this discussion, \cite{Glazebrook2024} recently reported a very massive (log$_{10}(M_*/\mathrm{M_\odot}) \simeq11.2)$ quiescent galaxy at $z=3.2$, which appears to have formed most of its stars at $z\sim11$ (only 400 Myr after the Big Bang), bringing our current understanding of star formation physics at high redshift into discussion \citep[e.g.,][]{Dekel2023, Dekel2025}. Various simulations are currently being updated in reflection of these results \citep[e.g.,][]{Lagos2024}.

Disentangling the complexity of quiescent galaxy evolution in the early Universe relies on a deep understanding of quenching mechanisms. In order for star formation in a galaxy to be suppressed, the gas needs to be prevented from cooling and condensing to form stars. This can either be achieved by heating the gas reservoir, or by expelling it from the galaxy.

A leading candidate for quenching early massive galaxies is an Active Galactic Nucleus (AGN) phase (e.g., \citealt{Croton2006, Wellons2015}). This results from accretion onto a central supermassive black hole, a process in which vast amounts of energy are deposited into the interstellar medium (ISM). In extreme cases, AGN-driven winds can drive gas out of the galaxy, leading to a rapid shut-down in star formation \citep{Fabian2012, Maiolino2012}.

Since the advent of \textit{JWST}, several studies have confirmed signatures of AGN-driven outflows in high-redshift quenched galaxies \citep{Deugenio2024, Davies2024, Belli2024, Wu2024, Bugiani2024, Valentino2025}, furthering evidence for AGN feedback as a primary quenching mechanism in the early Universe. Signatures of outflows have also been detected in the absence of an AGN (and vice versa), most likely due to the remnants of an AGN phase (or the very beginning of one; \citealt{Taylor2024}).

Another pathway to releasing large amounts of energy into the ISM is supernovae, often referred to in this context as stellar feedback. This is thought to suppress star formation in low-mass galaxies at all redshifts, which have shallower gravitational potential wells and therefore lose their gas reservoirs more easily than their more-massive counterparts \citep[e.g.,][]{Man2018}. However, at high redshift the extreme conditions prevalent throughout (high star-formation-rate surface densities and low metallicities) could allow stellar feedback to also play a role in quenching more massive galaxies (e.g., \citealt{Hopkins2010, Grudic2019}). Young massive stars can also contribute to radiation driven winds, which, like AGN feedback, can act on faster timescales than supernovae, and are a candidate mechanism for rapidly quenching low-mass systems in the early Universe (e.g., \citealt{Looser2023, Strait2023, Gelli2023, Baker2025c, Covelo-Paz2025}).


Another candidate quenching mechanism is galactic mergers, a chaotic process that can destabilise the gas within galaxies, hence often leading to a starburst phase (e.g., \citealt{Wellons2015, Rodriguez2019}). However, \cite{Ellison2022} show that after coalescence, accretion of gas onto the central black hole can induce AGN feedback that rapidly shuts down star formation, creating a `post-starburst' quiescent galaxy. It seems clear, therefore, that galactic mergers play an important role in perpetuating the colour bimodality across cosmic time \citep{Lambas2012}. Furthermore, major mergers between already-quenched systems may be crucial to the assembly of the most-massive objects at high redshift (e.g., \citealt{Ito2025}), as is already known to be the case in the local Universe (e.g., \citealt{Emsellem2011,Khochfar2011}). This is in addition to the apparently ubiquitous gradual growth of massive quiescent galaxies via minor mergers (e.g., \citealt{mclure2013, hamadouche2022, suess2023, Hamadouche2025}).


For massive galaxies to be quenched so early on in the history of the Universe both a rapid and powerful quenching mechanism is required. This immediately suggests AGN feedback as a likely candidate. Quenching mechanisms acting on longer timescales, such as strangulation or gas exhaustion \citep{Schawinski2014, Peng2015, Trussler2020} are disfavoured simply due to the young age of the Universe. However, with the advent of \textit{JWST} we can now move beyond such circumstantial evidence by using ultra-deep NIRSpec spectroscopy to directly probe the physics of $z>3$ massive quiescent galaxies. For example, the presence of an AGN can be confirmed via the relative strengths of emission lines (e.g., \citealt{Kewley2006}), and other key parameters such as stellar ages and metallicities can be probed via continuum fitting (e.g., \citealt{Carnall2019b, Belli2019, Leung2024}).

In this work we address the, still-uncertain, number density of massive quiescent galaxies at high redshift, as well as investigating their quenching physics. We first utilise a combination of imaging from PRIMER \citep{Dunlop2021} and JADES \citep{Eisenstein2023b}
to construct a homogeneous sample of quiescent galaxies at $z>2$ spanning a sky area of $\simeq$320 sq. arcmin in the fields UDS, COSMOS and GOODS South, representing one of the largest area studies to date (e.g., \citealt{Valentino2023, Russel2024, Baker2025b}). Crucially, we use only the area in these fields that benefits from optical \textit{HST} ACS data as well as \textit{JWST} NIRCam data, which is important for selecting the most robust samples, as we demonstrate in Section \ref{section:sample:hst}. 
 
Using the {\sc Bagpipes} spectral fitting software \citep{Carnall2018} we focus on selecting a mass-complete sample of quiescent candidates down to log$_{10}(M_*/\mathrm{M_\odot})=10$ and estimate the level of contamination from a combination of spectroscopic, X-ray and millimetre data. In particular, we crossmatch to archival \textit{JWST} NIRSpec data, obtaining spectra for $\simeq$ 40 per cent of our photometric galaxy sample (83 out of 225 objects). We calculate (contamination-corrected) number densities in three redshift bins: $2<z<3$, $3<z<4$ and $4<z<5$. Running this calculation over such a large area lends a significant statistical advantage, greatly diminishing the effects of statistical and cosmic variance.

We then cross-match our quiescent sample with the EXCELS survey \citep{Carnall2024}, which provides $1-5$ $\mu$m ultra-deep medium resolution ($R\simeq1000$) spectra for 14 objects at $z>3$. We then calculate emission-line fluxes for our EXCELS $z>3$ sample to investigate the presence of AGN in high-redshift quiescent galaxies. In a companion paper, we conduct full-spectral fitting on these objects to probe their star-formation histories (SFHs) and stellar metallicities (Leung et al. in prep).

The paper is structured as follows. In Section \ref{section:data} we introduce the PRIMER and JADES imaging datasets used in this work, as well as the multiple public spectroscopic surveys we employ, including EXCELS, and public ALMA observations from A$^3$COSMOS. In Section \ref{section:methods} we lay out the methodologies used for defining our mass-completeness limit, photometric Spectral Energy Distribution (SED) fitting, selection of quiescent galaxy candidates and emission line analysis. In Section \ref{section:sample} we present our photometric candidate sample, in particular investigating the level of contamination in Section \ref{section:sample:contamination}. In Sections \ref{section:densities} and \ref{section:lineflux} we present and discuss our results, including massive quiescent number density calculations and an investigation into the presence of AGN. We present our conclusions in Section \ref{section:conclusions}. All magnitudes are quoted in the AB system. For cosmological calculations, we adopt $\mathrm{\Omega_m} = 0.3$, $\mathrm{\Omega_\Lambda} = 0.7$, and $\mathrm{H_0 = 70 km s^{-1}Mpc^{-1}}$. 

\section{Data}\label{section:data}

\subsection{PRIMER and JADES imaging}\label{section:data:imaging}

Our primary dataset is built from two public \textit{JWST} NIRCam deep imaging surveys, the Public Release Imaging for Extragalactic Research (PRIMER; Dunlop et al. in prep) survey and the \textit{JWST} Advanced Deep Extragalactic Survey (JADES; \citealt{Eisenstein2023b}). Specifically, we include the Ultra-Deep Survey (UDS) and Cosmic Evolution Survey (COSMOS) fields from PRIMER, and the Great Observatories Origins Deep Survey South (GOODS South) field from JADES.

We use the v0.7 internal release of the PRIMER UDS imaging and the v0.8 internal release of COSMOS. These were processed using the PRIMER Enhanced NIRCam Image-processing Library (PENCIL) software (Magee et al. in prep), which is built on top of the standard \textit{JWST} imaging pipeline (v1.10.2) and provides enhanced routines for snowball, wisp and 1/f noise correction. For GOODS South we make use of the public imaging from JADES Data Release 2 (DR2; \citealt{Eisenstein2023c})\footnote{\href{https://archive.stsci.edu/hlsp/jades}{https://archive.stsci.edu/hlsp/jades}}.

Both surveys have imaging available in eight NIRCam filters (F090W, F115W, F150W, F200W, F277W, F356W, F410M and F444W). In addition, \textit{Hubble Space Telescope} (\textit{HST}) Advanced Camera for Surveys (ACS) imaging is also available over some of the NIRCam area in three filters: F435W, F606W, F814W \citep{Grogin2011, Koekemoer2011, Illingworth2016, Whitaker2019}, extending the wavelength coverage to $0.4 - 5$ $\mu$m. We investigate the effect of this \textit{HST} data in Section \ref{section:sample:hst}, finding that significantly more-complete and less-contaminated samples are selected with \textit{HST} data than without. We therefore only make use of the area over which all 11 bands are available (8 \textit{JWST} + 3 \textit{HST}) in our survey footprint, which results in a total effective sky area of 317.7 sq. arcmin across the three fields. Subsets of JADES GOODS South also have imaging in NIRCam F335M, as well as \textit{HST} ACS F775W and F850LP, which we also make use of where available. 

While a subset of the area used in this work also benefits from \textit{JWST} MIRI coverage, we do not include this data in our primary analysis. We investigated the effect of MIRI F770W on the selection of massive quiescent galaxies in Appendix \ref{appendix:miri}, concluding that, while MIRI coverage improves the selection of quiescent galaxies where only NIRCam imaging is otherwise available, it has a negligible effect on selection based on a combination of both \textit{HST} ACS and NIRCam imaging.

The photometric catalogues used in this work are identical to those described in \cite{Begley2025}. These are constructed by first PSF-homogenising each image to the F444W band, using empirical PSFs derived from stacks of bright stars in the mosaic images. {\sc SourceExtractor} \citep{Bertin1996} is then employed in dual-image mode to construct photometric catalogues for each field, with F356W as the detection image. This filter was chosen as it probes longer rest-frame wavelengths, which are more sensitive to stellar mass rather than ongoing star formation (F444W was not used as it is significantly shallower). Fluxes are measured in $0.5\ {\rm arcsec}$ diameter apertures.

An adaptive aperture run was also completed on the F356W images by enabling FLUX\_AUTO \citep{Kron1980}, which better traces the total flux of extended sources. The flux ratio between the adaptive-aperture run and the F356W fixed-aperture run was then used to scale up the aperture fluxes to total values in every filter. As demonstrated by \cite{McLeod2024}, the Kron aperture does not fully capture all object flux, and so a further 10 per cent correction factor was then added to all fluxes. For cases in which the ratio of FLUX\_AUTO to the aperture flux was smaller than the expected correction for a point source (based on our empirical PSFs), a point-source aperture correction was instead applied.


\begin{figure*}
	\includegraphics{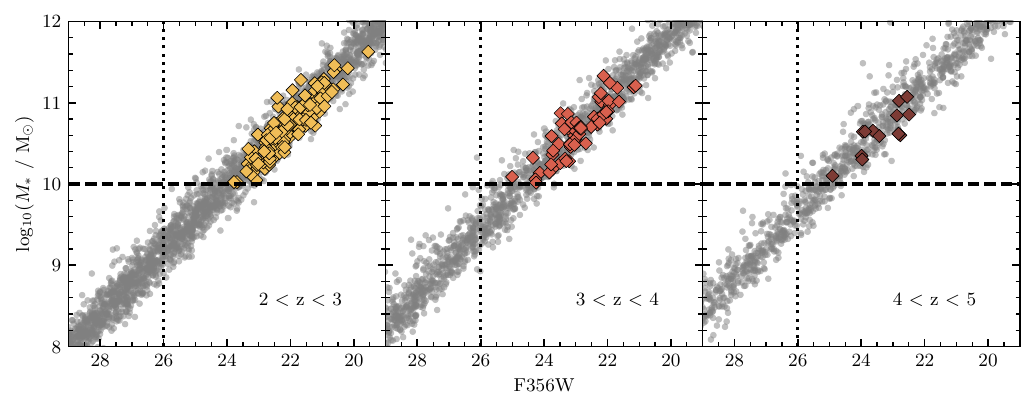}
    \caption{The F356W magnitudes and stellar masses of our {\sc Bagpipes} simulated quiescent galaxy population (grey), constructed as described in Section \ref{section:methods:masslimit}. Our massive quiescent candidates resulting from our selection process described in Section \ref{section:methods:selection} are plotted as diamonds and coloured by their redshift bin. Our chosen stellar mass limit of $M_*=10^{10}\,\mathrm{M_\odot}$ is included as a dashed line, and the corresponding magnitude limit for $>99$ per cent completeness, $\mathrm{F356W}=26$,  is included as a dotted line.}
    \label{figure:masslimit}
\end{figure*}

\subsection{Spectroscopy}\label{section:data:spectroscopy}

\subsubsection{The EXCELS survey}\label{section:data:spectroscopy:excels}

We use the \textit{JWST} EXCELS survey (GO 3543; PIs: Carnall, Cullen) as the primary spectroscopic companion to our photometric dataset. EXCELS observed ultra-deep, medium-resolution ($R=1000$) spectra in the UDS field using \textit{JWST} NIRSpec. Full details of the EXCELS target selection process and observing strategy are given in \cite{Carnall2024}. 

Massive quiescent galaxies at $z>3$ are the primary science targets for EXCELS, and a total of 14 candidates were selected to be observed in gratings G140M, G235M and G395M, covering a wavelength range of $1-5$ $\mu$m. All 14 of these candidates were recovered via our primary photometric selection process, described in Section \ref{section:methods:selection}.

In this work we make use of the updated reductions from Leung et al. (in prep.) for the 14 EXCELS $z>3$ objects. Briefly, these reductions use the \textit{JWST} pipeline with a custom \texttt{clean\_flicker\_noise} step to reduce the level 1b data products. Level 2a products are then reduced with the default \texttt{calwebb\_spec2} pipeline, except that  contaminated nods used for background subtraction were removed from the process. Pixels with certain quality bitmasks were automatically masked out and various other contaminants were manually masked out. Final 2D object spectra are then produced using the default \texttt{calwebb\_spec3} pipeline. Leung et al. (in prep.) use an optimal extraction method with a kernel that varies as a function of wavelength \citep{Horne1986} to extract 1D spectra, and join multiple gratings by downsampling, equalizing and averaging the overlapping regions. Each joined 1D spectrum is scaled to the aperture-corrected photometry by a Chebyshev polynomial of order 15, which is fitted to the ratio of the piecewise median to the best fit {\sc Bagpipes} SED.

A number of other, lower-redshift quiescent candidates in our sample (see Section \ref{section:methods:selection}) also have EXCELS spectra in either 1 or 2 of the above gratings. These generally provide spectroscopic redshifts for these objects, but in most cases do not provide access to the full suite of key rest-frame UV-optical emission and absorption features available for the 14 objects at $z>3$. We therefore only use the $z>3$ objects in our emission-line analysis in Section \ref{section:lineflux}.

For objects in our sample with EXCELS spectra, we adopt the redshifts measured by \cite{Carnall2024}. The improved reduction by Leung et al. (in prep.) also allows us to measure a redshift of $z=3.20$ with confidence flag 3 for PRIMER-EXCELS-55742, for which a spectroscopic redshift could not previously be measured.

\subsubsection{Archival spectroscopy}\label{section:data:spectroscopy:archival}

In addition to the EXCELS survey, we also cross-match  our quiescent sample (see Section \ref{section:methods:selection}) to archival \textit{JWST} NIRSpec data from the DAWN \textit{JWST} Archive (DJA)\footnote{\href{https://dawn-cph.github.io/dja}{https://dawn-cph.github.io/dja}}. The DJA products are reduced with in-house pipelines ({\sc grizli}\footnote{\href{https://github.com/gbrammer/grizli}{https://github.com/gbrammer/grizli}} and {\sc msaexp}\footnote{\href{https://github.com/gbrammer/msaexp}{https://github.com/gbrammer/msaexp}}). We use v4 of the DJA public NIRSpec data compilation, the full reduction process for which is described in \cite{deGraaf2024} and \cite{Heintz2024}.

We use a 0.3 arcsec search radius to cross-match against our sample, and search for all matches to each of our objects in the DJA catalogue. This produces a mixture of prism, medium-resolution and high-resolution grating NIRSpec data, often with multiple products available for each object. We perform our own independent visual inspections to confirm the spectroscopic redshifts made available by DJA using all available spectra for each object. The robustness of our spectroscopic redshifts for each object are flagged with an integer tag: 1 (uncertain), 2 (probable), 3 (secure) and 4 (very secure).

The archival spectroscopic data we make use of come from the following surveys: ``NIRCam-NIRSpec galaxy assembly survey - GOODS-S - part \#1'' (JADES, GTO 1180, PI: Daniel Eisenstein  \citealt{Eisenstein2017, dEugenio2025}); ``MIRI in the Hubble Ultra-Deep Field'' (SMILES, GTO 1207, PI: George Rieke, \citealt{Rieke2017, Alberts2024b}); ``NIRSpec WIDE MOS Survey - GOODS-S'' (GTO 1212, PI: Nora Luetzgendorf, \citealt{Ferruit2017a}); ``NIRSpec WIDE MOS Survey - COSMOS'' (GTO 1214, PI: Nora Luetzgendorf, \citealt{Ferruit2017b}); ``NIRSpec WIDE MOS Survey - UDS'' (GTO 1215, PI: Nora Luetzgendorf, \citealt{Ferruit2017c}); ``NIRCam-NIRSpec galaxy assembly survey - GOODS-S - part \#2'' (GTO 1286, PI: Nora Luetzgendorf, \citealt{Ferruit2017d}); ``Emission line galaxies beyond the limits of the Hubble UDF: Physical conditions in ultra-faint star forming galaxies'' (GO 1671, PI: Michael Maseda, \citealt{Maseda2021, Maseda2023}); ``The Stellar and Gas Content of Galaxies at Cosmic Noon'' (Blue Jay, GO 1810, PI: Sirio Belli, \citealt{Belli2021}, \citeyear{Belli2024}, \citealt{Davies2024}); ``Quiescent or dusty? Unveiling the nature of extremely red galaxies at $z>3$'' (GO 2198, PI: Laia Barrufet, \citealt{Barrufet2021,Barrufet2025}); ``How many quiescent galaxies are there at $3<z<4$ really?'' (GO 2565, PI: Karl Glazebrook, \citealt{Glazebrook2021, Nanayakkara2024}); ``Unveiling the Redshift Frontier with \textit{JWST}'' (JADES, GO 3215, PI: Daniel Eisenstein, \citealt{Eisenstein2023a, Eisenstein2023c}); ``A complete census of the rare, extreme and red: a NIRCam-selected extragalactic community survey with \textit{JWST}/NIRSpec'' (RUBIES, GO 4233, PI: Anna de Graaff, \citealt{deGraaff2023,deGraaf2024}); ``The CANDELS-Area Prism Epoch of Reionization Survey'' (CAPERS, GO 6368, PI: Mark Dickinson, \citealt{Dickinson2024}); ``The High-z Menagerie: A Rare Chance to Study the Early and Exotic Transient Universe'' (DD 6585, PI: David Coulter, \citealt{Coulter2024}).

\begin{table*}
  \caption{A breakdown of the free parameters \textsc{Bagpipes} uses to fit a model to our photometric data (see Section \ref{section:methods}). This includes the range and shape of their prior distributions. Logarithmic priors are all applied in base ten. For the Gaussian prior on $\delta$, the mean is $\mu=0$ and the standard deviation is $\sigma=0.1$. Visual representations of each prior distribution are shown on a linear x-axis and are not to scale.}
\begingroup
\setlength{\tabcolsep}{9pt} 
\renewcommand{\arraystretch}{1.2} 
\begin{tabular}{llllll}
\hline
Component & Parameter & Symbol / Unit & Range & \multicolumn{2}{l}{Prior} \ \\
\hline
General & Redshift & $z$ & (0, 10) & Uniform & \tophat\\
& Total stellar mass formed & $M_*\ /\ \mathrm{M_\odot}$ & (1, $10^{13}$) & Logarithmic & \logarithm\\[5pt]
& Stellar and gas-phase metallicities & $Z\ /\ \mathrm{Z_\odot}$ & (0.2, 3.5) & Logarithmic & \logarithm\\
\hline
Star-formation history & Double-power-law falling slope & $\alpha$ & (0.1, 1000) & Logarithmic & \logarithm\\[5pt]
& Double-power-law rising slope & $\beta$ & (0.1, 1000) & Logarithmic & \logarithm \\[5pt]
& Double power law turnover time & $\tau$ / Gyr & (0.1, 15) & Uniform & \tophat  \\
\hline
Dust attenuation & $V-$band attenuation & $A_V$ & (0, 8) & Uniform & \tophat\\[5pt]
& Deviation from \cite{Calzetti2000} slope & $\delta$ & ($-0.3$, 0.3) & Gaussian & \gaussian \\[5pt]
& Strength of 2175\AA\ bump & $B$ & (0, 5) & Uniform &  \tophat \\
\hline
\end{tabular}
\endgroup
\label{table:bagpipes}
\end{table*}

\subsection{ALMA interferometry}\label{section:data:alma}

We investigate observations of our massive quiescent galaxy sample taken by the Atacama Large Millimetre/submillimetre Array (ALMA), in order to assess any contamination from dusty star-forming galaxies (DSFGs). We make use of the A$^3$COSMOS database\footnote{\href{https://sites.google.com/view/a3cosmos}{https://sites.google.com/view/a3cosmos}}, the result of a project that homogenously reduces archival ALMA data in the COSMOS and GOODS South fields. A$^3$COSMOS provides uniform RMS continuum images and corresponding primary beam images for every ALMA pointing in each field. For more information about the data reduction pipeline see \cite{Liu2019} and \cite{Adscheid2024}. We use data release version 20250312, which includes all ALMA data made public before the 12\textsuperscript{th} March 2025.

We matched our massive quiescent galaxy sample to the A$^3$COSMOS database, noting all ALMA pointings that provide coverage of each of our passive candidates. For each candidate with coverage, we derive both an observed flux density and uncertainty separately for all covering images. We perform aperture photometry using an elliptical aperture with major and minor axes equal to double the clean (synthesized) beam FWHM axes. An aperture of this size encompasses 96 per cent of the flux from an unresolved source.

We homogenise each detected flux density and limiting flux density to ALMA band 6 (1.3 mm) by assuming a power-law relationship with wavelength. Considering the redshift regime of our sample, we assume the SEDs have a $\nu^3$ dependence.

We then reject any ALMA images which (i) have a clean beam FWHM major axis greater than 1.5 arcsec, to reduce the possibility of confusion with nearby sources, and (ii) have uncertainties greater than twice that of the deepest image available for each galaxy, to avoid the inclusion of excessively noisy data. We then combine fluxes and uncertainties from multiple ALMA observations for each galaxy via an inverse-variance weighted mean method. This returns a single 1.3 mm flux and uncertainty for each of our galaxies with good quality ALMA coverage.

\section{Methods}\label{section:methods}

\subsection{Stellar mass completeness limit}\label{section:methods:masslimit}

In targeting massive galaxies at high redshift, we decided on a desirable lower stellar-mass limit of $\mathrm{log_{10}}(M_*/\mathrm{M_\odot}) = 10$. This limit is widely used to distinguish massive and lower-mass galaxies and hence facilitates direct comparison with similar observation- or simulation-based studies. It also ensures an $\mathrm{SNR}\gtrsim20$ in F356W for more than 90 per cent of objects. We perform a simulation to convert this stellar-mass threshold into a limiting magnitude in F356W, in order to firstly confirm that our catalogues are complete down to this stellar mass limit, and to secondly avoid fitting all fainter objects in the PRIMER and JADES catalogues.

We derive this magnitude limit by first generating a million mock galaxies with {\sc Bagpipes}, using the same model configuration later used to fit each galaxy's SED (see Section \ref{section:methods:sedfit}). We make one modification to the priors listed in Table \ref{table:bagpipes}, setting an exponential prior on $A_\mathrm{V}$ with a scale parameter of 0.5. This is more physically realistic than the uninformative uniform prior used later on for fitting, as the vast majority of high-redshift massive quiescent galaxies have been shown to be relatively dust-poor (e.g., \citealt{Belli2019, Carnall2019b, Valentino2020, Carnall2023b}). We finally limit our mock population to quiescent objects only by requiring sSFR $< 0.2/t_\mathrm{H}(z)$ (see Section \ref{section:methods:selection}).

We show our mock galaxy population with gray points in Fig. \ref{figure:masslimit}. In practice, virtually all objects in our observed sample are at $z<5$, hence for the redshift bins $2<z<3$, $3<z<4$ and $4<z<5$, we derive a F356W magnitude limit below which we achieve $>99$ per cent completeness for objects at $\mathrm{log_{10}}(M_*/\mathrm{M_\odot}) > 10$. The largest of these limits is $\mathrm{F356W} \simeq 25.91$ at $4 < z <5$, therefore allowing us to confidently limit our sample to $\mathrm{F356W} < 26$ while ensuring $>99$ per cent completeness of the quiescent galaxy population at $\mathrm{log_{10}}(M_*/\mathrm{M_\odot}) > 10$ and $z<5$. Our mass and magnitude limits are shown with dashed and dotted lines respectively in Fig. \ref{figure:masslimit}.

Our F356W limit is more than 2 magnitudes brighter than the depth of our F356W imaging data, meaning that photometric incompleteness is negligible. We also include our observed quiescent sample from Section \ref{section:methods:selection} as coloured points in Fig. \ref{figure:masslimit}. It can be seen that these lie well within the region of our mock population, demonstrating that this is a realistic representation. 

\begin{figure*}

\hspace*{1.4cm}\includegraphics[width=0.925\textwidth]{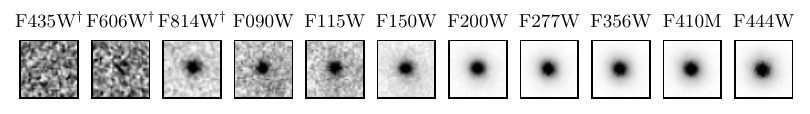}\\
\vspace{-0.3cm}
\includegraphics[width=\textwidth]{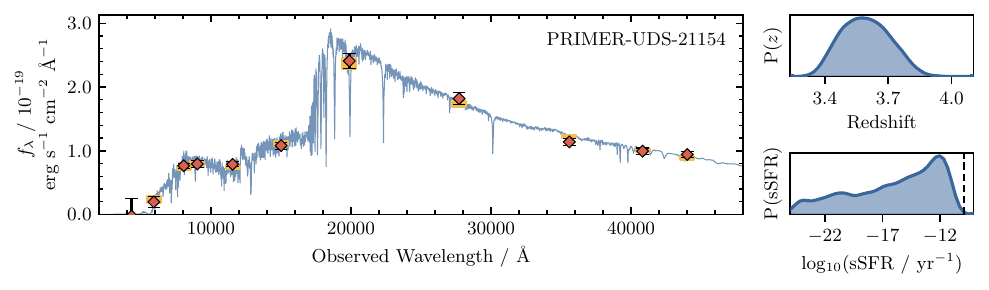}\\

\hspace*{1.4cm}\includegraphics[width=0.925\textwidth]{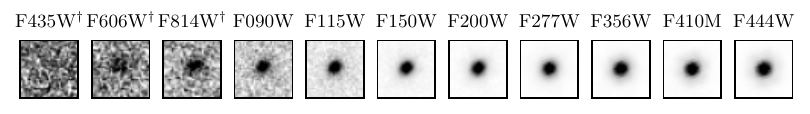}\\
\vspace{-0.3cm}
\includegraphics[width=\textwidth]{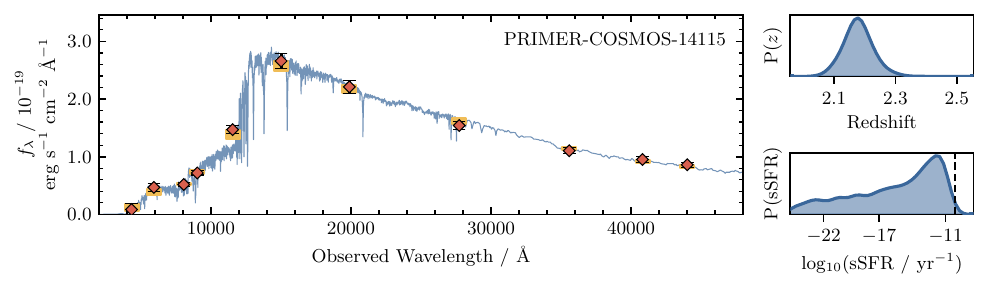}\\

\hspace*{1.4cm}\includegraphics[width=0.925\textwidth]{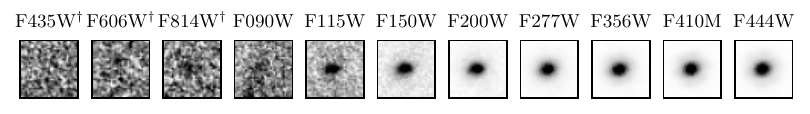}\\
\vspace{-0.3cm}
\includegraphics[width=\textwidth]{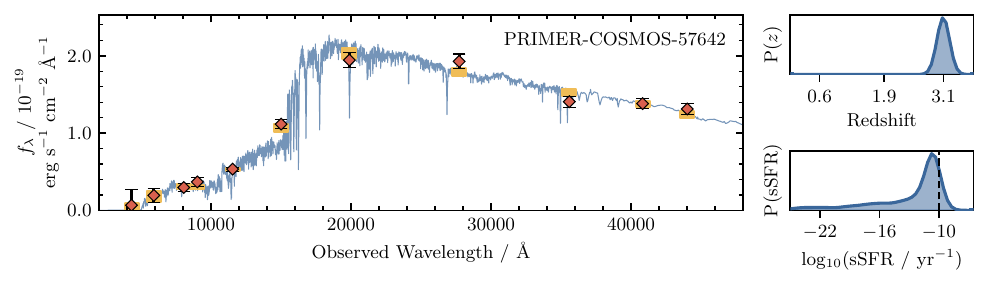}\\

\caption{Example SED fits to NIRCam imaging for candidates in our sample, along with posterior probability distributions for the redshift and sSFR parameters. We include robust candidates PRIMER-UDS-21154 (top), PRIMER-COSMOS-14115 (middle) and a non-robust candidate PRIMER-COSMOS-57642 (bottom). At the top of each example, galaxy imaging is shown in 2$\times$2 arcsec cutouts, where a dagger denotes an \textit{HST} ACS filter and its absence denotes a NIRCam filter. The observed photometry points are shown in red and our posterior median model is shown in blue. The yellow rectangles cover the 16th to 84th percentiles of the photometry posteriors, and the dashed line in the sSFR posterior represents the sSFR threshold used in selection (see Section \ref{section:methods:selection}).}
\label{figure:sedfit}
\end{figure*}

\subsection{Spectral energy distribution fitting}\label{section:methods:sedfit}

We fit the SED of each galaxy in the catalogues described in Section \ref{section:data:imaging} with $\mathrm{F356W}<26$ using the {\sc Bagpipes} spectral-fitting code \citep{Carnall2018}. We follow a similar parameter and prior structure to previous studies on high-redshift massive quiescent galaxies (\citealt{Carnall2020, Carnall2023a}, \citeyear{Carnall2024}). For a full list of the parameters and priors used for our {\sc Bagpipes} spectral fitting see Table \ref{table:bagpipes}. 

The model fitted by {\sc Bagpipes} uses stellar population models from \cite{Chevallard2016} (an updated version of the models from \citealt{Bruzual2003}) which itself uses stellar spectra from the MILES library \citep{Sanchez2006, Falcon2011} and stellar evolutionary tracks from \cite{Bressen2012} and \cite{Marigo2013}. 

The contribution from nebular emission is estimated  using the {\sc cloudy} photoionization code (\citealt{Ferland2017}; see section 3 of \citealt{Carnall2018}), assuming a stellar birth cloud lifetime of 10 Myr and ionisation parameter of $U=10^{-3}$. 

Dust attenuation is modelled using the approach from \cite{Salim2018}. This model includes a variable slope parametrized by a power-law deviation, $\delta$, from the \cite{Calzetti2000} model. Attenuation in the $V$-band, $A_{V}$, is allowed to vary from $0 - 8$ magnitudes, and we assume that attenuation is doubled for starlight that originates in stellar birth clouds (which we assume to have a lifetime of 10 Myr) and for nebular emission, compared to light from older stars inhabiting the wider ISM (e.g., \citealt{Charlot2000}). 

The star formation history is modelled via a double power law following \cite{Carnall2018, Carnall2019a}. Stellar and nebular metallicities are assumed to be the same and are varied with a logarithmic prior from $-0.7 < \mathrm{log}_{10}(Z/Z_{\sun}) < 0.4$. Absorption due to the intergalactic medium (IGM) is modelled using the prescription of \cite{Inoue2014}. Redshift is varied over a uniform prior covering the range $z = 0 - 10$. To fit our {\sc Bagpipes} model to the data, the {\sc pymultinest} package \citep{Buchner2014} is used to implement the {\sc multinest} nested sampling algorithm \citep{Skilling2004, Feroz2019}.

\subsection{Selection of quiescent candidates}\label{section:methods:selection}

First found as early-type galaxies in the local universe by \cite{Hubble1926}, the definition of  quenched/quiescent/passive galaxies are those which exhibit minimal star formation compared to their previously assembled stellar mass. Their resulting red colours give rise to the colour bimodality of the galaxy population \citep{Strateva2001}. For decades, various selection methods have been in use on multi-band photometric surveys to separate the quiescent galaxy population from the star-forming population. One of the most popular approaches in recent years has been to select quiescent galaxies via their rest-frame UVJ colours based on an empirical separation \citep{Williams2009}. 

More recently, powerful SED fitting codes have been developed \citep{Pacifici2012, Carnall2018, Johnson2021} which obtain Bayesian posterior probability distributions for galaxy physical parameters, including specific star formation rate (sSFR). This allows a more-direct separation of quiescent and star-forming galaxies, as well as indicating the robustness of individual classifications for specific objects.



In our selection of quiescent galaxies we explore two methods, both of which have been used extensively in the literature:

\begin{enumerate}
  \setlength\itemsep{1em}

    \item A specific star formation rate (sSFR) threshold, $\mathrm{sSFR} < 0.2/t_\mathrm{H}(z)$, where $t_\mathrm{H}(z)$ is the age of the Universe as a function of observed redshift (see \citealt{Gallazzi2014, Pacifici2016}). The criterion is applied by first calculating whether each sample in the sSFR posterior is below its respective threshold, which is dependent on redshift. We then require that $>50$ per cent of the joint posterior samples are both below their individual sSFR thresholds and above $z = 2$. We denote the latter criterion as $z_{50} > 2$. 
      
    \item A demarcation in UVJ space derived by \cite{Williams2009}. We select galaxies above $z_{50}=2$ as quiescent by requiring $(\mathrm{U-V}) > 0.88 \times (\mathrm{V-J}) + 0.69$. This is an empirical criterion based on $\sim 30,000$ galaxies, primarily in the local Universe. This particular expression was first proposed for galaxies at $z<0.5$ but was shown to be valid up to $z=3.75$ in \cite{Carnall2018} using pre-\textit{JWST} data from UltraVISTA. We also drop the horizontal and vertical cuts in UVJ space from \cite{Williams2009}, which are less appropriate at higher redshifts (e.g., \citealt{Park2023}). We use the posterior median U $-$ V and V $-$ J colours of each galaxy, as found by {\sc Bagpipes}, to determine its candidacy as quiescent.

\end{enumerate}

Next, we cut our photometric sample to galaxies with $\mathrm{log_{10}}(M_*/\mathrm{M_\odot}) > 10$, above which we have shown that our $\mathrm{F356W}~<~26$ sample is $>99$ per cent mass complete (see Section \ref{section:methods:masslimit}). In summary, our primary quiescent sample used throughout the rest of this work was selected as follows:

\begin{itemize}
    \item F356W < 26
    \item $z_{50} > 2$
    \item sSFR $< 0.2\ /\ t_\mathrm{H}$
    \item $\mathrm{log_{10}}(M_*\ /\ \mathrm{M_\odot}) > 10$
\end{itemize}

For method (i), which is our primary approach used throughout the rest of this paper, we further define a robust sub-sample of quiescent galaxies by requiring $>95$ per cent of the joint posterior samples to be below their individual sSFR thresholds and above $z = 1.75$. A redshift threshold of $z=1.75$ was chosen instead of $z=2$ to avoid the case where a redshift posterior that extends slightly below $z=2$ narrowly disqualifies an edge case. We will compare our primary method to the UVJ selection approach in method (ii) in Section \ref{section:sample:uvj}.

We finally visually inspected the \textsc{Bagpipes} fit and imaging data for each candidate (see Fig. \ref{figure:sedfit}). We exclude any poor fits to the photometry and cases in which the imaging is badly contaminated by artefacts or partly lies off the edge of the detector. This process removes 38 sources and finally results in a sample of 225 candidates, which is presented and discussed in Section \ref{section:sample}.

\begin{table}
\caption{A breakdown of the free parameters used to fit Equation \ref{equation:linemodel} to the continuum subtracted spectra for our 14 EXCELS $z>3$ galaxies. This includes the range and shape of their prior distributions. Logarithmic priors are all applied in base ten.}
\begingroup
\setlength{\tabcolsep}{9pt} 
\renewcommand{\arraystretch}{1.2} 
\begin{tabular}{llll}
\hline
Parameter / Unit & Range & \multicolumn{2}{l}{Prior} \\
\hline
$\sigma\ /\ \mathrm{\mathring{A}}$ & (0, 10) & Uniform & \tophat \\[5pt]
$v_{\mathrm{gas}}\ /\ \mathrm{km\ s^{-1}}$ & ($-$500, 500) & Uniform & \tophat \\[5pt]
$F_i\ /\ \mathrm{erg\ s^{-1}\ cm^{-2}}$ & ($10^{-21}$, $10^{-10}$) & Logarithmic & \logarithm \\
\hline
\end{tabular}
\endgroup
\label{table:linemodel}
\end{table}

\begin{table*}
\centering
\caption{An excerpt from our full massive quiescent galaxy candidate list, selected as described in Section \ref{section:methods:selection}, the whole of which is available as supplementary online material. The objects PRIMER-UDS-21154, PRIMER-COSMOS-14115 and PRIMER-COSMOS-57462 are visualised as examples in Figure \ref{figure:sedfit}.}
\label{table:excerpt}
\begin{tabular}{lcccccccc}\\\hline
Survey & ID & RA / deg & DEC / deg & $z_{50}$ & $\mathrm{log}_{10}({M_*\ /\ \mathrm{M_\odot}})$ &  F356W &Robust & Comments \\
\hline
PRIMER-UDS&19817 & 34.306810 & -5.294570 & $2.37_{-0.10}^{+0.13}$ & $10.57_{-0.06}^{+0.06}$ & $22.7_{-0.1}^{+0.1}$ & True \\[5pt]
PRIMER-UDS&21154 & 34.344910 & -5.292580 & $3.59_{-0.11}^{+0.12}$ & $10.84_{-0.02}^{+0.03}$ & $22.2_{-0.1}^{+0.1}$ & True \\[5pt]
PRIMER-UDS&26000 & 34.264260 & -5.284750 & $2.40_{-0.08}^{+0.12}$ & $10.54_{-0.05}^{+0.05}$ & $22.4_{-0.1}^{+0.1}$ & False \\[5pt]
PRIMER-UDS&26528 & 34.406730 & -5.284430 & $2.69_{-0.09}^{+0.08}$ & $10.52_{-0.05}^{+0.06}$ & $22.3_{-0.1}^{+0.1}$ & False \\[5pt]
PRIMER-UDS&26591 & 34.233620 & -5.283810 & $3.93_{-0.25}^{+0.19}$ & $10.68_{-0.06}^{+0.05}$ & $23.2_{-0.1}^{+0.1}$ & False \\[5pt]
... & ... & ... & ... & ... & ... & ... & ... & ...\\
\hline
\end{tabular}
\end{table*}

\subsection{Emission-line flux determination}\label{section:methods:lineflux}

\subsubsection{Continuum subtraction}\label{section:methods:lineflux:subtraction}

In the latter part of this work (Section \ref{section:lineflux}) we focus on emission-line diagnostics for 14 EXCELS $z>3$ quiescent galaxy spectra, in order to investigate the presence of AGN in our quiescent sample. To calculate emission line-fluxes, we first construct continuum-subtracted spectra. We make use of the EXCELS reductions described in Section \ref{section:data:spectroscopy:excels} and subtract the continuum models from Leung et al. (in prep). In brief, Leung et al. (in prep) construct continuum models by fitting the full spectrum of each EXCELS $z>3$ galaxy with {\sc Bagpipes}, following a similar prior structure to our photometric fits described in Section \ref{section:methods:sedfit}. We use these fits to re-generate the spectral model posterior for each EXCELS galaxy with the nebular emission component deactivated. We take the posterior median value at each wavelength as the best fit continuum flux, and the standard deviation as the uncertainty on the model. We subtract these continuum models from the calibrated and reduced 1D spectra, and combine the uncertainties on the data and continuum model in quadrature. Typically, the observed flux uncertainty is much larger, however the model uncertainty can sometimes dominate, especially around very faint emission lines.

\subsubsection{Emission line modelling}\label{section:methods:lineflux:modelling}

To model emission lines in continuum-subtracted spectra, we use a Gaussian emission-line model with components for each individual emission line, allowing each line to be fitted simultaneously. We fit the model to rest-frame spectra, which are found by first de-redshifting the observed wavelengths given the galaxy's spectroscopic redshift, $z$ (see Section \ref{section:data:spectroscopy:excels}), $\lambda = \lambda^{\mathrm{obs}} / (1+z)$, and then multiplying the observed $f_\lambda$ by $(1+z)$ when calculating line fluxes. The total emission line model is given by 
\begin{equation}\label{equation:linemodel}
    f_\lambda = \sum_{i}\frac{F_{i}}{\sqrt{2\pi}\sigma}\times \mathrm{exp}\left[-\frac{1}{2}\frac{(\lambda - \lambda_{i})^2}{\sigma^2}\right],
\end{equation}

\noindent where $F_i$ is the total flux of each emission line and $\lambda_{i}$ is its central wavelength, given by
\begin{equation}
    \lambda_{i} = \lambda_{i}^{\mathrm{rest}} \times \left[1 + v_{\mathrm{gas}}/c\right],
\end{equation}

\noindent where $\lambda_{i}^{\mathrm{rest}}$ is the rest-frame wavelength of the emission line, $c$ is the speed of light and $v_{\mathrm{gas}}$ is a line velocity shift. This last parameter acts as a small perturbation to the emission line central wavelengths to allow some flexibility for both the gas velocity and spectroscopic redshift. The model is fitted with a common line width, $\sigma$, for each Gaussian component. The flux of each line is a free parameter in our model, hence our model contains $2 + n_{\mathrm{lines}}$ free parameters.

For the purpose of investigating the presence of AGN, we include emission lines H$\beta$, {\sc [Oiii]}$\lambda4959$, {\sc [Oiii]}$\lambda5007$, H$\alpha$, {\sc [Nii]}$\lambda6548$ and {\sc [Nii]}$\lambda6584$. We fix the strength of {\sc [Oiii]}$\lambda4959$ to $1/2.98$ the strength of {\sc [Oiii]}$\lambda5007$, and {\sc [Nii]}$\lambda6548$ to $1/3.049$ the strength of {\sc [Nii]}$\lambda6584$. We fit the model within wavelength ranges of width 400$\ $Å roughly centred on the emission lines; 4800$\ $Å $< \lambda_{\mathrm{rest}} <$ 5200$\ $Å and 6350$\ $Å $< \lambda_{\mathrm{rest}} <$ 6750$\ $Å. Our resulting model has 6 free parameters, which we sample using the {\sc nautilus}\footnote{\href{https://github.com/johannesulf/nautilus}{https://github.com/johannesulf/nautilus}} nested sampling tool \citep{Lange2023}. We use the priors described in Table \ref{table:linemodel} to fit our emission line model. Where no emission line is apparent in the data we set the upper limit for the flux at the 3-$\sigma$ level: the flux value at the 99.7\textsuperscript{th} percentile of the posterior.

Towards placing the 14 $z>3$ EXCELS galaxies in the context of emission line diagnostic diagrams (e.g., BPT diagram; \citealt{Baldwin1981}), we also construct posteriors for the flux ratios {\sc [Oiii]}$\lambda5007$ / H$\beta$ and {\sc [Nii]}$\lambda6584$ / H$\alpha$ from the ratio of our fitted line-flux posteriors. We also make a determination of the H$\alpha$ equivalent width of each galaxy, EW$_\mathrm{H\alpha}$, by dividing its flux by the level of the rest-frame continuum. To estimate the continuum under H$\alpha$, $c_{\mathrm{H}\alpha}$, we average the flux from two equal-width bins of width 150$\ $Å either side of the H$\alpha$ line; 6350$\ $Å $< \lambda_{\mathrm{rest}} <$ 6500$\ $Å and 6600$\ $Å $< \lambda_{\mathrm{rest}} <$ 6750$\ $Å. We calculate the rest-frame equivalent width by dividing the fitted $\mathrm{H\alpha}$ flux by the continuum, \begin{equation}\label{equation:ew} \mathrm{EW}_{\mathrm{H}\alpha} = \frac{F_{\mathrm{H}\alpha}\ /\ \mathrm{erg\  s^{-1}\ cm^{-2}}}{c_{\mathrm{H}\alpha}\ /\ \mathrm{erg\  s^{-1}\ cm^{-2}\  \text{\AA}^{-1}}}. \end{equation}

\begin{figure}
\includegraphics[width=\columnwidth]{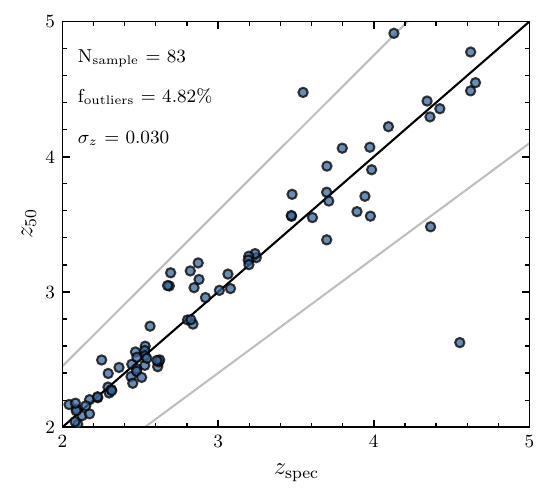}
\caption{A comparison between photometric and spectroscopic redshifts for the 83 of our massive quiescent galaxy candidates with robust spectroscopic redshifts (flag 3 or 4; see Section \ref{section:data:spectroscopy}). The value of $\sigma_z$ and the fraction of catastrophic outliers are provided in the legend. Our \textsc{Bagpipes} photometric redshifts can be seen to be highly robust.}
\label{figure:zzplot}
\end{figure}

\begin{figure*}
\includegraphics[width=\textwidth]{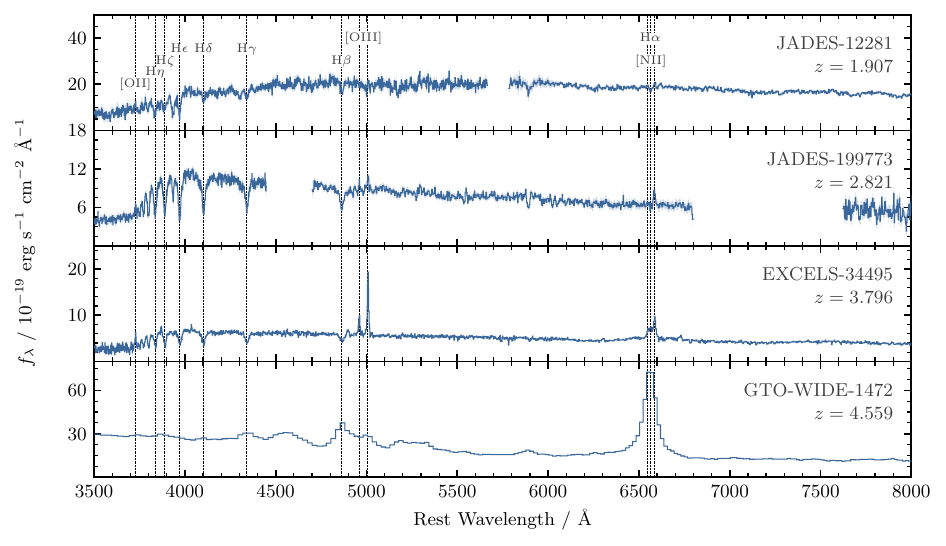}
\caption{A representative sub-sample of the 10 objects from our photometric sample of massive quiescent galaxy candidates with both X-ray detections and spectroscopic data (see Section \ref{section:sample:contamination:xray}). Fluxes and wavelengths are displayed in the rest frame. We include the spectroscopic campaign, spectroscopic ID and spectroscopic redshift for each object. From top-down, these correspond to our photometric IDs: JADES-GOODSS-74806, JADES-GOODSS-55942, PRIMER-UDS-35941, PRIMER-UDS-116064. It is clear that some sources are stellar dominated in the rest-frame optical, despite having X-ray counterparts.}
\label{figure:xrayspectra}
\end{figure*}

\section{Photometric Sample}\label{section:sample}

Focusing on our primary sample of sSFR selected $z>2$ quiescent galaxies (see Section \ref{section:methods:selection}), we find 225 objects of which 150 belong to our robust sub-sample. From these 225 objects, 152 (113 robust) are identified at $2<z<3$, 58 (31 robust) at $3<z<4$ and 13 (6 robust) at $4<z<5$. Two non-robust candidates were identified at $z>5$. The reliability of these extreme-redshift candidates is discussed in Section \ref{section:sample:zgt5objects}.

We provide the coordinates, F356W magnitudes and photometric redshifts for our massive quiescent galaxy candidates in Table \ref{table:excerpt} (the full version of which is available as supplementary online material), alongside other physical properties such as stellar mass and sSFR. We present the SEDs and cutout images for three example galaxy candidates in Fig. \ref{figure:sedfit}. The top panels show a $2\times2$ arcsec imaging cutout of each galaxy in every photometric filter used in the SED fit, where a dagger denotes an \textit{HST} ACS filter. The main panels show the {\sc Bagpipes} fit to the photometry, alongside posterior probability distributions for the redshift and sSFR parameters, where the median sSFR threshold for quiescence is included as a dashed line.

\subsection{Checking for contamination}\label{section:sample:contamination}

With the aim of constructing the purest possible sample of massive quiescent galaxy candidates, in this section we investigate the presence of possible contaminants. We approach this using three separate methods, each designed to tackle a common contaminant for massive quiescent galaxy samples: redshift outliers, AGN and DSFGs. 

The first method, described in Section \ref{section:sample:contamination:spectra}, involves a crossmatch of our photometric sample to all public \textit{JWST} NIRSpec spectroscopy using the DJA (see Section \ref{section:data:spectroscopy:archival}). The second, described in Section \ref{section:sample:contamination:xray}, involves a crossmatch to \textit{Chandra} X-ray source catalogues to investigate the presence of AGN. The third, described in Section \ref{section:sample:contamination:alma}, involves a crossmatch to ALMA imaging to investigate the fraction of DSFG interlopers in our passive catalogue. We also considered a crossmatch to radio source catalogues to investigate potential contamination by AGN in our sample, however it is well established that no strong link exists between radio activity and AGN domination of galaxy optical spectra (e.g., \citealt{Best2005a, Best2005b}). At the end of this process, in Section \ref{section:sample:contamination:combine}, we calculate a total contamination fraction for use in our number density analysis in Section~\ref{section:densities}.

\subsubsection{Spectroscopic data}\label{section:sample:contamination:spectra}

\begin{figure*}
\includegraphics[width=\textwidth]{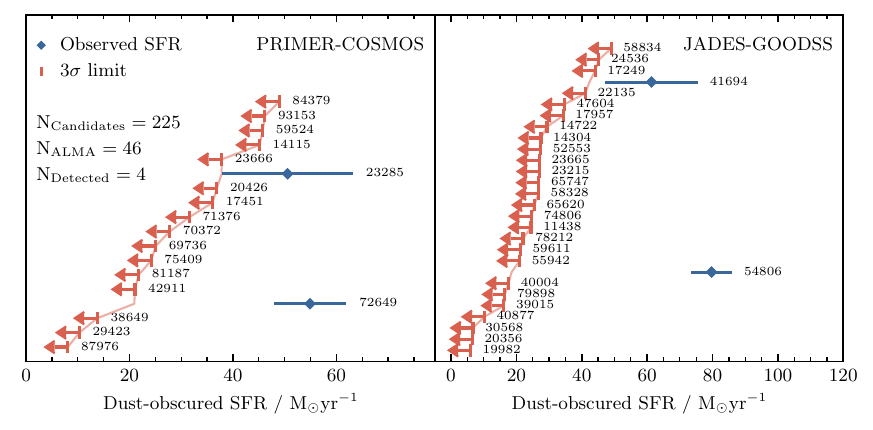}
\caption{The SFRs of a sub-sample of our massive quiescent galaxy sample, as derived from deep ALMA coverage (see Section \ref{section:data:alma}), which we define as $3\sigma$ sensitivity $<50\ \mathrm{M_\odot\ yr^{-1}}$ (see Section \ref{section:sample:contamination:alma}). This is available for 46 out of 225 objects. We include our 4 detections (defined as SFR $> 3\sigma$) as blue points and non-detections as red upper limits at their respective $3\sigma$ limits.}
\label{figure:almadepths}
\end{figure*}

As described in Section \ref{section:data:spectroscopy} we obtain good quality (flag 3 or 4) spectroscopic redshifts for 83 objects in our photometric sample of 225 massive quiescent galaxy candidates. For these 83 objects, we present a comparison between photometric and spectroscopic redshifts in Fig. \ref{figure:zzplot}. To test the quality of our photometric redshifts, we include a calculation of $\sigma_z$ and the catastrophic outlier rate. The value of $\sigma_z$ is defined as $1.483 \times \mathrm{MAD(dz)}$ where MAD(dz) is the Median Absolute Deviation of $\mathrm{dz} = (z_{50} - z_{\mathrm{spec}})/(1+z_{\mathrm{spec}})$. We classify any object with $|\mathrm{dz}| > 0.15$ to be a catastrophic outlier (and indeed all 4 such objects fail our visual inspection as described below).  Using the above definitions, we calculate $\sigma_z=0.03$ and a catastrophic outlier fraction of 4.82 per cent (4 out of 83 objects). An outlier fraction of $<5$ per cent and small $\sigma_z$ confirms that our photometric redshifts are highly robust.

In addition to identifying any redshift outliers, we also visually identify any spectra that clearly do not correspond to passive galaxies.  Of the 83 massive quiescent galaxy candidates with spectroscopic data, we visually identify six contaminants, showing no evidence for a Balmer break (3/6 are robust). From these, four are the redshift outliers identified above. This allows us to estimate a contamination fraction from spectral validation of 7.2 per cent (6 out of 83 objects).

\subsubsection{X-ray data}\label{section:sample:contamination:xray}

Sources for which bright AGN emission dominates the rest-frame optical can also contaminate samples of massive quiescent galaxies. An approach for estimating the fraction of such AGN in our sample is to search for sources of X-rays, which are produced in the hot corona surrounding an accreting black hole. We use source catalogues from the X-ray surveys X-UDS \citep{Kocevski2018}, $Chandra$ COSMOS-Legacy \citep{Civano2016} and $Chandra$ DFS-7Ms \citep{Luo2017}.

We crossmatch our full $\mathrm{F356W}<26$ photometric catalogue with these X-ray source catalogues within a radius of 1.5 arcsec \citep[e.g.,][]{Almaini2025}, retaining only the closest match. We find 25 matches with candidates from our massive quiescent galaxy sample of 225 objects. Out of these 25 massive quiescent galaxies, 14 are robust. In addition, there are proportionally far more matches in GOODS South due to the availability of ultra-deep $Chandra$ imaging; 14 out of 25 X-ray source matches are in GOODS South, whereas only 42 of our 225 quiescent candidates are in GOODS south.

This would correspond to an AGN contamination fraction of 11 per cent. However, upon inspecting the X-ray detected passive galaxy candidates with corresponding spectroscopic data (10 out of 25), the picture becomes less clear. In Fig. \ref{figure:xrayspectra}, we show the spectra for a representative sub-sample of these 10 objects. It can be seen that an X-ray detection does not imply AGN dominance in the rest-frame optical. Some X-ray sources are clearly AGN dominated, to the point where a passive description is not appropriate (e.g., GTO-WIDE-1472 in the bottom panel), however some of the spectra show what would otherwise be clearly identified passive galaxies, with strong Balmer breaks and no emission lines (e.g., JADES-12281 and JADES-199773 in the top two panels, the latter of which is also included in the quiescent sample of \citealt{Baker2025a} despite being X-ray detected). Some others, while bearing emission lines, are clearly still dominated by their stellar continuum component (e.g., EXCELS-34495 in the third panel).

Taking this into account, we choose not to treat all objects with X-ray detections as contaminants, to avoid the exclusion of clearly robust high-redshift massive quiescent galaxies that also contain an X-ray source. The X-ray sources for which the rest-optical spectra are clearly AGN dominated (e.g., GTO-WIDE-1472) were all previously flagged during our visual inspection process in Section \ref{section:sample:contamination:spectra}.

\subsubsection{Millimetre data}\label{section:sample:contamination:alma}

Dusty star-forming galaxies are also a possible source of contamination in our passive sample, due to the degeneracy between the optical SEDs of DSFGs and passive galaxies (e.g., \citealt{Santini2019, Santini2021}). To quantify the contamination from DSFGs, we investigate mm/sub-mm observations of our passive galaxy sample in the A$^3$COSMOS database (see Section \ref{section:data:alma}). These wavelengths probe the regime in which reprocessing of UV light by dust is dominant. In order to derive a DSFG contamination fraction, we seek to know how many of our massive quiescent galaxies are detected as strong mm/sub-mm sources, and how many have deep enough ALMA coverage to reliably be considered as non-detections. 

We begin with the 1.3mm flux densities for each of our massive quiescent galaxy candidates with ALMA coverage (see Section \ref{section:data:alma}). To convert these into SFRs, we adopt the relationship from \cite{Dunlop2017},
\begin{equation}\label{equation:almasfr}
    \mathrm{SFR\ /\ M_\odot\ yr}^{-1} \simeq 0.3 \times (F_{1.3\mathrm{mm}}\ /\ \mu \mathrm{Jy}).
\end{equation}

At this point, we reject any objects for which we lack the required sensitivity for a reliable detection or non-detection, specifically requiring objects to have $3\times\mathrm{SFR_{1\sigma}} < 50\,\mathrm{M_\odot yr^{-1}}$.

This results in four ALMA detected galaxies at a level above $3\sigma$ out of 46 galaxies imaged with sufficient depth, resulting in an ALMA detection rate of 8.7 per cent (1/4 is a robust candidate). This sample is shown in Fig. \ref{figure:almadepths}. PRIMER-COSMOS-72649 and JADES-GOODSS-54806 are detected well above the $3\sigma$ level, whereas PRIMER-COSMOS-23285 and JADES-GOODSS-41694 are only marginally detected. Despite the two marginal detections, we choose to treat all four as contaminants, to make the most conservative estimate of our DSFG contamination fraction. The detection of PRIMER-COSMOS-72649 is consistent with the COSMOS DSFG sample from \cite{Liu2025}, in which it also appears.

\subsubsection{Total contamination rate}\label{section:sample:contamination:combine}

In this sub-section, we have investigated the prevalence of contaminants in our 225-object photometric sample of $z>2$ massive quiescent galaxy candidates. In Section \ref{section:sample:contamination:spectra} we identified six contaminants from a sub-population of 83 objects with spectroscopic data, and in Section \ref{section:sample:contamination:alma} we have identified four contaminants  from a sub-population of 46 objects with ALMA observations. Considering both of these sub-populations together, 20 objects have both spectroscopy and ALMA data, only one of which is identified as a contaminant in both. Here we compute the implied total contamination fraction for our sample arising from these results.

To compute the total contamination fraction we adopt an approximate Bayesian computation methodology, an approach from the field of simulation-based inference. We first draw samples from prior distributions for P(S), the probability of an object in our photometric sample being a spectroscopic contaminant, P(A), the probability of an object being an ALMA contaminant, and P(A $\cap$ S), the probability of an object being both. The former two are allowed to vary between 0 and 1, whereas the latter is allowed to vary across the whole range of possible intersections for two events: at minimum the larger of 0 and P(A) $+$ P(S) $-$ 1, and at maximum the smaller of P(A) and P(S). We draw 50 million samples from these priors.

\begin{figure}
    \includegraphics[width=\columnwidth]{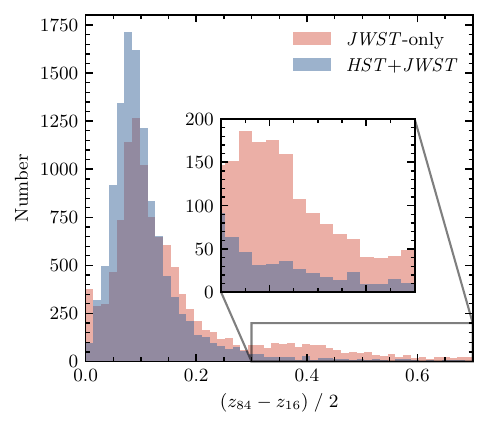}
    \caption{Histograms over the fitted redshift posterior width for galaxies with $z_{50} > 2$ in our main \textit{HST}+\textit{JWST} run (regardless of star-forming or passive classification). We include redshift posterior data from analyses using only \textit{JWST} NIRCam data (\textit{JWST}-only; red) and \textit{JWST} NIRCam plus \textit{HST} ACS data (\textit{HST}+\textit{JWST}; blue). We zoom into the wide-posterior region where `\textit{JWST}-only' galaxies are $\simeq4\times$ more numerous than `\textit{HST}+\textit{JWST}' galaxies.}
    \label{figure:zposteriordist}
\end{figure}

\begin{figure*}
    \includegraphics[width=\textwidth]{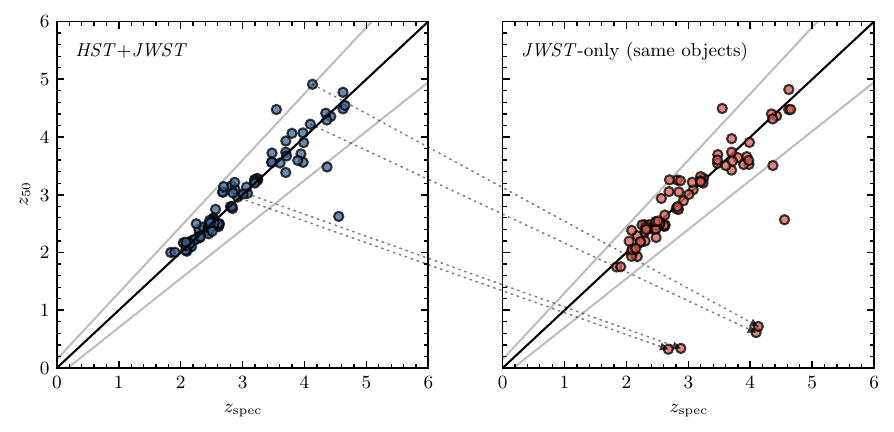}
    \caption{A comparison between the photometric redshifts derived in our `\textit{JWST}-only' and `\textit{HST}+\textit{JWST}' runs for the same spectroscopic objects shown in Fig. \ref{figure:zzplot}. In both plots we use the same sample as described in Section \ref{section:sample:contamination:spectra}, those selected as high-redshift massive quiescent galaxy candidates in our main \textit{HST}+\textit{JWST} run that have corresponding spectroscopic data. We use dashed arrows to annotate the 4 galaxies that move to $z_{50} < 2$ in the \textit{JWST}-only run, and which would therefore have been missed in the absence of \textit{HST} data.}
    \label{figure:hstzzplot}
\end{figure*}

For each draw we then simulate a population of 225 objects, assigning objects as contaminants at random, following the prior probabilities for that draw. We then assign 83 mock objects at random to be observed spectroscopically, and 46 at random to be observed by ALMA, enforcing an overlap of 20 objects between the two sub-samples. From these mock populations we then discard all except those which result in the same number of identified spectroscopic contaminants (six objects), ALMA contaminants (four objects), and the same overlap between the two (one object) as our observed sample.

The values of P(A), P(S) and P(A $\cap$ S) corresponding to these retained populations represent posterior probability distributions for these quantities in our observed sample. We obtain P(A) = $9.7^{+4.6}_{-3.6}$ per cent, P(S) = $8.0^{+3.1}_{-2.6}$ per cent, and P(A~$\cap$~S) = $4.0^{+2.9}_{-1.9}$ per cent. Finally, we obtain a total contamination fraction for our photometric sample of $12.9^{+4.0}_{-3.1}$ per cent.

We repeat the above analysis for our non-robust and robust sub-samples separately, deriving contamination fractions of $24.7^{+6.8}_{-6.9}$ and $8.7^{+4.0}_{-2.7}$ per cent respectively. These results are in good agreement with the sample contamination fractions that would be predicted from the photometric quiescent probabilities for each object derived from our joint posterior probability distributions for redshift and sSFR (see Section \ref{section:methods:selection}). The average quiescent probability is 0.86 in our full sample, 0.73 in the non-robust sub-sample, and 0.97 in the robust sub-sample. The largest discrepancy is in the robust sub-sample, where we estimate a true ALMA+spectroscopic contamination fraction more than twice as large as our prediction from photometric fitting alone, though the expectation from photometric fitting is still within $\simeq2\sigma$.

\subsection{The impact of \textit{HST} data on $\mathbf{z>2}$ quiescent galaxy selection}\label{section:sample:hst}

In Section \ref{section:data:imaging}, we restricted our photometric galaxy catalogue to only those with full coverage in the \textit{HST} ACS bands F435W, F606W and F814W, in addition to our eight NIRCam bands. This is contrary to many similar studies in the recent literature that use much larger imaging areas, much of which only benefits from NIRCam photometry. In this section we investigate the impact of \textit{HST} data on the selection of high-redshift massive quiescent galaxies, by comparing our fiducial results (`\textit{HST}+\textit{JWST}') with an identical analysis on the same catalogue except for the removal of these filters. It might be expected that the lack of \textit{HST} ACS optical filters may lead to (i) poorer $z_{50}$ accuracy and (ii) less sensitivity to UV star formation. In the following analysis, we explore both of these effects, finding both to be significant.

We complete a `\textit{JWST}-only' SED-fitting run over the same area excluding the \textit{HST} photometric points, and select massive quiescent galaxies using the exact same method as was used in our `\textit{HST}+\textit{JWST}' run (see Section \ref{section:methods:selection}).

For the \textit{JWST}-only run we identified 202 (124 robust) passive candidates, compared to 225 (150 robust) in the \textit{HST}+\textit{JWST} run, representing a 10 per cent decrease of $z>2$ massive quiescent galaxy candidates (23 objects), and a 17 per cent decrease in the robust sub-sample. In addition, out of these 202 passive galaxies, only 185 are in common with the \textit{HST}+\textit{JWST} run, meaning that in reality 18 per cent (40 objects) are lost as a result of the lack of \textit{HST} optical filters. In addition, the remaining 17 candidates identified solely in the \textit{JWST}-only run are most likely spurious, representing a false positive fraction of $\simeq 8$ per cent.

If we consider only our robust sub-sample, the \textit{JWST}-only run is able to reproduce 110 of the 150 robust candidates from our \textit{HST}+\textit{JWST} run, representing a 27 per cent decrease of truly robust candidates. The remaining 14 of the 124 robust candidates in the \textit{JWST}-only run are incorrectly identified as robust, and three of these are entirely misidentified as quiescent, belonging to the sub-sample of 17 above that are not selected at all in our \textit{HST}+\textit{JWST} run.

In the following sub-sections we consider the objects both lost and gained in the \textit{JWST}-only run in more detail. We first discuss the 40 objects that appear in our \textit{HST}+\textit{JWST} run but not in our \textit{JWST}-only run in Section \ref{section:sample:hst:candidateslost}. We then investigate the 17 likely spurious candidates that appear in our \textit{JWST}-only run but not our \textit{HST}+\textit{JWST} run in Section \ref{section:sample:hst:candidatesgained}.

\subsubsection{Objects in \textit{HST}+\textit{JWST} but not \textit{JWST}-only}\label{section:sample:hst:candidateslost}

A main contributor to the loss of candidates in the \textit{JWST}-only run is the tendency for galaxies to be placed at lower redshift compared with the \textit{HST}+\textit{JWST} run, hence falling below our $z_{50} > 2$ cut. 

\begin{figure*}
    \includegraphics[width=\textwidth]{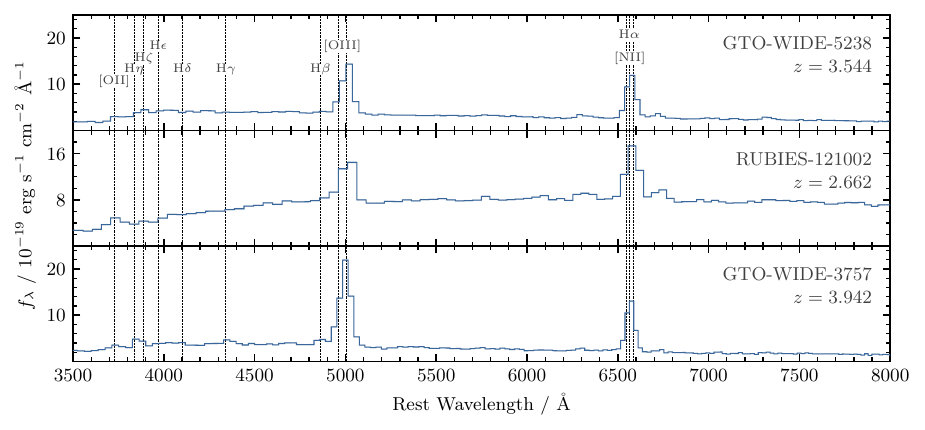}
    \caption{Spectra of galaxies selected as quiescent by our `\textit{JWST}-only' run but not selected as quiescent in our flagship `\textit{HST}+\textit{JWST}' run (see Section \ref{section:sample:hst}). Each spectrum contains strong emission-lines and shows no evidence justifying a passive description. This provides further evidence that the omission of \textit{HST} ACS bands introduces a significant amount of contamination into high-redshift quiescent galaxy samples. From the top down these correspond to photometric IDs PRIMER-UDS-62083, PRIMER-UDS-93415 and PRIMER-UDS-96929.}
    \label{figure:spuriousspectra}
\end{figure*}

\begin{figure*}
    \includegraphics[width=\textwidth]{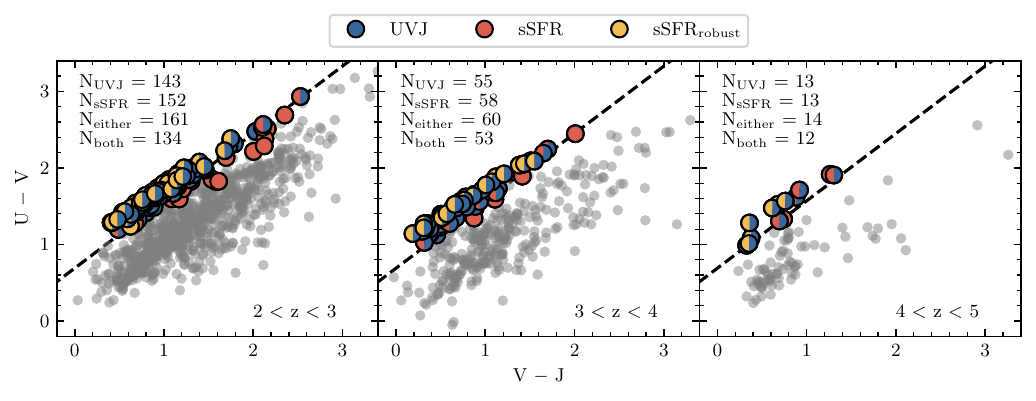}\\
    \vspace{-1.1cm}
    \includegraphics[width=\textwidth]{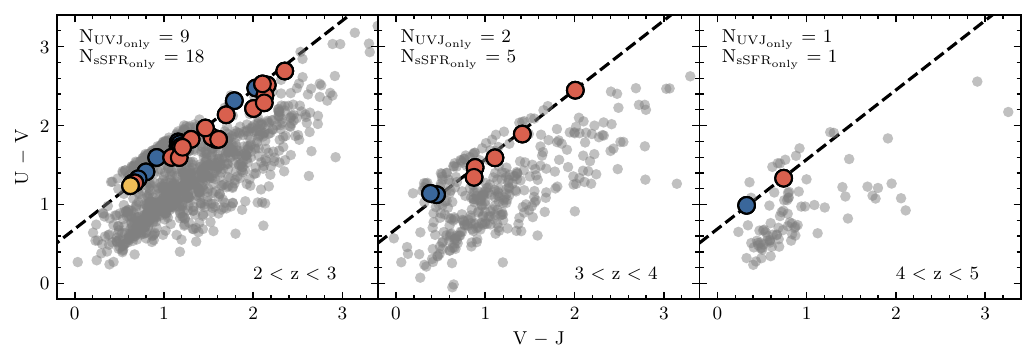}
    \caption{UVJ colour-space diagrams. The entire $\mathrm{F356W}<26$ photometric catalogue described in Section \ref{section:data:imaging} is plotted as grey points. In the upper panel we highlight any galaxies selected as quiescent by one or multiple of the UVJ, sSFR and sSFR$_\mathrm{robust}$ selection methods described in Section \ref{section:methods:selection}. Where multiple methods selected the same galaxy, the marker is bicoloured (all sSFR$_\mathrm{robust}$ objects also qualify for the normal sSFR selection). The black dashed lines denotes the UVJ criterion we use, which is the $z=0-0.5$ selection from \protect\cite{Williams2009} without their horizontal and vertical cuts. We include the redshift range and the total number of galaxies selected via each/both/either methods. In the lower panel we highlight quiescent candidates that are selected by only one of the UVJ and sSFR selection methods. Overall, the two methods agree extremely well ($\gtrsim90$ per cent common objects) across our whole redshift range.}
    \label{figure:uvjselection}
\end{figure*}

This is likely in part due to an increase in redshift uncertainty throughout the sample. In Fig. \ref{figure:zposteriordist} we compare the redshift posterior distribution widths (16\textsuperscript{th}$-$84\textsuperscript{th} percentiles) for both runs, including all objects with $z_{50} > 2$ in the \textit{HST}+\textit{JWST} run regardless of star-forming vs passive classification. It is clear that the redshift parameter in the \textit{JWST}-only run is much less well constrained. As shown in the zoom-in panel of Fig. \ref{figure:zposteriordist}, the wide-posterior region is dominated by the \textit{JWST}-only run. Indeed, 21 out of 40 objects appearing in \textit{HST}+\textit{JWST} but not \textit{JWST}-only are fitted with $z_{50} < 2$ in the \textit{JWST}-only run. The other 19 objects, while still fitted with $z_{50} > 2$, were classified as star-forming in the \textit{JWST}-only run.

To further compare the accuracy of our \textit{JWST}-only photometric redshifts to those derived in our \textit{HST}+\textit{JWST} run, we consider the same spectroscopic sample as described in Section \ref{section:sample:contamination:alma}. We present this comparison in Fig. \ref{figure:hstzzplot} where we highlight 4 objects which would not have been selected in our \textit{JWST}-only run, as their \textit{JWST}-only $z_{50}$ values are below $z=2$. Given this, we can state that the inclusion of \textit{HST} ACS filters F435W, F606W and F814W is necessary for selecting the most robust and complete samples of high-redshift massive quiescent galaxy candidates.

\subsubsection{Objects in \textit{JWST}-only but not \textit{HST}+\textit{JWST}}\label{section:sample:hst:candidatesgained}

We then spectroscopically validate the \textit{JWST}-only run in exactly the same way as set out in Section \ref{section:sample:contamination:spectra} for our main \textit{HST}+\textit{JWST} run. This returns spectra for 68 of the 202 \textit{JWST}-only galaxies (compared to 83 in the \textit{HST}+\textit{JWST} run), 65 of which are common to both runs. The remaining three spectra belong to the population of 17 likely spurious candidates selected solely by the \textit{JWST}-only run. We include the spectra of each in Fig. \ref{figure:spuriousspectra}, where it is immediately obvious that all are strong line-emitting contaminants.

The top panel of Fig. \ref{figure:spuriousspectra} features PRIMER-UDS-62083. This object was selected as a high-redshift massive quiescent galaxy candidate in \cite{Schreiber2018}, before being shown to have strong emission lines in their MOSFIRE spectroscopic data (their ID for this object is ZF-UDS-8197). Several subsequent \textit{JWST} spectroscopic surveys confirm the presence of strong emission lines, disqualifying its candidacy as a massive quiescent galaxy. The fact that this object can be confidently identified as a contaminant from our combined \textit{HST}+\textit{JWST} dataset clearly demonstrates the improved constraining power of the \textit{JWST} imaging compared with previous ground-based near-IR datasets (though only when used in combination with \textit{HST}-optical imaging).

These three objects provide further evidence that when selecting high-redshift quiescent galaxy candidates for expensive spectroscopic follow-up, the cleanest samples can only be selected from \textit{HST} optical imaging in combination with \textit{JWST} NIRCam.

\subsection{Comparison of sSFR and UVJ selection methods}\label{section:sample:uvj}

As discussed in Section \ref{section:methods:selection}, we selected our primary massive quiescent galaxy sample with a sSFR criterion, but also selected an additional quiescent sample via a demarcation in UVJ space following \cite{Williams2009} (Using their $z=0-0.5$ diagonal colour cut and removing the horizontal and vertical cuts).  The purpose of this additional, UVJ-selected sample is to investigate the similarity of both selection methods, and hence shed light on any intrinsic bias when comparing the results from this work with those in the literature derived from a UVJ selection (e.g., \citealt{Valentino2023}).

The resulting UVJ catalogue contains 212 massive quiescent galaxies in total, slightly lower than the 225 selected in our main sSFR selection. In Fig. \ref{figure:uvjselection} we place our entire $\mathrm{F356W}<26$ photometric catalogue in UVJ space, and highlight any galaxies selected as quiescent by either the UVJ colour selection (blue), or the sSFR and robust sSFR selections (red and yellow respectively; the latter is a subset of the former). In the lower panel of Fig. \ref{figure:uvjselection}, we show any galaxies selected as quiescent by only one of these methods. It is clear that the UVJ and sSFR selection methods agree extremely well; 94 per cent of UVJ selected objects at $2<z<5$ are also sSFR selected, and 89 per cent vice-versa. There is only slight disagreement occurring around the `green valley', just below the UVJ cut (e.g., \citealt{Schawinski2014}). In general, our results argue that sSFR and UVJ selection methods broadly select the same candidate samples, only disagreeing at a $\simeq10$ per cent level. Thus, results stemming from either selection method are generally comparable, though it should be noted that a wide variety of different UVJ criteria are in common usage in the literature.

An important caveat to consider, however, is the whether this result, the outcome of a fully Bayesian SED modelling approach, is consistent with results obtained whilst using different fitting methods to derive UVJ colours. Our fully Bayesian approach provides physical solutions, and our UVJ colours and sSFR values are derived from same round of SED fitting. It is therefore perhaps not surprising to find such a good agreement between our respective selection methods. However, a variety of studies in the literature have instead used {\sc eazy}\footnote{\href{https://github.com/gbrammer/eazy-photoz}{https://github.com/gbrammer/eazy-photoz}} to estimate UVJ colours and hence select quiescent galaxy candidates (e.g., \citealt{Valentino2023, Baker2025a}), which is designed to produce rapid property estimates using a mixture of the minimal optimal set of SED templates.

Hence, to evaluate the validity of comparison to such works, we fitted our primary sample of 225 massive quiescent galaxy candidates with {\sc eazy}. We used the default fitting parameters and templates, and applied an iterative zero-point correction. As mentioned above, in a fully Bayesian approach, a UVJ selection only loses $\simeq10$ per cent of galaxy candidates selected via their sSFR. In using {\sc eazy} to derive UVJ colours, we found an additional $\simeq 7$ per cent loss of massive quiescent galaxy candidates. Thus, while these results motivate a fully Bayesian approach, its clear that studies using rapid template fitting methods are generally comparable, only disagreeing at a $\leq20$ per cent level.

\subsection{High-redshift candidates}\label{section:sample:zgt5objects}

Our massive quiescent galaxy candidate sample, selected in Section \ref{section:methods:selection}, contains two non-robust candidates at $z_{50} > 5$. The first, PRIMER-UDS-77426, is at a posterior median redshift of 5.353 and has a posterior median stellar mass only slightly beyond our lower limit of $10^{10}\,\mathrm{M_\odot}$, with $M_*=10^{10.03}\,\mathrm{M_\odot}$. We find no conclusive reason to reject the validity of this candidate based on its photometry and imaging. The second, PRIMER-UDS-39790, is at a median redshift of 7.736, by far the highest redshift in our sample, and has a posterior median stellar mass of $M_*=10^{10.11}\,\mathrm{M_\odot}$. From its red and compact nature in NIRCam imaging, its likely that PRIMER-UDS-39790 is a little red dot (LRD), though it does not appear in the LRD catalogue from \cite{Kocevski2025}. One G140M spectrum exists for PRIMER-UDS-39790, taken in the EXCELS survey, however the S/N is too poor for any conclusions to be drawn (as such this spectrum was omitted from our analysis in Section \ref{section:sample:contamination:spectra}).

It is finally worth noting that the object RUBIES-UDS-QG-z7 recently reported by \cite{Weibel2024} is not included in our sample, despite having a reported stellar mass and redshift consistent with our selection criteria, because it lies just outside the footprint of the \textit{HST} ACS imaging in the UDS field.

\section{Quiescent number densities}\label{section:densities}

\begin{figure}
\includegraphics[width=\columnwidth]{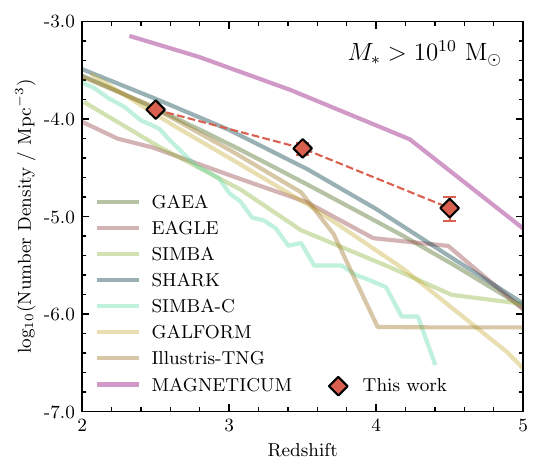}
\caption{A comparison of the number densities calculated in this work (red diamonds) with those calculated from state-of-the-art Semi-Analytical Models (SAMs) and Hydrodynamical simulations of galaxy evolution. All simulations use the passive selection $\mathrm{sSFR < 0.2/t_\mathrm{H}(z)}$, except GAEA which uses $\mathrm{log_{10}(sSFR\ /\ yr^{-1}) < -10\ }$ (the former criterion remains within 0.3 dex of this value across the whole redshift range from $2<z<5$). We homogenise all values to only include galaxies with $\mathrm{log}(M_*\ /\ \mathrm{M_\odot}) > 10$.}
\label{figure:simdensities}
\end{figure}

We present our massive quiescent number densities in Table \ref{table:literaturedensities}, which contains `raw' number densities calculated directly after our selection process in Section \ref{section:methods:selection}, as well as `contamination-corrected' number densities. We calculate these contamination-corrected number densities by constructing the joint probability distribution between the Poisson distributed uncertainty on the number counts and the contamination fraction posterior probability distribution calculated in Section \ref{section:sample:contamination:combine}. We take the median of this joint distribution as our contamination-corrected number density and use the 16\textsuperscript{th} and 84\textsuperscript{th} percentiles to trace our uncertainty. The following discussion in Section \ref{section:densities:literature} and Section \ref{section:densities:sims} include only our sSFR-selected, contamination-corrected number densities. These are shown alongside recent literature values in Fig. \ref{figure:literaturedensitites} and simulation predictions in Fig. \ref{figure:simdensities}. 


\subsection{Number densities in the literature}\label{section:densities:literature}

\begin{figure}
\includegraphics[width=\columnwidth]{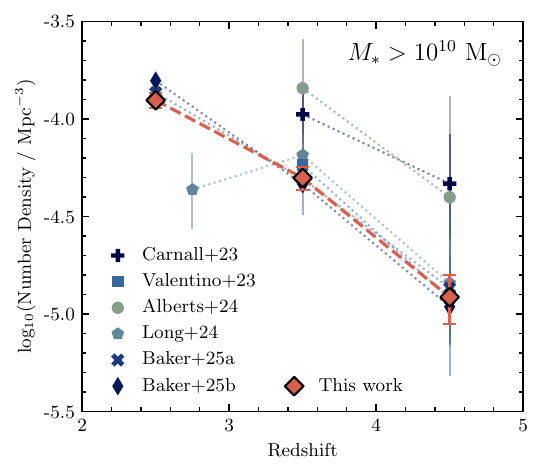}
\caption{A comparison of the number densities calculated in this work (red diamonds) with those reported in the recent literature using \textit{JWST}, all of which are tabulated in Table \ref{table:literaturedensities}. We homogenise all values to $\mathrm{log}(M_*\ /\ \mathrm{M_\odot}) > 10$ and include results from \protect\cite{Carnall2023a,Alberts2024a,Valentino2023,Long2023,Baker2025a,Baker2025b}.}
\label{figure:literaturedensitites}
\end{figure}

In Table \ref{table:literaturedensities} we also compare our massive quiescent galaxy number densities to those from recent \textit{JWST} studies by \cite{Carnall2023a}, \cite{Valentino2023}, \cite{Long2023}, \cite{Alberts2024a} and \cite{Baker2025a, Baker2025b}. We homogenise each dataset under the condition $\mathrm{log}(M_*\ /\ \mathrm{M_\odot}) > 10$. All of the above studies use a fainter F356W limit than we have derived in Section \ref{section:methods:masslimit}, meaning they should be mass complete down to this limit. We present the results of this comparison in Fig \ref{figure:literaturedensitites}. Our results fall below the large number densities reported in studies focusing on a single small field such as CEERS \citep{Carnall2023a} or the JADES MIRI parallel area \citep{Alberts2024a}, both of which contain an over-density at $3<z<4$ (see \citealt{Jin2024}, \citealt{Shah2024}), and suffer from small-number statistics at $4<z<5$.

The more-conservative number densities we report in this work are in good agreement with results from more recent, larger area investigations (e.g., \citealt{Valentino2023}, \citealt{Baker2025b}). This contrast with smaller-scale studies highlights the intrinsic uncertainty that arises from small sample sizes and field-to-field variations.

\begin{table}
\centering
\caption{A compendium of \textit{JWST} high-redshift massive quiescent galaxy number densities reported in the literature compared to those in this work. All values are homogenised to galaxies with $\mathrm{log}(M_*\ /\ \mathrm{M_\odot}) > 10$.}
\label{table:literaturedensities}
\begin{tabular}{lcccc}
\hline
Study & Area& \multicolumn{3}{c}{Number Density / $10^{-5}\;\mathrm{Mpc^{-3}}$ }\\[5pt]
&  / arcmin$^2$ & $2 < z < 3$ & $3 < z < 4$ & $4 < z < 5$\\
\hline
 & & This work &  &\\[5pt] 
Raw & 317.7 & 14.4$_{-1.1}^{+1.1}$ & 5.8$_{-0.8}^{+0.7}$ & 1.4$_{-0.4}^{+0.4}$\\[5pt]
Corrected & 317.7 & 12.5$_{-1.1}^{+1.2}$ & 5.0$_{-0.7}^{+0.7}$ & 1.2$_{-0.3}^{+0.4}$\\[5pt]
\hline
Carnall+23 & 30 & --- & 10.6$_{-3.3}^{+4.5}$ & 4.7$_{-2.2}^{+3.7}$\\[5pt]
Valentino+23 & 145.1 & --- & 5.8$_{-1.1}^{+1.4}$ & 1.2$_{-0.5}^{+0.8}$\\[5pt]
Alberts+24 & 8.8 & --- & 14.4$_{-6.9}^{+11.4}$ & 4.0$_{-3.3}^{+9.1}$\\[5pt]
Long+24 & 97 & 4.3$_{-1.6}^{+2.3}$$^\dagger$  & 6.5$_{-1.4}^{+1.8}$ & 1.4$_{-0.7}^{+1.1}$\\[5pt]
Baker+25a & 77.1 & 13.6$_{-2.6}^{+3.1}$ & 4.9$_{-1.7}^{+2.4}$ & 1.4$_{-0.9}^{+1.8}$\\[5pt]
Baker+25b & 816.9 & 15.7$_{-0.8}^{+0.8}$ & 4.6$_{-0.4}^{+0.5}$ & 1.1$_{-0.2}^{+0.3}$ \\[5pt]
\hline
\multicolumn{5}{l}{\footnotesize $^\dagger$ Computed in the redshift range $2.5 < z < 3$.}
\end{tabular}
\end{table}

We also run a stand-alone number density calculation under the condition $\mathrm{F200W} < 24.5$ without a stellar mass limit, in order to directly compare our passive number densities to pre-\textit{JWST} estimates from \cite{Schreiber2018}, which are limited to $\mathrm{K_s} < 24.5$. In the redshift range $3<z<4$, \cite{Schreiber2018} report a number density of $1.4^{+0.3}_{-0.3} \times 10^{-5}\;\mathrm{Mpc^{-3}}$, whereas in this work we calculate a massive quiescent galaxy number density of $4.2^{+0.6}_{-0.6} \times 10^{-5}\;\mathrm{Mpc^{-3}}$, which is $\simeq3$ times greater at a tension of $\simeq 4.2\sigma$. If we only consider our robust sub-sample, our resulting number density is $2.5^{+0.5}_{-0.5} \times 10^{-5}\;\mathrm{Mpc^{-3}}$, around double that reported in  \cite{Schreiber2018}, at a tension of $\simeq 1.9\sigma$.

Thus, in comparison with the literature, we find passive number densities at $3<z<4$ that are three times (at least double) pre-\textit{JWST} estimates. We report more conservative passive number densities compared to early work using \textit{JWST}, however if we account for known over-densities our results are generally in-line with the literature, especially with more-recent and larger-area \textit{JWST} studies. It is worth noting that our analysis in Section \ref{section:sample:hst} does not imply significantly different number densities should be obtained between our work and work including area that does not have \textit{HST} data available. This is because, in that test case, the 18 per cent of objects lost from our sample by the removal of \textit{HST} photometry are partially balanced by the addition of 8 per cent of new spurious sources.

\subsection{Number densities in simulations}\label{section:densities:sims}

With the aim of placing our results in the context of modern galaxy evolution theory, we compare our massive quiescent number densities to predictions from a compendium of state-of-the-art galaxy evolution simulations. We use results from \cite{Lagos2025} to include massive quiescent number densities from {\sc GAEA} \citep{deLucia2024}, {\sc Eagle} \citep{Schaye2015, Crain2015}, {\sc Simba} \citep{Dave2019}, {\sc Shark} \citep{Lagos2024}, {\sc Galform} \citep{Lacey2016} and {\sc Illustris-TNG} \citep{Springel2018, Pillepich2018}. We include predictions from {\sc Magneticum} \citep{Dolag2025} by using results based on \citeauthor{Kimmig2025} (\citeyear{Kimmig2025}, priv. comm.). We include predictions from {\sc Simba-C} \citep{Hough2023} by using results based on \citeauthor{Szpila2025} (\citeyear{Szpila2025}, priv. comm.). All results are homogenised to a stellar mass limit of $\mathrm{log_{10}}(M_*\ /\ \mathrm{M_\odot}) > 10$ and a passive selection criterion of $\mathrm{sSFR<0.2/t_\mathrm{H}(z)}$, except GAEA which uses $\mathrm{log_{10}(sSFR\ /\ yr^{-1}) < -10}$ (the former criterion remains within 0.3 dex of this value across the whole redshift range from $2<z<5$). 

We present a comparison between our observed massive quiescent number densities and results from the above simulations in Fig. \ref{figure:simdensities}. At $2<z<3$, a good proportion of simulations are generally able to reproduce our observed massive quiescent number density. However, at $z>3$ there is an increasing discrepancy between observations and simulations. At $4<z<5$, our observed massive quiescent number density is $\simeq0.5-1$ dex greater than densities reported in all but one simulation, leading to the conclusion that, in general, simulations do not produce enough massive quiescent galaxies at this early point in the Universe.

Contrary to the above, at $4<z<5$ {\sc Magneticum} predicts a density $\simeq0.5$ dex greater than our observations. While {\sc Magneticum} is able to grow enough massive quiescent galaxies at $4<z<5$, it suffers from a great over-abundance in lower redshift bins e.g., the number density of massive quiescent galaxies is $\simeq 1$ dex higher than observations at $2<z<3$. As shown in \cite{Lustig2023}, this discrepancy at lower redshifts is due to an over-abundance of galaxies with masses in the range $M_* = 10^{10}-10^{11}\ \mathrm{M_\odot}$. \cite{Lagos2025} argue that this arises from the combination of an over-efficient AGN feedback model and resolution effects leading to over-efficient quenching around $M_* = 10^{10}\ \mathrm{M_\odot}$.

Thus in summary, whilst the situation at $2<z<3$ has much improved in recent years, observed massive quiescent galaxy number densities at $4 < z < 5$ are still broadly underestimated by simulations, and there is not one simulation at present that can well reproduce the observed number density evolution from $2 < z < 5$.

\begin{figure}
    \includegraphics[width=\columnwidth]{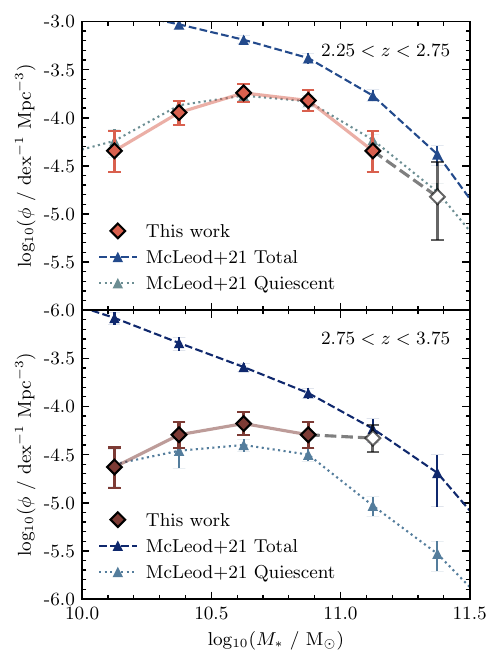}
    \caption{The quiescent galaxy mass function as calculated in this work (orange / red) compared with results from \protect\cite{McLeod2021} (blue). We use the same redshift bins as \protect\cite{McLeod2021} to aid comparison, which include $2.25 < z < 2.75$ (top) and $2.75 < z < 3.75$ (bottom). We do not use the contamination fraction from Section \ref{section:sample:contamination:combine} as contamination is likely to vary as a function of stellar mass. Due to the small sample size in each mass bin, we opt instead to just remove the nine spectroscopic/DSFG contaminants identified in Section \ref{section:sample:contamination}. The highest mass bins are left uncoloured as we expect a large amount of contamination within these bins (see Section \ref{section:densities:massfunction}).}
    \label{figure:massfunction}
\end{figure}

\subsection{Quiescent Mass function}\label{section:densities:massfunction}

To extend our number density analysis, we construct mass functions for our sample of sSFR selected passives over the range $10 < \mathrm{log}(M_*\ /\ \mathrm{M_\odot}) < 11.5$ in 0.25 dex bins. Instead of applying the total contamination fraction derived in Section \ref{section:sample:contamination:combine}, which will in reality vary between mass bins, we simply remove the 9 galaxies that were found to be contaminants in either spectroscopic data or ALMA imaging. We compare our results with pre-\textit{JWST}, ground-based mass functions reported in \cite{McLeod2021} (other examples include e.g., \citealt{Santini2021,Weaver2023,Baker2025b}). We adopt the same bounds as their two highest redshift bins: $2.25<z<2.75$ and $2.75<z<3.75$. We present a comparison of both mass functions in Fig. \ref{figure:massfunction}.

\begin{figure*}
    \includegraphics[width=\textwidth]{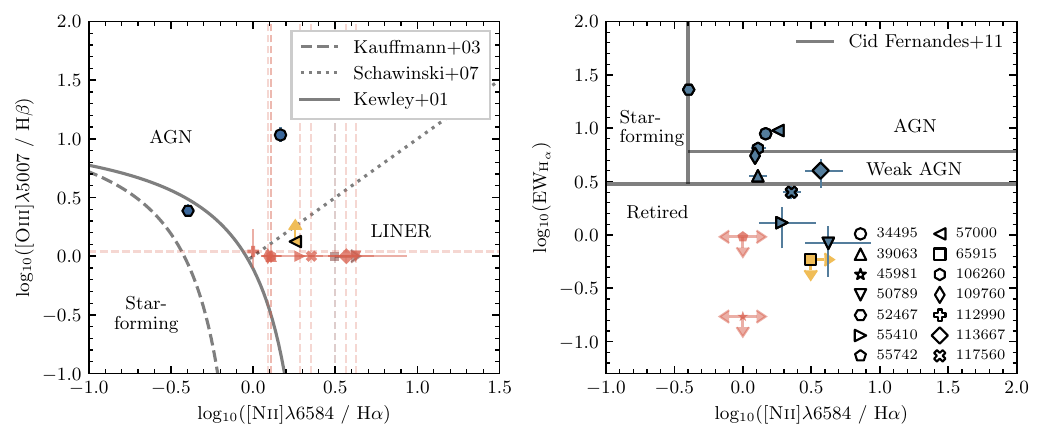}
    \caption{Emission line diagnostic diagrams for the EXCELS $z>3$ massive quiescent galaxies. Galaxies are coloured as blue if all pertinent emission lines are detected, yellow if one emission line is undetected, red if two are undetected and dark-red if three are undetected. To the left we include the BPT diagram \protect\citep{Baldwin1981}, including classification lines from \protect\cite{Kewley2001}, \protect\cite{Kauffmann2003} and \protect\cite{Schawinski2007}. Any galaxies constrained by only one line ratio are plotted at the zero line of the unconstrained axis with corresponding dashed lines to show their possible distribution. To the right is the WHaN diagram \citep{CidFernandes2011} with its classification lines included. We also include the legend for both plots, with the numbers on the right panel corresponding to each galaxy's EXCELS ID.}
    \label{figure:diagnostics}
\end{figure*}

It is likely that some level of contamination still exists in our sample, which cannot be accurately modelled due to the small sample sizes in each mass bin. This is particularly true of our highest-mass bins, which are most likely to contain further bright AGN/DSFGs for which we do not have spectroscopic data (similar to GTO-WIDE-1472 from Fig. \ref{figure:xrayspectra}, which has been removed from the highest-mass bin in the top panel). This is because AGN and (even quite dusty) star-forming galaxies tend to be brighter than faint old stellar populations, which tends to lead to high stellar masses for AGN/DSFGs that are misidentified as quiescent galaxies. We therefore highlight the most massive bins in Fig. \ref{figure:massfunction} as likely more-significantly contaminated.

Our mass function profiles generally agree well with \cite{McLeod2021}, with both turning over at $\simeq 10^{10.5} \mathrm{M_\odot}$. However, our results show generally higher massive quiescent number densities in the redshift bin $2.75 < z < 3.75$ than calculated by \cite{McLeod2021}, as expected due to the addition of \textit{JWST} data (e.g., \citealt{Carnall2023a}).

We further find an over-abundance of extremely massive passive galaxies, $\mathrm{log}(M_*/\mathrm{M_\odot})$ > 11, at $2.75 < z < 3.75$. We find 13 objects in this mass/redshift range, one of which is removed as a spectroscopic and ALMA contaminant (JADES-GOODSS-54806). From the 12 remaining galaxies, five are spectroscopically confirmed. It is likely that further contaminants exist among the seven objects that do not have spectroscopic data. From the five objects that are spectroscopically confirmed, three have (relatively weak) emission lines in addition to clearly visible old stellar components. If these emission lines originate from an AGN component, then, as discussed above, the contribution of AGN continuum emission to the galaxy SED could result in an over-estimation of stellar mass. Therefore, to investigate the presence of AGN in our sample, in the next section we focus on classifying some of our massive quiescent galaxies by their emission line diagnostics, specifically those at $z>3$ targeted in the EXCELS survey (see Section \ref{section:data:spectroscopy:excels}).

\section{Emission Line Analyses}\label{section:lineflux}

As noted in Section \ref{section:sample:contamination:xray}, 25 of our 225 passive galaxies are X-ray detected. In the same section we investigated the spectra of these objects and found that an X-ray detection does not necessarily rule out a passive classification. Several such objects do however display faint emission lines, which are also fairly common in our whole 83-object spectroscopic sample. It appears clear then that to probe further into the presence of AGN in our massive quiescent galaxy sample we must turn to emission line diagnostic diagrams. Indeed, such an investigation will also probe any ongoing star formation and hence further test the robustness of our quiescent selection.

Therefore, we make use of the EXCELS survey (see Section \ref{section:data:spectroscopy:excels}) to probe emission line diagnostics for a sub-sample of our 225 massive quiescent candidates. Specifically, we focus on 14 $z>3$ massive quiescent galaxies with deep medium resolution spectroscopy in the wavelength range $1-5$ $\mu$m. Two of these are X-ray detected: 34495 and 106260 (in this section we refer to objects by their EXCELS IDs from \citealt{Carnall2024}, our corresponding photometric IDs are included in Table \ref{table:excerpt}).

As discussed in Section \ref{section:sample:contamination:spectra}, it is not uncommon for passive galaxies to display emission lines (e.g., \citealt{Belli2024}). These lines can be due to residual star formation that is low-level enough for the galaxy to still be classified as quiescent, an AGN phase, or hot low-mass evolved stars (HOLMES, e.g., \citealt{Yan2012}). We use emission line fluxes obtained in Section \ref{section:methods:lineflux:modelling} to place each EXCELS $z>3$ galaxy on both the BPT diagram \citep{Baldwin1981} and the WHaN diagram \citep{CidFernandes2011}.

We show the BPT diagram for our 14 EXCELS $z>3$ galaxies in the left panel of Fig. \ref{figure:diagnostics}. No galaxies are placed well within the star-forming region, which confirms the quiescent classification of the vast majority of these galaxies. EXCELS-52467 exists within the composite region of the diagram, suggesting a combination of both low-level star-formation and AGN activity (see below). Only EXCELS-34495 is placed confidently within the AGN region, the spectrum of which is shown in Fig. \ref{figure:xrayspectra}. Beyond these conclusions, the remaining picture is made less clear by the widespread non-detection of both {\sc [Oiii]}$\lambda5007$ and H$\beta$. Over half of our EXCELS $z>3$ galaxies are hence only constrained by their {\sc [Nii]}$\lambda6584$ / H$\alpha$ ratio. Two of the EXCELS $z>3$ galaxies, EXCELS-45981 and EXCELS-55742, have no detected flux for the pertinent emission lines and are hence unconstrained in any axis on the BPT diagram, showing no evidence for either AGN or star-formation activity.

Consequently, we then explore the WHaN diagram \citep[e.g.,][]{CidFernandes2011}, shown on the right side of Fig. \ref{figure:diagnostics}. This diagram is useful for distinguishing between different sources of non-stellar line emission, particularly in cases where relatively few lines are detected. The WHaN diagram identifies AGN based only on the H$\alpha$ and {\sc[Nii]}$\lambda6584$ features, in particular separating these from galaxies that emit only due to HOLMES. Galaxies that only have line emission due to HOLMES are sometimes referred to as `retired' galaxies, and fall into the lower section of the WHaN diagram, below log$_{10}(\mathrm{EW_{H\alpha}}) = 0.5$.

Similarly to the BPT diagram, the WHaN diagram does not place any galaxies in our sample decisively in the star-forming region. Unlike the BPT diagram, the WHaN diagram confirms the widespread presence of AGN, both strong and weak, in our massive quiescent galaxy sample. A total of six EXCELS $z>3$ galaxies appear to host an AGN, equating to $\simeq 50$ per cent (after disqualifying EXCELS-112990, which has a chip-gap over H$\alpha$ and is therefore inconclusive). A further $\simeq 50$ per cent lie in the retired galaxy region, suggesting no sources of ionization beyond the HOLMES that are a component of any galaxy stellar population beyond a certain age.

Interestingly, in both diagrams EXCELS-52467 exists between the star-forming and AGN regions, suggesting the possibility of both ongoing star formation and an actively accreting black hole. It should be noted however that the \cite{Kauffmann2003} and \cite{Schawinski2007} lines in Fig. \ref{figure:diagnostics} are both derived from local galaxy populations. At higher redshift, star-forming galaxies typically have harder ionizing spectra due to their alpha enhanced stellar populations, which moves them further towards the AGN region of the BPT diagram (e.g., \citealt{Shapley2025}). This suggests that an AGN component might not be required to explain the position of EXCELS-52467 on the BPT diagram. We calculate an H$\alpha$ flux for EXCELS-52467 of $10.5_{-0.3}^{+0.3}\times10^{-18}\ \mathrm{erg\ s^{-1}\ cm^{-2}}$, which we translate into a maximum star-formation rate (assuming no AGN contribution) using the relationship from \cite{Kennicutt2012}, $\mathrm{SFR}=5.3_{-0.2}^{+0.2}\ \mathrm{M_\odot\ yr^{-1}}$. This is $\simeq0.1-0.2$ dex higher than the sSFR threshold for quiescence for this object, demonstrating this galaxy is likely very close to the star-forming vs quiescent boundary. This suggests that EXCELS-52467 might either be observed in the process of quenching, or potentially whilst undergoing a small rejuvenation event.

Thus, results from the BPT diagram and WHaN diagram both suggest a surprisingly high fraction of weak AGN in our sample of massive quiescent galaxies, at a $\simeq50$ per cent level. We compare our results to those from \cite{Bugiani2024} and \cite{Skarbinski2025}, which both focus on similar, relatively small samples of quiescent galaxies at $1\lesssim z\lesssim 3$, as well as \citealt{Baker2025a} which focuses on a small sample at $2< z< 5$. All studies also find an AGN fraction of $\simeq50$ per cent, agreeing well with our result. It is interesting that such a large fraction of massive quiescent galaxies at these redshifts appear to display residual, low-level AGN accretion after the quenching of star formation. This is consistent with the concept of `maintenance-mode' AGN feedback (e.g., \citealt{Barisic2019}), widely invoked in cosmological simulations to keep quenched galaxies from re-igniting star formation by continuous injection of energy into the circumgalactic medium. The relationship between AGN and quenching in the EXCELS sample will be further investigated in upcoming work by Taylor et al. (in prep), in which we probe the relationship between AGN activity and gas outflows.

As we have shown in Section \ref{section:sample:contamination:spectra}, less than 40 per cent of our massive quiescent galaxy candidates have spectroscopic data, the majority of which is in the form of low-resolution prism spectroscopy. Deep, medium-resolution spectroscopy of larger samples of high-redshift massive quiescent galaxies would be of great value in shedding further light on the role of AGN in quenching and maintaining quiescence in early massive galaxies.

\section{Conclusions}\label{section:conclusions}

In summary, we construct a massive quiescent galaxy catalogue of 225 objects at $z>2$, selected from PRIMER and JADES photometry spanning a total area of 317.7 sq. arcmin. We restrict our selection to galaxies with $M_* > 10^{10}\ \mathrm{M_\odot}$ and $\mathrm{sSFR} < 0.2/t_\mathrm{H}$, and only include galaxies with coverage in three \textit{HST} ACS optical filters as well as eight NIRCam near-infrared filters. 

We investigate the contamination in our sample of massive quiescent galaxy candidates in Section \ref{section:sample:contamination} via (i) \textit{JWST} NIRSpec spectroscopic validation, to find clearly non-passive contaminants and redshift outliers; (ii) $Chandra$ X-ray imaging, to find AGN contaminants; and (iii) ALMA interferometry, to find DSFG contaminants. We find NIRSpec spectroscopic data for 83 ($\simeq 40$ per cent) of our candidates, from which we identify six contaminants. We conclude that an X-ray detection does not necessarily disqualify a galaxy as being quiescent, as the spectra we have for X-ray detected galaxies often remain stellar-dominated, showing no evidence for being contaminants. We find four DSFGs out of 46 candidates with deep ALMA coverage, one of which is also a spectroscopic contaminant. We use simulation-based inference to combine the contamination fractions from spectroscopic and DSFG contaminants, resulting in a modest estimated contamination fraction of $12.9_{-3.1}^{+4.0}$ per cent.

We investigate the effect of restricting our analysis to only galaxies with full coverage in three \textit{HST} ACS filters in addition to eight NIRCam filters in Section \ref{section:sample:hst}. We conclude that the inclusion of these optical photometric bands is very important for selecting the most complete and clean samples of massive quiescent galaxy candidates. We show that the absence of such data not only introduces a significant amount of star-forming galaxy contamination ($\simeq10$ per cent), but also results in a $\simeq20$ per cent loss of candidates that are identified in the presence of \textit{HST} data. We also compare our quiescent selection, sSFR $< 0.2/t_\mathrm{H}$, against a more classical demarcation in UVJ space in Section \ref{section:sample:uvj} and Fig. \ref{figure:uvjselection}. We find that both selection methods broadly produce the same sample of candidates, only disagreeing at the $\simeq5-10$ per cent level.

We calculate the number density of our massive quiescent galaxy candidate sample in Section \ref{section:densities}, corrected for contamination, in three bins spanning $2<z<5$. We first compare our number densities to a representative pre-\textit{JWST} estimate from \cite{Schreiber2018}, finding that our results are greater by a factor of three. We then place our results in the context of similar literature using \textit{JWST} observations (see Fig. \ref{figure:literaturedensitites} and Table \ref{table:literaturedensities}), agreeing well with more-recent and larger-area studies, but reporting more conservative densities compared to earlier and smaller-scale \textit{JWST} studies. The areas used in these smaller-scale studies are known to host over-densities at $3<z<4$, explaining this discrepancy and highlighting the intrinsic uncertainty from cosmic variation in small-scale studies.

We also place our observed massive quiescent number densities alongside results from state-of-the-art galaxy evolution simulations. This includes massive quiescent number densities from both SAMs and hydrodynamical simulations, which are homogenised to our selection criteria of $M_* > 10^{10}\ \mathrm{M_\odot}$ and $\mathrm{sSFR}<0.2/t_\mathrm{H}$. In general, simulations are able to reproduce the observed massive quiescent number density at $2<z<3$. However, at $3<z<5$ most simulations are increasingly unable to produce enough massive quiescent galaxies, with discrepancies of up to $\simeq 1$ dex. In contrast to all of the other simulations, {\sc magneticum} consistently over-predicts the number density of massive quiescent galaxies, especially towards $z=2$, but comes closest to reproducing our observed number density at the highest redshifts in our study ($4 < z < 5)$. Thus we conclude that simulations generally still struggle to reproduce observed massive quiescent number densities at $z>3$.

Towards further investigating the presence of AGN in our sample of massive quiescent galaxies, in Section \ref{section:lineflux} we perform an emission line analysis using the EXCELS survey, which provides medium-resolution spectroscopy at $1-5\,\mu$m for 14 of our galaxies at $z~>~3$. We measure emission line fluxes for H$\alpha$, H$\beta$, {\sc [Oiii]}$\lambda5007$, and {\sc [Nii]}$\lambda6584$ using continuum-subtracted spectra. We then place each EXCELS $z>3$ massive quiescent galaxy on the BPT and WHaN diagrams (see Fig. \ref{figure:diagnostics}). We find a high incidence of AGN in our sample at a $\simeq50$ per cent level, in agreement with recent studies at $1 \lesssim z \lesssim 3$. This result is consistent with the concept of maintenance-mode feedback, in which low-level AGN accretion episodes post-quenching provide an ongoing injection of energy into the circumgalactic medium, preventing the re-ignition of star-formation activity. We will further investigate the role of AGN in quenching and maintaining quiescence in the EXCELS sample in Taylor et al. (in prep).

We are quickly entering an era in which the \textit{JWST} archive provides larger and larger imaging datasets from which precise number densities can be extracted. However, it is clear that \textit{HST} ACS optical bands are of critical importance for selecting robust samples at intermediate redshifts. More coverage in these filters overlapping existing NIRCam data would therefore be highly valuable. Whilst the NIRSpec archive is also growing, only $\simeq40$ per cent of our candidates have spectroscopic observations, and much of this is low-resolution prism data, which cannot be used to extract detailed physical properties, or medium-resolution data that does not cover the whole rest-frame UV-optical wavelength range. Significantly larger spectroscopic samples will be required to properly characterise the evolution of early massive quiescent galaxies, including the role of AGN, hence our remaining massive quiescent galaxy candidates are excellent targets for future NIRSpec spectroscopy.

\section*{Acknowledgements}

SDS, ACC, H-HL and ET acknowledge support from a UKRI Frontier Research Guarantee Grant (PI Carnall; grant reference EP/Y037065/1). FC, TMS, KZA-C and DS acknowledge support from a UKRI Frontier Research Guarantee Grant (PI Cullen; grant reference: EP/X021025/1). Support for Program number JWST-GO-03543.014 was provided through a grant from the STScI under NASA contract NAS5-03127. We gratefully acknowledge support from the NASA Astrophysics Data Analysis Program (ADAP) under grant 80NSSC23K0495. Some of the data products presented herein were retrieved from the Dawn \textit{JWST} Archive (DJA). DJA is an initiative of the Cosmic Dawn Center (DAWN), which is funded by the Danish National Research Foundation under grant DNRF140.

\section*{Data Availability}
All \textit{JWST} and \textit{HST} data products are available via the Mikulski Archive for Space Telescopes (\href{https://mast.stsci.edu}{https://mast.stsci.edu}). Photometric data and fitted model posteriors are available upon request. A detailed catalogue of the full sample of massive quiescent galaxies selected in Section \ref{section:methods:selection} is available as supplementary online material.



\bibliographystyle{mnras}
\bibliography{bib} 

@ARTICLE{Baggen2024,
       author = {{Baggen}, Josephine F.~W. and {van Dokkum}, Pieter and {Brammer}, Gabriel and {de Graaff}, Anna and {Franx}, Marijn and {Greene}, Jenny and {Labb{\'e}}, Ivo and {Leja}, Joel and {Maseda}, Michael V. and {Nelson}, Erica J. and {Rix}, Hans-Walter and {Wang}, Bingjie and {Weibel}, Andrea},
        title = "{The Small Sizes and High Implied Densities of ``Little Red Dots'' with Balmer Breaks Could Explain Their Broad Emission Lines without an Active Galactic Nucleus}",
      journal = {\apjl},
         year = 2024,
        month = dec,
       volume = {977},
       number = {1},
          eid = {L13},
        pages = {L13},
          doi = {10.3847/2041-8213/ad90b8},
archivePrefix = {arXiv},
       eprint = {2408.07745},
 primaryClass = {astro-ph.GA},
       adsurl = {https://ui.adsabs.harvard.edu/abs/2024ApJ...977L..13B}
}

@ARTICLE{Hopkins2010,
       author = {{Hopkins}, Philip F. and {Murray}, Norman and {Quataert}, Eliot and {Thompson}, Todd A.},
        title = "{A maximum stellar surface density in dense stellar systems}",
      journal = {\mnras},
         year = 2010,
        month = jan,
       volume = {401},
       number = {1},
        pages = {L19-L23},
          doi = {10.1111/j.1745-3933.2009.00777.x},
archivePrefix = {arXiv},
       eprint = {0908.4088},
 primaryClass = {astro-ph.CO},
       adsurl = {https://ui.adsabs.harvard.edu/abs/2010MNRAS.401L..19H}
}

@ARTICLE{Illingworth2016,
       author = {{Illingworth}, Garth and {Magee}, Daniel and {Bouwens}, Rychard and {Oesch}, Pascal and {Labbe}, Ivo and {van Dokkum}, Pieter and {Whitaker}, Katherine and {Holden}, Bradford and {Franx}, Marijn and {Gonzalez}, Valentino},
        title = "{The Hubble Legacy Fields (HLF-GOODS-S) v1.5 Data Products: Combining 2442 Orbits of GOODS-S/CDF-S Region ACS and WFC3/IR Images}",
      journal = {arXiv e-prints},
         year = 2016,
        month = jun,
          eid = {arXiv:1606.00841},
        pages = {arXiv:1606.00841},
          doi = {10.48550/arXiv.1606.00841},
archivePrefix = {arXiv},
       eprint = {1606.00841},
 primaryClass = {astro-ph.GA},
       adsurl = {https://ui.adsabs.harvard.edu/abs/2016arXiv160600841I}
}

@ARTICLE{Whitaker2019,
       author = {{Whitaker}, Katherine E. and {Ashas}, Mohammad and {Illingworth}, Garth and {Magee}, Daniel and {Leja}, Joel and {Oesch}, Pascal and {van Dokkum}, Pieter and {Mowla}, Lamiya and {Bouwens}, Rychard and {Franx}, Marijn and {Holden}, Bradford and {Labb{\'e}}, Ivo and {Rafelski}, Marc and {Teplitz}, Harry and {Gonzalez}, Valentino},
        title = "{The Hubble Legacy Field GOODS-S Photometric Catalog}",
      journal = {\apjs},
         year = 2019,
        month = sep,
       volume = {244},
       number = {1},
          eid = {16},
        pages = {16},
          doi = {10.3847/1538-4365/ab3853},
archivePrefix = {arXiv},
       eprint = {1908.05682},
 primaryClass = {astro-ph.GA},
       adsurl = {https://ui.adsabs.harvard.edu/abs/2019ApJS..244...16W}
}

@ARTICLE{Koekemoer2011,
       author = {{Koekemoer}, Anton M. and {Faber}, S.~M. and {Ferguson}, Henry C. and {Grogin}, Norman A. and {Kocevski}, Dale D. and {Koo}, David C. and {Lai}, Kamson and {Lotz}, Jennifer M. and {Lucas}, Ray A. and {McGrath}, Elizabeth J. and {Ogaz}, Sara and {Rajan}, Abhijith and {Riess}, Adam G. and {Rodney}, Steve A. and {Strolger}, Louis and {Casertano}, Stefano and {Castellano}, Marco and {Dahlen}, Tomas and {Dickinson}, Mark and {Dolch}, Timothy and {Fontana}, Adriano and {Giavalisco}, Mauro and {Grazian}, Andrea and {Guo}, Yicheng and {Hathi}, Nimish P. and {Huang}, Kuang-Han and {van der Wel}, Arjen and {Yan}, Hao-Jing and {Acquaviva}, Viviana and {Alexander}, David M. and {Almaini}, Omar and {Ashby}, Matthew L.~N. and {Barden}, Marco and {Bell}, Eric F. and {Bournaud}, Fr{\'e}d{\'e}ric and {Brown}, Thomas M. and {Caputi}, Karina I. and {Cassata}, Paolo and {Challis}, Peter J. and {Chary}, Ranga-Ram and {Cheung}, Edmond and {Cirasuolo}, Michele and {Conselice}, Christopher J. and {Roshan Cooray}, Asantha and {Croton}, Darren J. and {Daddi}, Emanuele and {Dav{\'e}}, Romeel and {de Mello}, Duilia F. and {de Ravel}, Loic and {Dekel}, Avishai and {Donley}, Jennifer L. and {Dunlop}, James S. and {Dutton}, Aaron A. and {Elbaz}, David and {Fazio}, Giovanni G. and {Filippenko}, Alexei V. and {Finkelstein}, Steven L. and {Frazer}, Chris and {Gardner}, Jonathan P. and {Garnavich}, Peter M. and {Gawiser}, Eric and {Gruetzbauch}, Ruth and {Hartley}, Will G. and {H{\"a}ussler}, Boris and {Herrington}, Jessica and {Hopkins}, Philip F. and {Huang}, Jia-Sheng and {Jha}, Saurabh W. and {Johnson}, Andrew and {Kartaltepe}, Jeyhan S. and {Khostovan}, Ali A. and {Kirshner}, Robert P. and {Lani}, Caterina and {Lee}, Kyoung-Soo and {Li}, Weidong and {Madau}, Piero and {McCarthy}, Patrick J. and {McIntosh}, Daniel H. and {McLure}, Ross J. and {McPartland}, Conor and {Mobasher}, Bahram and {Moreira}, Heidi and {Mortlock}, Alice and {Moustakas}, Leonidas A. and {Mozena}, Mark and {Nandra}, Kirpal and {Newman}, Jeffrey A. and {Nielsen}, Jennifer L. and {Niemi}, Sami and {Noeske}, Kai G. and {Papovich}, Casey J. and {Pentericci}, Laura and {Pope}, Alexandra and {Primack}, Joel R. and {Ravindranath}, Swara and {Reddy}, Naveen A. and {Renzini}, Alvio and {Rix}, Hans-Walter and {Robaina}, Aday R. and {Rosario}, David J. and {Rosati}, Piero and {Salimbeni}, Sara and {Scarlata}, Claudia and {Siana}, Brian and {Simard}, Luc and {Smidt}, Joseph and {Snyder}, Diana and {Somerville}, Rachel S. and {Spinrad}, Hyron and {Straughn}, Amber N. and {Telford}, Olivia and {Teplitz}, Harry I. and {Trump}, Jonathan R. and {Vargas}, Carlos and {Villforth}, Carolin and {Wagner}, Cory R. and {Wandro}, Pat and {Wechsler}, Risa H. and {Weiner}, Benjamin J. and {Wiklind}, Tommy and {Wild}, Vivienne and {Wilson}, Grant and {Wuyts}, Stijn and {Yun}, Min S.},
        title = "{CANDELS: The Cosmic Assembly Near-infrared Deep Extragalactic Legacy Survey{\textemdash}The Hubble Space Telescope Observations, Imaging Data Products, and Mosaics}",
      journal = {\apjs},
         year = 2011,
        month = dec,
       volume = {197},
       number = {2},
          eid = {36},
        pages = {36},
          doi = {10.1088/0067-0049/197/2/36},
archivePrefix = {arXiv},
       eprint = {1105.3754},
 primaryClass = {astro-ph.CO},
       adsurl = {https://ui.adsabs.harvard.edu/abs/2011ApJS..197...36K}
}

@ARTICLE{Grogin2011,
       author = {{Grogin}, Norman A. and {Kocevski}, Dale D. and {Faber}, S.~M. and {Ferguson}, Henry C. and {Koekemoer}, Anton M. and {Riess}, Adam G. and {Acquaviva}, Viviana and {Alexander}, David M. and {Almaini}, Omar and {Ashby}, Matthew L.~N. and {Barden}, Marco and {Bell}, Eric F. and {Bournaud}, Fr{\'e}d{\'e}ric and {Brown}, Thomas M. and {Caputi}, Karina I. and {Casertano}, Stefano and {Cassata}, Paolo and {Castellano}, Marco and {Challis}, Peter and {Chary}, Ranga-Ram and {Cheung}, Edmond and {Cirasuolo}, Michele and {Conselice}, Christopher J. and {Roshan Cooray}, Asantha and {Croton}, Darren J. and {Daddi}, Emanuele and {Dahlen}, Tomas and {Dav{\'e}}, Romeel and {de Mello}, Du{\'\i}lia F. and {Dekel}, Avishai and {Dickinson}, Mark and {Dolch}, Timothy and {Donley}, Jennifer L. and {Dunlop}, James S. and {Dutton}, Aaron A. and {Elbaz}, David and {Fazio}, Giovanni G. and {Filippenko}, Alexei V. and {Finkelstein}, Steven L. and {Fontana}, Adriano and {Gardner}, Jonathan P. and {Garnavich}, Peter M. and {Gawiser}, Eric and {Giavalisco}, Mauro and {Grazian}, Andrea and {Guo}, Yicheng and {Hathi}, Nimish P. and {H{\"a}ussler}, Boris and {Hopkins}, Philip F. and {Huang}, Jia-Sheng and {Huang}, Kuang-Han and {Jha}, Saurabh W. and {Kartaltepe}, Jeyhan S. and {Kirshner}, Robert P. and {Koo}, David C. and {Lai}, Kamson and {Lee}, Kyoung-Soo and {Li}, Weidong and {Lotz}, Jennifer M. and {Lucas}, Ray A. and {Madau}, Piero and {McCarthy}, Patrick J. and {McGrath}, Elizabeth J. and {McIntosh}, Daniel H. and {McLure}, Ross J. and {Mobasher}, Bahram and {Moustakas}, Leonidas A. and {Mozena}, Mark and {Nandra}, Kirpal and {Newman}, Jeffrey A. and {Niemi}, Sami-Matias and {Noeske}, Kai G. and {Papovich}, Casey J. and {Pentericci}, Laura and {Pope}, Alexandra and {Primack}, Joel R. and {Rajan}, Abhijith and {Ravindranath}, Swara and {Reddy}, Naveen A. and {Renzini}, Alvio and {Rix}, Hans-Walter and {Robaina}, Aday R. and {Rodney}, Steven A. and {Rosario}, David J. and {Rosati}, Piero and {Salimbeni}, Sara and {Scarlata}, Claudia and {Siana}, Brian and {Simard}, Luc and {Smidt}, Joseph and {Somerville}, Rachel S. and {Spinrad}, Hyron and {Straughn}, Amber N. and {Strolger}, Louis-Gregory and {Telford}, Olivia and {Teplitz}, Harry I. and {Trump}, Jonathan R. and {van der Wel}, Arjen and {Villforth}, Carolin and {Wechsler}, Risa H. and {Weiner}, Benjamin J. and {Wiklind}, Tommy and {Wild}, Vivienne and {Wilson}, Grant and {Wuyts}, Stijn and {Yan}, Hao-Jing and {Yun}, Min S.},
        title = "{CANDELS: The Cosmic Assembly Near-infrared Deep Extragalactic Legacy Survey}",
      journal = {\apjs},
         year = 2011,
        month = dec,
       volume = {197},
       number = {2},
          eid = {35},
        pages = {35},
          doi = {10.1088/0067-0049/197/2/35},
archivePrefix = {arXiv},
       eprint = {1105.3753},
 primaryClass = {astro-ph.CO},
       adsurl = {https://ui.adsabs.harvard.edu/abs/2011ApJS..197...35G}
}

@ARTICLE{Kron1980,
       author = {{Kron}, R.~G.},
        title = "{Photometry of a complete sample of faint galaxies.}",
      journal = {\apjs},
         year = 1980,
        month = jun,
       volume = {43},
        pages = {305-325},
          doi = {10.1086/190669},
       adsurl = {https://ui.adsabs.harvard.edu/abs/1980ApJS...43..305K}
}

@ARTICLE{Begley2025,
       author = {{Begley}, R. and {McLure}, R.~J. and {Cullen}, F. and {McLeod}, D.~J. and {Dunlop}, J.~S. and {Carnall}, A.~C. and {Stanton}, T.~M. and {Shapley}, A.~E. and {Cochrane}, R. and {Donnan}, C.~T. and {Ellis}, R.~S. and {Fontana}, A. and {Grogin}, N.~A. and {Koekemoer}, A.~M.},
        title = "{The evolution of [O III] + H{\ensuremath{\beta}} equivalent width from z ≃ 3-8: implications for the production and escape of ionizing photons during reionization}",
      journal = {\mnras},
         year = 2025,
        month = mar,
       volume = {537},
       number = {4},
        pages = {3245-3264},
          doi = {10.1093/mnras/staf211},
archivePrefix = {arXiv},
       eprint = {2410.10988},
 primaryClass = {astro-ph.GA},
       adsurl = {https://ui.adsabs.harvard.edu/abs/2025MNRAS.537.3245B}
}

@ARTICLE{Khochfar2011,
       author = {{Khochfar}, Sadegh and {Emsellem}, Eric and {Serra}, Paolo and {Bois}, Maxime and {Alatalo}, Katherine and {Bacon}, R. and {Blitz}, Leo and {Bournaud}, Fr{\'e}d{\'e}ric and {Bureau}, M. and {Cappellari}, Michele and {Davies}, Roger L. and {Davis}, Timothy A. and {de Zeeuw}, P.~T. and {Duc}, Pierre-Alain and {Krajnovi{\'c}}, Davor and {Kuntschner}, Harald and {Lablanche}, Pierre-Yves and {McDermid}, Richard M. and {Morganti}, Raffaella and {Naab}, Thorsten and {Oosterloo}, Tom and {Sarzi}, Marc and {Scott}, Nicholas and {Weijmans}, Anne-Marie and {Young}, Lisa M.},
        title = "{The ATLAS$^{3D}$ project - VIII. Modelling the formation and evolution of fast and slow rotator early-type galaxies within {\ensuremath{\Lambda}}CDM}",
      journal = {\mnras},
         year = 2011,
        month = oct,
       volume = {417},
       number = {2},
        pages = {845-862},
          doi = {10.1111/j.1365-2966.2011.19486.x},
archivePrefix = {arXiv},
       eprint = {1107.5059},
 primaryClass = {astro-ph.CO},
       adsurl = {https://ui.adsabs.harvard.edu/abs/2011MNRAS.417..845K}
}

@ARTICLE{Emsellem2011,
       author = {{Emsellem}, Eric and {Cappellari}, Michele and {Krajnovi{\'c}}, Davor and {Alatalo}, Katherine and {Blitz}, Leo and {Bois}, Maxime and {Bournaud}, Fr{\'e}d{\'e}ric and {Bureau}, Martin and {Davies}, Roger L. and {Davis}, Timothy A. and {de Zeeuw}, P.~T. and {Khochfar}, Sadegh and {Kuntschner}, Harald and {Lablanche}, Pierre-Yves and {McDermid}, Richard M. and {Morganti}, Raffaella and {Naab}, Thorsten and {Oosterloo}, Tom and {Sarzi}, Marc and {Scott}, Nicholas and {Serra}, Paolo and {van de Ven}, Glenn and {Weijmans}, Anne-Marie and {Young}, Lisa M.},
        title = "{The ATLAS$^{3D}$ project - III. A census of the stellar angular momentum within the effective radius of early-type galaxies: unveiling the distribution of fast and slow rotators}",
      journal = {\mnras},
         year = 2011,
        month = jun,
       volume = {414},
       number = {2},
        pages = {888-912},
          doi = {10.1111/j.1365-2966.2011.18496.x},
archivePrefix = {arXiv},
       eprint = {1102.4444},
 primaryClass = {astro-ph.CO},
       adsurl = {https://ui.adsabs.harvard.edu/abs/2011MNRAS.414..888E}
}

@ARTICLE{Dekel2025,
       author = {{Dekel}, Avishai and {Mandelker}, Nir and {Li}, Zhaozhou and {Yao}, Zhiyuan and {Zhu}, Bocheng and {Lapiner}, Sharon and {Dutta Chowdhury}, Dhruba and {Ginzburg}, Omri},
        title = "{From FFB starbursts at cosmic dawn to quenching at cosmic morning: Hi-z galaxy bimodality}",
      journal = {\mnras},
         year = 2025,
        month = nov,
       volume = {544},
       number = {1},
        pages = {160-180},
          doi = {10.1093/mnras/staf1692},
archivePrefix = {arXiv},
       eprint = {2506.11664},
 primaryClass = {astro-ph.GA},
       adsurl = {https://ui.adsabs.harvard.edu/abs/2025MNRAS.544..160D}
}

@ARTICLE{Croton2006,
       author = {{Croton}, Darren J. and {Springel}, Volker and {White}, Simon D.~M. and {De Lucia}, G. and {Frenk}, C.~S. and {Gao}, L. and {Jenkins}, A. and {Kauffmann}, G. and {Navarro}, J.~F. and {Yoshida}, N.},
        title = "{The many lives of active galactic nuclei: cooling flows, black holes and the luminosities and colours of galaxies}",
      journal = {\mnras},
         year = 2006,
        month = jan,
       volume = {365},
       number = {1},
        pages = {11-28},
          doi = {10.1111/j.1365-2966.2005.09675.x},
archivePrefix = {arXiv},
       eprint = {astro-ph/0508046},
 primaryClass = {astro-ph},
       adsurl = {https://ui.adsabs.harvard.edu/abs/2006MNRAS.365...11C}
}

@ARTICLE{Wellons2015,
       author = {{Wellons}, Sarah and {Torrey}, Paul and {Ma}, Chung-Pei and {Rodriguez-Gomez}, Vicente and {Vogelsberger}, Mark and {Kriek}, Mariska and {van Dokkum}, Pieter and {Nelson}, Erica and {Genel}, Shy and {Pillepich}, Annalisa and {Springel}, Volker and {Sijacki}, Debora and {Snyder}, Gregory and {Nelson}, Dylan and {Sales}, Laura and {Hernquist}, Lars},
        title = "{The formation of massive, compact galaxies at z = 2 in the Illustris simulation}",
      journal = {\mnras},
         year = 2015,
        month = may,
       volume = {449},
       number = {1},
        pages = {361-372},
          doi = {10.1093/mnras/stv303},
archivePrefix = {arXiv},
       eprint = {1411.0667},
 primaryClass = {astro-ph.GA},
       adsurl = {https://ui.adsabs.harvard.edu/abs/2015MNRAS.449..361W}
}

@ARTICLE{Valentino2025,
       author = {{Valentino}, F. and {Heintz}, K.~E. and {Brammer}, G. and {Ito}, K. and {Kokorev}, V. and {Whitaker}, K.~E. and {Gallazzi}, A. and {de Graaff}, A. and {Weibel}, A. and {Frye}, B.~L. and {Kamieneski}, P.~S. and {Jin}, S. and {Ceverino}, D. and {Faisst}, A. and {Farcy}, M. and {Fujimoto}, S. and {Gillman}, S. and {Gottumukkala}, R. and {Hamadouche}, M. and {Harrington}, K.~C. and {Hirschmann}, M. and {Jespersen}, C.~K. and {Kakimoto}, T. and {Kubo}, M. and {Lagos}, C. d. P. and {Lee}, M. and {Magdis}, G.~E. and {Man}, A.~W.~S. and {Onodera}, M. and {Rizzo}, F. and {Shimakawa}, R. and {Setton}, D.~J. and {Tanaka}, M. and {Toft}, S. and {Wu}, P.-F. and {Zhu}, P.},
        title = "{Gas outflows in two recently quenched galaxies at z = 4 and 7}",
      journal = {\aap},
         year = 2025,
        month = jul,
       volume = {699},
          eid = {A358},
        pages = {A358},
          doi = {10.1051/0004-6361/202553908},
archivePrefix = {arXiv},
       eprint = {2503.01990},
 primaryClass = {astro-ph.GA},
       adsurl = {https://ui.adsabs.harvard.edu/abs/2025A&A...699A.358V}
}

@ARTICLE{Ito2025,
       author = {{Ito}, K. and {Valentino}, F. and {Farcy}, M. and {De Lucia}, G. and {Lagos}, C.~D.~P. and {Hirschmann}, M. and {Brammer}, G. and {de Graaff}, A. and {Bl{\'a}nquez-Ses{\'e}}, D. and {Ceverino}, D. and {Faisst}, A.~L. and {Fontanot}, F. and {Gillman}, S. and {Hamadouche}, M.~L. and {Heintz}, K.~E. and {Jin}, S. and {Jespersen}, C.~K. and {Kubo}, M. and {Lee}, M. and {Magdis}, G. and {Man}, A.~W.~S. and {Onodera}, M. and {Rizzo}, F. and {Shimakawa}, R. and {Tanaka}, M. and {Toft}, S. and {Whitaker}, K.~E. and {Xie}, L. and {Zhu}, P.},
        title = "{A merging pair of massive quiescent galaxies at z = 3.44 in the Cosmic Vine}",
      journal = {\aap},
         year = 2025,
        month = may,
       volume = {697},
          eid = {A111},
        pages = {A111},
          doi = {10.1051/0004-6361/202453211},
archivePrefix = {arXiv},
       eprint = {2503.01953},
 primaryClass = {astro-ph.GA},
       adsurl = {https://ui.adsabs.harvard.edu/abs/2025A&A...697A.111I}
}

@ARTICLE{Grudic2019,
       author = {{Grudi{\'c}}, Michael Y. and {Hopkins}, Philip F.},
        title = "{The elephant in the room: the importance of the details of massive star formation in molecular clouds}",
      journal = {\mnras},
         year = 2019,
        month = sep,
       volume = {488},
       number = {2},
        pages = {2970-2975},
          doi = {10.1093/mnras/stz1820},
archivePrefix = {arXiv},
       eprint = {1809.08344},
 primaryClass = {astro-ph.GA},
       adsurl = {https://ui.adsabs.harvard.edu/abs/2019MNRAS.488.2970G}
}

@ARTICLE{McLure2013,
       author = {{McLure}, R.~J. and {Pearce}, H.~J. and {Dunlop}, J.~S. and {Cirasuolo}, M. and {Curtis-Lake}, E. and {Bruce}, V.~A. and {Caputi}, K.~I. and {Almaini}, O. and {Bonfield}, D.~G. and {Bradshaw}, E.~J. and {Buitrago}, F. and {Chuter}, R. and {Foucaud}, S. and {Hartley}, W.~G. and {Jarvis}, M.~J.},
        title = "{The sizes, masses and specific star formation rates of massive galaxies at 1.3 < z < 1.5: strong evidence in favour of evolution via minor mergers}",
      journal = {\mnras},
         year = 2013,
        month = jan,
       volume = {428},
       number = {2},
        pages = {1088-1106},
          doi = {10.1093/mnras/sts092},
archivePrefix = {arXiv},
       eprint = {1205.4058},
 primaryClass = {astro-ph.CO},
       adsurl = {https://ui.adsabs.harvard.edu/abs/2013MNRAS.428.1088M}
}

@ARTICLE{Hamadouche2022,
       author = {{Hamadouche}, M.~L. and {Carnall}, A.~C. and {McLure}, R.~J. and {Dunlop}, J.~S. and {McLeod}, D.~J. and {Cullen}, F. and {Begley}, R. and {Bolzonella}, M. and {Buitrago}, F. and {Castellano}, M. and {Cucciati}, O. and {Fontana}, A. and {Gargiulo}, A. and {Moresco}, M. and {Pozzetti}, L. and {Zamorani}, G.},
        title = "{A combined VANDELS and LEGA-C study: the evolution of quiescent galaxy size, stellar mass, and age from z = 0.6 to z = 1.3}",
      journal = {\mnras},
         year = 2022,
        month = may,
       volume = {512},
       number = {1},
        pages = {1262-1274},
          doi = {10.1093/mnras/stac535},
archivePrefix = {arXiv},
       eprint = {2201.10576},
 primaryClass = {astro-ph.GA},
       adsurl = {https://ui.adsabs.harvard.edu/abs/2022MNRAS.512.1262H}
}

@ARTICLE{Suess2023,
       author = {{Suess}, Katherine A. and {Williams}, Christina C. and {Robertson}, Brant and {Ji}, Zhiyuan and {Johnson}, Benjamin D. and {Nelson}, Erica and {Alberts}, Stacey and {Hainline}, Kevin and {D'Eugenio}, Francesco and {{\"U}bler}, Hannah and {Rieke}, Marcia and {Rieke}, George and {Bunker}, Andrew J. and {Carniani}, Stefano and {Charlot}, Stephane and {Eisenstein}, Daniel J. and {Maiolino}, Roberto and {Stark}, Daniel P. and {Tacchella}, Sandro and {Willott}, Chris},
        title = "{Minor Merger Growth in Action: JWST Detects Faint Blue Companions around Massive Quiescent Galaxies at 0.5 {\ensuremath{\leq}} z {\ensuremath{\leq}} 3.0}",
      journal = {\apjl},
         year = 2023,
        month = oct,
       volume = {956},
       number = {2},
          eid = {L42},
        pages = {L42},
          doi = {10.3847/2041-8213/acf5e6},
archivePrefix = {arXiv},
       eprint = {2307.14209},
 primaryClass = {astro-ph.GA},
       adsurl = {https://ui.adsabs.harvard.edu/abs/2023ApJ...956L..42S}
}

@ARTICLE{Finkelstein2025,
       author = {{Finkelstein}, Steven L. and {Bagley}, Micaela B. and {Arrabal Haro}, Pablo and {Dickinson}, Mark and {Ferguson}, Henry C. and {Kartaltepe}, Jeyhan S. and {Kocevski}, Dale D. and {Koekemoer}, Anton M. and {Lotz}, Jennifer M. and {Papovich}, Casey and {P{\'e}rez-Gonz{\'a}lez}, Pablo G. and {Pirzkal}, Nor and {Somerville}, Rachel S. and {Trump}, Jonathan R. and {Yang}, Guang and {Yung}, L.~Y. Aaron and {Fontana}, Adriano and {Grazian}, Andrea and {Grogin}, Norman A. and {Kewley}, Lisa J. and {Kirkpatrick}, Allison and {Larson}, Rebecca L. and {Pentericci}, Laura and {Ravindranath}, Swara and {Wilkins}, Stephen M. and {Almaini}, Omar and {Amor{\'\i}n}, Ricardo O. and {Barro}, Guillermo and {Bhatawdekar}, Rachana and {Bisigello}, Laura and {Brooks}, Madisyn and {Buat}, V{\'e}ronique and {Buitrago}, Fernando and {Burgarella}, Denis and {Calabr{\`o}}, Antonello and {Castellano}, Marco and {Cheng}, Yingjie and {Cleri}, Nikko J. and {Cole}, Justin W. and {Cooper}, M.~C. and {Cooper}, Olivia R. and {Costantin}, Luca and {Cox}, Isa G. and {Croton}, Darren and {Daddi}, Emanuele and {Davis}, Kelcey and {Dekel}, Avishai and {Elbaz}, David and {Fern{\'a}ndez}, Vital and {Fujimoto}, Seiji and {Gandolfi}, Giovanni and {Gardner}, Jonathan P. and {Gawiser}, Eric and {Giavalisco}, Mauro and {G{\'o}mez-Guijarro}, Carlos and {Guo}, Yuchen and {Gupta}, Ansh R. and {Hathi}, Nimish P. and {Harish}, Santosh and {Henry}, Aur{\'e}lien and {Hirschmann}, Michaela and {Hu}, Weida and {Hutchison}, Taylor A. and {Iyer}, Kartheik G. and {Jaskot}, Anne E. and {Jha}, Saurabh W. and {Jung}, Intae and {Kassin}, Susan A. and {Kokorev}, Vasily and {Kurczynski}, Peter and {Leung}, Gene C.~K. and {Llerena}, Mario and {Long}, Arianna S. and {Lucas}, Ray A. and {Lu}, Shiying and {McGrath}, Elizabeth J. and {McIntosh}, Daniel H. and {Merlin}, Emiliano and {Mobasher}, Bahram and {Morales}, Alexa M. and {Napolitano}, Lorenzo and {Pacucci}, Fabio and {Pandya}, Viraj and {Rafelski}, Marc and {Rodighiero}, Giulia and {Rose}, Caitlin and {Santini}, Paola and {Seill{\'e}}, Lise-Marie and {Simons}, Raymond C. and {Shen}, Lu and {Straughn}, Amber N. and {Tacchella}, Sandro and {Taylor}, Anthony J. and {Vanderhoof}, Brittany N. and {Vega-Ferrero}, Jes{\'u}s and {Weiner}, Benjamin J. and {Willmer}, Christopher N.~A. and {Zhu}, Peixin and {Bell}, Eric F. and {Wuyts}, Stijn and {Holwerda}, Benne W. and {Wang}, Xin and {Wang}, Weichen and {Zavala}, Jorge A. and {CEERS Collaboration}},
        title = "{The Cosmic Evolution Early Release Science Survey (CEERS)}",
      journal = {\apjl},
         year = 2025,
        month = apr,
       volume = {983},
       number = {1},
          eid = {L4},
        pages = {L4},
          doi = {10.3847/2041-8213/adbbd3},
archivePrefix = {arXiv},
       eprint = {2501.04085},
 primaryClass = {astro-ph.GA},
       adsurl = {https://ui.adsabs.harvard.edu/abs/2025ApJ...983L...4F}
}

@ARTICLE{Adamo2024,
       author = {{Adamo}, Angela and {Atek}, Hakim and {Bagley}, Micaela B. and {Ba{\~n}ados}, Eduardo and {Barrow}, Kirk S.~S. and {Berg}, Danielle A. and {Bezanson}, Rachel and {Brada{\v{c}}}, Maru{\v{s}}a and {Brammer}, Gabriel and {Carnall}, Adam C. and {Chisholm}, John and {Coe}, Dan and {Dayal}, Pratika and {Eisenstein}, Daniel J. and {Eldridge}, Jan J. and {Ferrara}, Andrea and {Fujimoto}, Seiji and {Graaff}, Anna de and {Habouzit}, Melanie and {Hutchison}, Taylor A. and {Kartaltepe}, Jeyhan S. and {Kassin}, Susan A. and {Kriek}, Mariska and {Labb{\'e}}, Ivo and {Maiolino}, Roberto and {Marques-Chaves}, Rui and {Maseda}, Michael V. and {Mason}, Charlotte and {Matthee}, Jorryt and {McQuinn}, Kristen B.~W. and {Meynet}, Georges and {Naidu}, Rohan P. and {Oesch}, Pascal A. and {Pentericci}, Laura and {P{\'e}rez-Gonz{\'a}lez}, Pablo G. and {Rigby}, Jane R. and {Roberts-Borsani}, Guido and {Schaerer}, Daniel and {Shapley}, Alice E. and {Stark}, Daniel P. and {Stiavelli}, Massimo and {Strom}, Allison L. and {Vanzella}, Eros and {Wang}, Feige and {Wilkins}, Stephen M. and {Williams}, Christina C. and {Willott}, Chris J. and {Wylezalek}, Dominika and {Nota}, Antonella},
        title = "{The first billion years according to JWST}",
      journal = {Nature Astronomy},
         year = 2025,
        month = aug,
       volume = {9},
        pages = {1134-1147},
          doi = {10.1038/s41550-025-02624-5},
archivePrefix = {arXiv},
       eprint = {2405.21054},
 primaryClass = {astro-ph.GA},
       adsurl = {https://ui.adsabs.harvard.edu/abs/2025NatAs...9.1134A}
}

@article{Valentino2023,
	author = {Valentino, Francesco and Brammer, Gabriel and Gould, Katriona M. L. and Kokorev, Vasily and Fujimoto, Seiji and Jespersen, Christian Kragh and Vijayan, Aswin P. and Weaver, John R. and Ito, Kei and Tanaka, Masayuki and Ilbert, Olivier and Magdis, Georgios E. and Whitaker, Katherine E. and Faisst, Andreas L. and Gallazzi, Anna and Gillman, Steven and Gim{\'e}nez-Arteaga, Clara and G{\'o}mez-Guijarro, Carlos and Kubo, Mariko and Heintz, Kasper E. and Hirschmann, Michaela and Oesch, Pascal and Onodera, Masato and Rizzo, Francesca and Lee, Minju and Strait, Victoria and Toft, Sune},
	doi = {10.3847/1538-4357/acbefa},
	issn = {0004-637X},
	journal = {\apj},
	month = apr,
	note = {Publisher: IOP ADS Bibcode: 2023ApJ...947...20V},
	pages = {20},
	title = {An {Atlas} of {Color}-selected {Quiescent} {Galaxies} at z {\textgreater} 3 in {Public} {JWST} {Fields}},
	url = {https://ui.adsabs.harvard.edu/abs/2023ApJ...947...20V},
	urldate = {2024-09-11},
	volume = {947},
	year = {2023}}

@ARTICLE{Carnall2019a,
       author = {{Carnall}, Adam C. and {Leja}, Joel and {Johnson}, Benjamin D. and {McLure}, Ross J. and {Dunlop}, James S. and {Conroy}, Charlie},
        title = "{How to Measure Galaxy Star Formation Histories. I. Parametric Models}",
      journal = {\apj},
         year = 2019,
        month = mar,
       volume = {873},
       number = {1},
          eid = {44},
        pages = {44},
          doi = {10.3847/1538-4357/ab04a2},
archivePrefix = {arXiv},
       eprint = {1811.03635},
 primaryClass = {astro-ph.GA},
       adsurl = {https://ui.adsabs.harvard.edu/abs/2019ApJ...873...44C}
}

@article{Russel2024,
    author = {Russell, Tobias A and Dobric, Neva and Adams, Nathan J and Conselice, Christopher J and Austin, Duncan and Harvey, Thomas and Trussler, James A A and Ferreira, Leonardo and Westcott, Lewi and Harris, Honor and Windhorst, Rogier A and Coe, Dan and Cohen, Seth H and Driver, Simon P and Frye, Brenda and Grogin, Norman A and Hathi, Nimish P and Jansen, Rolf A and Koekemoer, Anton M and Marshall, Madeline A and Ortiz III, Rafael and Pirzkal, Nor and Robotham, Aaron and Ryan, Russell E, Jr and Summers, Jake and D’Silva, Jordan C J and Willmer, Christopher N A and Yan, Haojing},
    title = {Cosmic stillness: high quiescent galaxy fractions across upper mass scales in the early Universe to z = 7 with JWST},
    journal = {\mnras},
    volume = {544},
    number = {4},
    pages = {4482-4504},
    year = {2025},
    month = {11},
    issn = {0035-8711},
    doi = {10.1093/mnras/staf1916},
    url = {https://doi.org/10.1093/mnras/staf1916},
    eprint = {https://academic.oup.com/mnras/article-pdf/544/4/4482/65236122/staf1916.pdf},
}

@article{Carnall2023a,
	author = {Carnall, A. C. and McLeod, D. J. and McLure, R. J. and Dunlop, J. S. and Begley, R. and Cullen, F. and Donnan, C. T. and Hamadouche, M. L. and Jewell, S. M. and Jones, E. W. and Pollock, C. L. and Wild, V.},
	doi = {10.1093/mnras/stad369},
	issn = {0035-8711},
	journal = {\mnras},
	month = apr,
	note = {Publisher: OUP ADS Bibcode: 2023MNRAS.520.3974C},
	pages = {3974--3985},
	title = {A surprising abundance of massive quiescent galaxies at 3 {\textless} z {\textless} 5 in the first data from {JWST} {CEERS}},
	url = {https://ui.adsabs.harvard.edu/abs/2023MNRAS.520.3974C},
	urldate = {2024-09-11},
	volume = {520},
	year = {2023}}

@article{Carnall2024,
	adsurl = {https://ui.adsabs.harvard.edu/abs/2024MNRAS.534..325C},
	archiveprefix = {arXiv},
	author = {{Carnall}, A.~C. and {Cullen}, F. and {McLure}, R.~J. and {McLeod}, D.~J. and {Begley}, R. and {Donnan}, C.~T. and {Dunlop}, J.~S. and {Shapley}, A.~E. and {Rowlands}, K. and {Almaini}, O. and {Arellano-C{\'o}rdova}, K.~Z. and {Barrufet}, L. and {Cimatti}, A. and {Ellis}, R.~S. and {Grogin}, N.~A. and {Hamadouche}, M.~L. and {Illingworth}, G.~D. and {Koekemoer}, A.~M. and {Leung}, H. -H. and {Lovell}, C.~C. and {P{\'e}rez-Gonz{\'a}lez}, P.~G. and {Santini}, P. and {Stanton}, T.~M. and {Wild}, V.},
	doi = {10.1093/mnras/stae2092},
	eprint = {2405.02242},
	journal = {\mnras},
	month = oct,
	number = {1},
	pages = {325-348},
	primaryclass = {astro-ph.GA},
	title = {{The JWST EXCELS survey: too much, too young, too fast? Ultra-massive quiescent galaxies at 3 < z < 5}},
	volume = {534},
	year = 2024}

@article{Weibel2024,
       author = {{Weibel}, Andrea and {de Graaff}, Anna and {Setton}, David J. and {Miller}, Tim B. and {Oesch}, Pascal A. and {Brammer}, Gabriel and {Lagos}, Claudia D.~P. and {Whitaker}, Katherine E. and {Williams}, Christina C. and {Baggen}, Josephine F.~W. and {Bezanson}, Rachel and {Boogaard}, Leindert A. and {Cleri}, Nikko J. and {Greene}, Jenny E. and {Hirschmann}, Michaela and {Hviding}, Raphael E. and {Kuruvanthodi}, Adarsh and {Labb{\'e}}, Ivo and {Leja}, Joel and {Maseda}, Michael V. and {Matthee}, Jorryt and {McConachie}, Ian and {Naidu}, Rohan P. and {Roberts-Borsani}, Guido and {Schaerer}, Daniel and {Suess}, Katherine A. and {Valentino}, Francesco and {van Dokkum}, Pieter and {Wang}, Bingjie},
        title = "{RUBIES Reveals a Massive Quiescent Galaxy at z = 7.3}",
      journal = {\apj},
         year = 2025,
        month = apr,
       volume = {983},
       number = {1},
          eid = {11},
        pages = {11},
          doi = {10.3847/1538-4357/adab7a},
archivePrefix = {arXiv},
       eprint = {2409.03829},
 primaryClass = {astro-ph.GA},
       adsurl = {https://ui.adsabs.harvard.edu/abs/2025ApJ...983...11W}
}

@article{Schreiber2018,
	author = {Schreiber, C. and Glazebrook, K. and Nanayakkara, T. and Kacprzak, G. G. and Labb{\'e}, I. and Oesch, P. and Yuan, T. and Tran, K. -V. and Papovich, C. and Spitler, L. and Straatman, C.},
	doi = {10.1051/0004-6361/201833070},
	issn = {0004-6361},
	journal = {\aap},
	month = oct,
	note = {ADS Bibcode: 2018A\&A...618A..85S},
	pages = {A85},
	title = {Near infrared spectroscopy and star-formation histories of 3 ≤ z ≤ 4 quiescent galaxies},
	url = {https://ui.adsabs.harvard.edu/abs/2018A&A...618A..85S},
	urldate = {2024-09-11},
	volume = {618},
	year = {2018}}

@article{Merlin2019,
	adsurl = {https://ui.adsabs.harvard.edu/abs/2019MNRAS.490.3309M},
	archiveprefix = {arXiv},
	author = {{Merlin}, E. and {Fortuni}, F. and {Torelli}, M. and {Santini}, P. and {Castellano}, M. and {Fontana}, A. and {Grazian}, A. and {Pentericci}, L. and {Pilo}, S. and {Schmidt}, K.~B.},
	doi = {10.1093/mnras/stz2615},
	eprint = {1909.07996},
	journal = {\mnras},
	month = dec,
	number = {3},
	pages = {3309-3328},
	primaryclass = {astro-ph.GA},
	title = {{Red and dead CANDELS: massive passive galaxies at the dawn of the Universe}},
	volume = {490},
	year = 2019}

@article{Glazebrook2024,
	author = {Glazebrook, Karl and Nanayakkara, Themiya and Schreiber, Corentin and Lagos, Claudia and Kawinwanichakij, Lalitwadee and Jacobs, Colin and Chittenden, Harry and Brammer, Gabriel and Kacprzak, Glenn G. and Labbe, Ivo and Marchesini, Danilo and Marsan, Z. Cemile and Oesch, Pascal A. and Papovich, Casey and Remus, Rhea-Silvia and Tran, Kim-Vy H. and Esdaile, James and Chandro-Gomez, Angel},
	doi = {10.1038/s41586-024-07191-9},
	issn = {0028-0836},
	journal = {Nature},
	month = apr,
	note = {ADS Bibcode: 2024Natur.628..277G},
	pages = {277--281},
	title = {A massive galaxy that formed its stars at z ≈ 11},
	url = {https://ui.adsabs.harvard.edu/abs/2024Natur.628..277G},
	urldate = {2024-09-11},
	volume = {628},
	year = {2024}}

@article{Fontana2009,
	adsurl = {https://ui.adsabs.harvard.edu/abs/2009A&A...501...15F},
	archiveprefix = {arXiv},
	author = {{Fontana}, A. and {Santini}, P. and {Grazian}, A. and {Pentericci}, L. and {Fiore}, F. and {Castellano}, M. and {Giallongo}, E. and {Menci}, N. and {Salimbeni}, S. and {Cristiani}, S. and {Nonino}, M. and {Vanzella}, E.},
	doi = {10.1051/0004-6361/200911650},
	eprint = {0901.2898},
	journal = {\aap},
	month = jul,
	number = {1},
	pages = {15-20},
	primaryclass = {astro-ph.GA},
	title = {{The fraction of quiescent massive galaxies in the early Universe}},
	volume = {501},
	year = 2009}

@article{Williams2009,
	adsurl = {https://ui.adsabs.harvard.edu/abs/2009ApJ...691.1879W},
	archiveprefix = {arXiv},
	author = {{Williams}, Rik J. and {Quadri}, Ryan F. and {Franx}, Marijn and {van Dokkum}, Pieter and {Labb{\'e}}, Ivo},
	doi = {10.1088/0004-637X/691/2/1879},
	eprint = {0806.0625},
	journal = {\apj},
	month = feb,
	number = {2},
	pages = {1879-1895},
	primaryclass = {astro-ph},
	title = {{Detection of Quiescent Galaxies in a Bicolor Sequence from Z = 0-2}},
	volume = {691},
	year = 2009}

@article{Carnall2018,
	adsurl = {https://ui.adsabs.harvard.edu/abs/2018MNRAS.480.4379C},
	archiveprefix = {arXiv},
	author = {{Carnall}, A.~C. and {McLure}, R.~J. and {Dunlop}, J.~S. and {Dav{\'e}}, R.},
	doi = {10.1093/mnras/sty2169},
	eprint = {1712.04452},
	journal = {\mnras},
	month = nov,
	number = {4},
	pages = {4379-4401},
	primaryclass = {astro-ph.GA},
	title = {{Inferring the star formation histories of massive quiescent galaxies with BAGPIPES: evidence for multiple quenching mechanisms}},
	volume = {480},
	year = 2018}

@article{Carnall2020,
	author = {Carnall, A. C. and Walker, S. and McLure, R. J. and Dunlop, J. S. and McLeod, D. J. and Cullen, F. and Wild, V. and Amorin, R. and Bolzonella, M. and Castellano, M. and Cimatti, A. and Cucciati, O. and Fontana, A. and Gargiulo, A. and Garilli, B. and Jarvis, M. J. and Pentericci, L. and Pozzetti, L. and Zamorani, G. and Calabro, A. and Hathi, N. P. and Koekemoer, A. M.},
	doi = {10.1093/mnras/staa1535},
	issn = {0035-8711},
	journal = {\mnras},
	month = jul,
	note = {Publisher: OUP ADS Bibcode: 2020MNRAS.496..695C},
	pages = {695--707},
	title = {Timing the earliest quenching events with a robust sample of massive quiescent galaxies at 2 {\textless} z {\textless} 5},
	url = {https://ui.adsabs.harvard.edu/abs/2020MNRAS.496..695C},
	urldate = {2024-09-11},
	volume = {496},
	year = {2020}}

@article{Lagos2024,
	author = {Lagos, Claudia del P. and Bravo, Mat{\'\i}as and Tobar, Rodrigo and Obreschkow, Danail and Power, Chris and Robotham, Aaron S. G. and Proctor, Katy L. and Hansen, Samuel and Chandro-G{\'o}mez, {\'A}ngel and Carrivick, Julian},
	doi = {10.1093/mnras/stae1024},
	issn = {0035-8711},
	journal = {\mnras},
	month = jul,
	note = {Publisher: OUP ADS Bibcode: 2024MNRAS.531.3551L},
	pages = {3551--3578},
	title = {Quenching massive galaxies across cosmic time with the semi-analytic model {SHARK} {V2}.0},
	url = {https://ui.adsabs.harvard.edu/abs/2024MNRAS.531.3551L},
	urldate = {2024-09-11},
	volume = {531},
	year = {2024}}

@article{Cecchi2019,
	adsurl = {https://ui.adsabs.harvard.edu/abs/2019ApJ...880L..14C},
	archiveprefix = {arXiv},
	author = {{Cecchi}, Rachele and {Bolzonella}, Micol and {Cimatti}, Andrea and {Girelli}, Giacomo},
	doi = {10.3847/2041-8213/ab2c80},
	eid = {L14},
	eprint = {1906.11842},
	journal = {\apjl},
	month = jul,
	number = {1},
	pages = {L14},
	primaryclass = {astro-ph.GA},
	title = {{Quiescent Galaxies at z {\ensuremath{\gtrsim}} 2.5: Observations versus Models}},
	volume = {880},
	year = 2019}

@article{Girelli2019,
	adsurl = {https://ui.adsabs.harvard.edu/abs/2019A&A...632A..80G},
	archiveprefix = {arXiv},
	author = {{Girelli}, Giacomo and {Bolzonella}, Micol and {Cimatti}, Andrea},
	doi = {10.1051/0004-6361/201834547},
	eid = {A80},
	eprint = {1910.07544},
	journal = {\aap},
	month = dec,
	pages = {A80},
	primaryclass = {astro-ph.GA},
	title = {{Massive and old quiescent galaxies at high redshift}},
	volume = {632},
	year = 2019}

@misc{Dunlop2021,
	adsurl = {https://ui.adsabs.harvard.edu/abs/2021jwst.prop.1837D},
	author = {{Dunlop}, James S. and {Abraham}, Roberto G. and {Ashby}, Matthew L.~N. and {Bagley}, Micaela and {Best}, Philip N. and {Bongiorno}, Angela and {Bouwens}, Rychard and {Bowler}, Rebecca A.~A. and {Brammer}, Gabriel and {Bremer}, Malcolm and {Calabro'}, Antonello and {Carnall}, Adam and {Castellano}, Marco and {Cirasuolo}, Michele and {Conselice}, Christopher and {Cullen}, Fergus and {Dave}, Romeel and {Dayal}, Pratika and {Dekel}, Avishai and {Dickinson}, Mark and {Duncan}, Kenneth James and {Elbaz}, David and {Ellis}, Richard S. and {Ferguson}, Harry C. and {Ferrara}, Andrea and {Finkelstein}, Steven L. and {Fontana}, Adriano and {Furlanetto}, Steven and {Fynbo}, Johan P.~U. and {Gallerani}, Simona and {Gardner}, Jonathan P. and {Giavalisco}, Mauro and {Grazian}, Andrea and {Grogin}, Norman and {Harikane}, Yuichi and {Hopkins}, Philip F. and {Ilbert}, Olivier and {Illingworth}, Garth D. and {Juneau}, Stephanie and {Jung}, Intae and {Kartaltepe}, Jeyhan and {Kassin}, Susan and {Kauffmann}, Olivier Benjamin and {Khochfar}, Sadegh and {Kirkpatrick}, Allison and {Kocevski}, Dale D. and {Koekemoer}, Anton M. and {Labbe}, Ivo and {Laporte}, Nicolas and {Larson}, Rebecca L. and {Lucas}, Ray A. and {Magee}, Daniel K. and {Mason}, Charlotte and {McCracken}, Henry Joy and {McLeod}, Derek and {McLure}, Ross and {Merlin}, Emiliano and {Mesinger}, Andrei and {Milvang-Jensen}, Bo and {Newman}, Jeffrey Allen and {Oesch}, Pascal and {Ouchi}, Masami and {Pacifici}, Camilla and {Papovich}, Casey and {Peacock}, John and {Peeples}, Molly and {Pentericci}, Laura and {Perez-Gonzalez}, Pablo G. and {Pirzkal}, Norbert and {Pope}, Alexandra and {Pye}, John P. and {Reddy}, Naveen A. and {Robertson}, Brant and {Salvato}, Mara and {Santini}, Paola and {Schaerer}, Daniel and {Shapley}, Alice E. and {Simons}, Raymond and {Smit}, Renske and {Smith}, Britton D. and {Snyder}, Greg and {Somerville}, Rachel S. and {Stanway}, Elizabeth R. and {Stefanon}, Mauro and {Tasca}, Lidia and {Tikkanen}, Tuomo and {Tresse}, Laurence and {Trump}, Jonathan R. and {Whitaker}, Katherine E. and {Wilkins}, Stephen Matthew and {Wright}, Gillian and {Wyithe}, J. Stuart B. and {van Dokkum}, Pieter and {van der Werf}, Paul},
	howpublished = {JWST Proposal. Cycle 1, ID. \#1837},
	month = mar,
	pages = {1837},
	title = {{PRIMER: Public Release IMaging for Extragalactic Research}},
	year = 2021}

@article{Eisenstein2023b,
	adsurl = {https://ui.adsabs.harvard.edu/abs/2023arXiv230602465E},
	archiveprefix = {arXiv},
	author = {{Eisenstein}, Daniel J. and {Willott}, Chris and {Alberts}, Stacey and {Arribas}, Santiago and {Bonaventura}, Nina and {Bunker}, Andrew J. and {Cameron}, Alex J. and {Carniani}, Stefano and {Charlot}, Stephane and {Curtis-Lake}, Emma and {D'Eugenio}, Francesco and {Endsley}, Ryan and {Ferruit}, Pierre and {Giardino}, Giovanna and {Hainline}, Kevin and {Hausen}, Ryan and {Jakobsen}, Peter and {Johnson}, Benjamin D. and {Maiolino}, Roberto and {Rieke}, Marcia and {Rieke}, George and {Rix}, Hans-Walter and {Robertson}, Brant and {Stark}, Daniel P. and {Tacchella}, Sandro and {Williams}, Christina C. and {Willmer}, Christopher N.~A. and {Baker}, William M. and {Baum}, Stefi and {Bhatawdekar}, Rachana and {Boyett}, Kristan and {Chen}, Zuyi and {Chevallard}, Jacopo and {Circosta}, Chiara and {Curti}, Mirko and {Danhaive}, A. Lola and {DeCoursey}, Christa and {de Graaff}, Anna and {Dressler}, Alan and {Egami}, Eiichi and {Helton}, Jakob M. and {Hviding}, Raphael E. and {Ji}, Zhiyuan and {Jones}, Gareth C. and {Kumari}, Nimisha and {L{\"u}tzgendorf}, Nora and {Laseter}, Isaac and {Looser}, Tobias J. and {Lyu}, Jianwei and {Maseda}, Michael V. and {Nelson}, Erica and {Parlanti}, Eleonora and {Perna}, Michele and {Pusk{\'a}s}, D{\'a}vid and {Rawle}, Tim and {Rodr{\'\i}guez Del Pino}, Bruno and {Sandles}, Lester and {Saxena}, Aayush and {Scholtz}, Jan and {Sharpe}, Katherine and {Shivaei}, Irene and {Silcock}, Maddie S. and {Simmonds}, Charlotte and {Skarbinski}, Maya and {Smit}, Renske and {Stone}, Meredith and {Suess}, Katherine A. and {Sun}, Fengwu and {Tang}, Mengtao and {Topping}, Michael W. and {{\"U}bler}, Hannah and {Villanueva}, Natalia C. and {Wallace}, Imaan E.~B. and {Whitler}, Lily and {Witstok}, Joris and {Woodrum}, Charity},
	doi = {10.48550/arXiv.2306.02465},
	eid = {arXiv:2306.02465},
	eprint = {2306.02465},
	journal = {arXiv e-prints},
	month = jun,
	pages = {arXiv:2306.02465},
	primaryclass = {astro-ph.GA},
	title = {{Overview of the JWST Advanced Deep Extragalactic Survey (JADES)}},
	year = 2023}

@ARTICLE{Barrufet2025,
       author = {{Barrufet}, L. and {Oesch}, P.~A. and {Marques-Chaves}, R. and {Arellano-Cordova}, K. and {Baggen}, J.~F.~W. and {Carnall}, A.~C. and {Cullen}, F. and {Dunlop}, J.~S. and {Gottumukkala}, R. and {Fudamoto}, Y. and {Illingworth}, G.~D. and {Magee}, D. and {McLure}, R.~J. and {McLeod}, D.~J. and {Micha{\l}owski}, M.~J. and {Stefanon}, M. and {van Dokkum}, P.~G. and {Weibel}, A.},
        title = "{Quiescent or dusty? Unveiling the nature of extremely red galaxies at z > 3}",
      journal = {\mnras},
         year = 2025,
        month = mar,
       volume = {537},
       number = {4},
        pages = {3453-3469},
          doi = {10.1093/mnras/staf013},
archivePrefix = {arXiv},
       eprint = {2404.08052},
 primaryClass = {astro-ph.GA},
       adsurl = {https://ui.adsabs.harvard.edu/abs/2025MNRAS.537.3453B}
}

@ARTICLE{dEugenio2025,
       author = {{D'Eugenio}, Francesco and {Cameron}, Alex J. and {Scholtz}, Jan and {Carniani}, Stefano and {Willott}, Chris J. and {Curtis-Lake}, Emma and {Bunker}, Andrew J. and {Parlanti}, Eleonora and {Maiolino}, Roberto and {Willmer}, Christopher N.~A. and {Jakobsen}, Peter and {Robertson}, Brant E. and {Johnson}, Benjamin D. and {Tacchella}, Sandro and {Cargile}, Phillip A. and {Rawle}, Tim and {Arribas}, Santiago and {Chevallard}, Jacopo and {Curti}, Mirko and {Egami}, Eiichi and {Eisenstein}, Daniel J. and {Kumari}, Nimisha and {Looser}, Tobias J. and {Rieke}, Marcia J. and {Rodr{\'\i}guez Del Pino}, Bruno and {Saxena}, Aayush and {{\"U}bler}, Hannah and {Venturi}, Giacomo and {Witstok}, Joris and {Baker}, William M. and {Bhatawdekar}, Rachana and {Bonaventura}, Nina and {Boyett}, Kristan and {Charlot}, Stephane and {Danhaive}, A. Lola and {Hainline}, Kevin N. and {Hausen}, Ryan and {Helton}, Jakob M. and {Ji}, Xihan and {Ji}, Zhiyuan and {Jones}, Gareth C. and {Juod{\v{z}}balis}, Ignas and {Maseda}, Michael V. and {P{\'e}rez-Gonz{\'a}lez}, Pablo G. and {Perna}, Michele and {Pusk{\'a}s}, D{\'a}vid and {Shivaei}, Irene and {Silcock}, Maddie S. and {Simmonds}, Charlotte and {Smit}, Renske and {Sun}, Fengwu and {Villanueva}, Natalia C. and {Williams}, Christina C. and {Zhu}, Yongda},
        title = "{JADES Data Release 3: NIRSpec/Microshutter Assembly Spectroscopy for 4000 Galaxies in the GOODS Fields}",
      journal = {\apjs},
         year = 2025,
        month = mar,
       volume = {277},
       number = {1},
          eid = {4},
        pages = {4},
          doi = {10.3847/1538-4365/ada148},
archivePrefix = {arXiv},
       eprint = {2404.06531},
 primaryClass = {astro-ph.GA},
       adsurl = {https://ui.adsabs.harvard.edu/abs/2025ApJS..277....4D}
}

@ARTICLE{dEugenio2023,
       author = {{D'Eugenio}, Francesco and {Maiolino}, Roberto and {Carniani}, Stefano and {Chevallard}, Jacopo and {Curtis-Lake}, Emma and {Witstok}, Joris and {Charlot}, Stephane and {Baker}, William M. and {Arribas}, Santiago and {Boyett}, Kristan and {Bunker}, Andrew J. and {Curti}, Mirko and {Eisenstein}, Daniel J. and {Hainline}, Kevin and {Ji}, Zhiyuan and {Johnson}, Benjamin D. and {Kumari}, Nimisha and {Looser}, Tobias J. and {Nakajima}, Kimihiko and {Nelson}, Erica and {Rieke}, Marcia and {Robertson}, Brant and {Scholtz}, Jan and {Smit}, Renske and {Sun}, Fengwu and {Venturi}, Giacomo and {Tacchella}, Sandro and {{\"U}bler}, Hannah and {Willmer}, Christopher N.~A. and {Willott}, Chris},
        title = "{JADES: Carbon enrichment 350 Myr after the Big Bang}",
      journal = {\aap},
         year = 2024,
        month = sep,
       volume = {689},
          eid = {A152},
        pages = {A152},
          doi = {10.1051/0004-6361/202348636},
archivePrefix = {arXiv},
       eprint = {2311.09908},
 primaryClass = {astro-ph.GA},
       adsurl = {https://ui.adsabs.harvard.edu/abs/2024A&A...689A.152D}
}

@article{Carnall2023b,
	author = {Carnall, Adam C. and McLure, Ross J. and Dunlop, James S. and McLeod, Derek J. and Wild, Vivienne and Cullen, Fergus and Magee, Dan and Begley, Ryan and Cimatti, Andrea and Donnan, Callum T. and Hamadouche, Massissilia L. and Jewell, Sophie M. and Walker, Sam},
	doi = {10.1038/s41586-023-06158-6},
	issn = {0028-0836},
	journal = {Nature},
	month = jul,
	note = {ADS Bibcode: 2023Natur.619..716C},
	pages = {716--719},
	title = {A massive quiescent galaxy at redshift 4.658},
	url = {https://ui.adsabs.harvard.edu/abs/2023Natur.619..716C},
	urldate = {2024-09-11},
	volume = {619},
	year = {2023}}

@article{Nanayakkara2024,
	author = {Nanayakkara, Themiya and Glazebrook, Karl and Jacobs, Colin and Kawinwanichakij, Lalitwadee and Schreiber, Corentin and Brammer, Gabriel and Esdaile, James and Kacprzak, Glenn G. and Labbe, Ivo and Lagos, Claudia and Marchesini, Danilo and Marsan, Z. Cemile and Oesch, Pascal A. and Papovich, Casey and Remus, Rhea-Silvia and Tran, Kim-Vy H.},
	doi = {10.1038/s41598-024-52585-4},
	issn = {2045-2322},
	journal = {Scientific Reports},
	month = feb,
	note = {ADS Bibcode: 2024NatSR..14.3724N},
	pages = {3724},
	title = {A population of faint, old, and massive quiescent galaxies at 3},
	url = {https://ui.adsabs.harvard.edu/abs/2024NatSR..14.3724N},
	urldate = {2024-09-11},
	volume = {14},
	year = {2024}}

@article{Gallazzi2014,
	adsurl = {https://ui.adsabs.harvard.edu/abs/2014ApJ...788...72G},
	archiveprefix = {arXiv},
	author = {{Gallazzi}, Anna and {Bell}, Eric F. and {Zibetti}, Stefano and {Brinchmann}, Jarle and {Kelson}, Daniel D.},
	doi = {10.1088/0004-637X/788/1/72},
	eid = {72},
	eprint = {1404.5624},
	journal = {\apj},
	month = jun,
	number = {1},
	pages = {72},
	primaryclass = {astro-ph.GA},
	title = {{Charting the Evolution of the Ages and Metallicities of Massive Galaxies since z = 0.7}},
	volume = {788},
	year = 2014}

@article{Dekel2023,
	author = {{Dekel}, Avishai and Sarkar, Kartick S. and Birnboim, Yuval and Mandelker, Nir and Li, Zhaozhou},
	doi = {10.1093/mnras/stad1557},
	issn = {0035-8711, 1365-2966},
	journal = {\mnras},
	month = jun,
	note = {arXiv:2303.04827 [astro-ph]},
	number = {3},
	pages = {3201--3218},
	title = {Efficient {Formation} of {Massive} {Galaxies} at {Cosmic} {Dawn} by {Feedback}-{Free} {Starbursts}},
	url = {http://arxiv.org/abs/2303.04827},
	urldate = {2024-09-11},
	volume = {523},
	year = {2023}}

@ARTICLE{Ferland2017,
       author = {{Ferland}, G.~J. and {Chatzikos}, M. and {Guzm{\'a}n}, F. and {Lykins}, M.~L. and {van Hoof}, P.~A.~M. and {Williams}, R.~J.~R. and {Abel}, N.~P. and {Badnell}, N.~R. and {Keenan}, F.~P. and {Porter}, R.~L. and {Stancil}, P.~C.},
        title = "{The 2017 Release Cloudy}",
      journal = {\rmxaa},
         year = 2017,
        month = oct,
       volume = {53},
        pages = {385-438},
          doi = {10.48550/arXiv.1705.10877},
archivePrefix = {arXiv},
       eprint = {1705.10877},
 primaryClass = {astro-ph.GA},
       adsurl = {https://ui.adsabs.harvard.edu/abs/2017RMxAA..53..385F}
}

@ARTICLE{Johnson2021,
       author = {{Johnson}, Benjamin D. and {Leja}, Joel and {Conroy}, Charlie and {Speagle}, Joshua S.},
        title = "{Stellar Population Inference with Prospector}",
      journal = {\apjs},
         year = 2021,
        month = jun,
       volume = {254},
       number = {2},
          eid = {22},
        pages = {22},
          doi = {10.3847/1538-4365/abef67},
archivePrefix = {arXiv},
       eprint = {2012.01426},
 primaryClass = {astro-ph.GA},
       adsurl = {https://ui.adsabs.harvard.edu/abs/2021ApJS..254...22J}
}

@article{Davies2024,
	author = {Davies, Rebecca L. and Belli, Sirio and Park, Minjung and Mendel, J. Trevor and Johnson, Benjamin D. and Conroy, Charlie and Benton, Chlo{\"e} and Bugiani, Letizia and Emami, Razieh and Leja, Joel and Li, Yijia and Maheson, Gabriel and Mathews, Elijah P. and Naidu, Rohan P. and Nelson, Erica J. and Tacchella, Sandro and Terrazas, Bryan A. and Weinberger, Rainer},
	bdsk-color = {7},
	doi = {10.1093/mnras/stae327},
	issn = {0035-8711},
	journal = {\mnras},
	month = mar,
	note = {Publisher: OUP ADS Bibcode: 2024MNRAS.528.4976D},
	pages = {4976--4992},
	title = {{JWST} reveals widespread {AGN}-driven neutral gas outflows in massive z 2 galaxies},
	url = {https://ui.adsabs.harvard.edu/abs/2024MNRAS.528.4976D},
	urldate = {2024-09-11},
	volume = {528},
	year = {2024}}

@ARTICLE{Wu2024,
       author = {{Wu}, Po-Feng},
        title = "{Ejective Feedback as a Quenching Mechanism in the First 1.5 Billion Years of the Universe: Detection of Neutral Gas Outflow in a z = 4 Recently Quenched Galaxy}",
      journal = {\apj},
         year = 2025,
        month = jan,
       volume = {978},
       number = {2},
          eid = {131},
        pages = {131},
          doi = {10.3847/1538-4357/ad98ef},
archivePrefix = {arXiv},
       eprint = {2409.00471},
 primaryClass = {astro-ph.GA},
       adsurl = {https://ui.adsabs.harvard.edu/abs/2025ApJ...978..131W}
}

@article{Belli2024,
	author = {Belli, Sirio and Park, Minjung and Davies, Rebecca L. and Mendel, J. Trevor and Johnson, Benjamin D. and Conroy, Charlie and Benton, Chlo{\"e} and Bugiani, Letizia and Emami, Razieh and Leja, Joel and Li, Yijia and Maheson, Gabriel and Mathews, Elijah P. and Naidu, Rohan P. and Nelson, Erica J. and Tacchella, Sandro and Terrazas, Bryan A. and Weinberger, Rainer},
	doi = {10.1038/s41586-024-07412-1},
	issn = {0028-0836},
	journal = {Nature},
	month = jun,
	note = {ADS Bibcode: 2024Natur.630...54B},
	pages = {54--58},
	title = {Star formation shut down by multiphase gas outflow in a galaxy at a redshift of 2.45},
	url = {https://ui.adsabs.harvard.edu/abs/2024Natur.630...54B},
	urldate = {2024-09-11},
	volume = {630},
	year = {2024}}

@article{Carnall2019b,
	adsurl = {https://ui.adsabs.harvard.edu/abs/2019MNRAS.490..417C},
	archiveprefix = {arXiv},
	author = {{Carnall}, A.~C. and {McLure}, R.~J. and {Dunlop}, J.~S. and {Cullen}, F. and {McLeod}, D.~J. and {Wild}, V. and {Johnson}, B.~D. and {Appleby}, S. and {Dav{\'e}}, R. and {Amorin}, R. and {Bolzonella}, M. and {Castellano}, M. and {Cimatti}, A. and {Cucciati}, O. and {Gargiulo}, A. and {Garilli}, B. and {Marchi}, F. and {Pentericci}, L. and {Pozzetti}, L. and {Schreiber}, C. and {Talia}, M. and {Zamorani}, G.},
	doi = {10.1093/mnras/stz2544},
	eprint = {1903.11082},
	journal = {\mnras},
	month = nov,
	number = {1},
	pages = {417-439},
	primaryclass = {astro-ph.GA},
	title = {{The VANDELS survey: the star-formation histories of massive quiescent galaxies at 1.0 < z < 1.3}},
	volume = {490},
	year = 2019}

@article{Dunlop1996,
	adsurl = {https://ui.adsabs.harvard.edu/abs/1996Natur.381..581D},
	author = {{Dunlop}, James and {Peacock}, John and {Spinrad}, Hyron and {Dey}, Arjun and {Jimenez}, Raul and {Stern}, Daniel and {Windhorst}, Rogier},
	doi = {10.1038/381581a0},
	journal = {\nat},
	month = jun,
	number = {6583},
	pages = {581-584},
	title = {{A 3.5-Gyr-old galaxy at redshift 1.55}},
	volume = {381},
	year = 1996}

@article{Cimatti2004,
	adsurl = {https://ui.adsabs.harvard.edu/abs/2004Natur.430..184C},
	archiveprefix = {arXiv},
	author = {{Cimatti}, A. and {Daddi}, E. and {Renzini}, A. and {Cassata}, P. and {Vanzella}, E. and {Pozzetti}, L. and {Cristiani}, S. and {Fontana}, A. and {Rodighiero}, G. and {Mignoli}, M. and {Zamorani}, G.},
	doi = {10.1038/nature02668},
	eprint = {astro-ph/0407131},
	journal = {\nat},
	month = jul,
	number = {6996},
	pages = {184-187},
	primaryclass = {astro-ph},
	title = {{Old galaxies in the young Universe}},
	volume = {430},
	year = 2004}

@article{Strateva2001,
	adsurl = {https://ui.adsabs.harvard.edu/abs/2001AJ....122.1861S},
	archiveprefix = {arXiv},
	author = {{Strateva}, Iskra and {Ivezi{\'c}}, {\v{Z}}eljko and {Knapp}, Gillian R. and {Narayanan}, Vijay K. and {Strauss}, Michael A. and {Gunn}, James E. and {Lupton}, Robert H. and {Schlegel}, David and {Bahcall}, Neta A. and {Brinkmann}, Jon and {Brunner}, Robert J. and {Budav{\'a}ri}, Tam{\'a}s and {Csabai}, Istv{\'a}n and {Castander}, Francisco Javier and {Doi}, Mamoru and {Fukugita}, Masataka and {Gy{\H{o}}ry}, Zsuzsanna and {Hamabe}, Masaru and {Hennessy}, Greg and {Ichikawa}, Takashi and {Kunszt}, Peter Z. and {Lamb}, Don Q. and {McKay}, Timothy A. and {Okamura}, Sadanori and {Racusin}, Judith and {Sekiguchi}, Maki and {Schneider}, Donald P. and {Shimasaku}, Kazuhiro and {York}, Donald},
	doi = {10.1086/323301},
	eprint = {astro-ph/0107201},
	journal = {\aj},
	month = oct,
	number = {4},
	pages = {1861-1874},
	primaryclass = {astro-ph},
	title = {{Color Separation of Galaxy Types in the Sloan Digital Sky Survey Imaging Data}},
	volume = {122},
	year = 2001}

@ARTICLE{Hubble1926,
       author = {{Hubble}, E.~P.},
        title = "{Extragalactic nebulae.}",
      journal = {\apj},
         year = 1926,
        month = dec,
       volume = {64},
        pages = {321-369},
          doi = {10.1086/143018},
       adsurl = {https://ui.adsabs.harvard.edu/abs/1926ApJ....64..321H}
}

@article{Man2018,
	adsurl = {https://ui.adsabs.harvard.edu/abs/2018NatAs...2..695M},
	archiveprefix = {arXiv},
	author = {{Man}, Allison and {Belli}, Sirio},
	doi = {10.1038/s41550-018-0558-1},
	eprint = {1809.00722},
	journal = {Nature Astronomy},
	month = sep,
	pages = {695-697},
	primaryclass = {astro-ph.GA},
	title = {{Star formation quenching in massive galaxies}},
	volume = {2},
	year = 2018}

@article{Lambas2012,
	adsurl = {https://ui.adsabs.harvard.edu/abs/2012A&A...539A..45L},
	archiveprefix = {arXiv},
	author = {{Lambas}, D.~G. and {Alonso}, S. and {Mesa}, V. and {O'Mill}, A.~L.},
	doi = {10.1051/0004-6361/201117900},
	eid = {A45},
	eprint = {1111.2291},
	journal = {\aap},
	month = mar,
	pages = {A45},
	primaryclass = {astro-ph.CO},
	title = {{Galaxy interactions. I. Major and minor mergers}},
	volume = {539},
	year = 2012}

@article{Park2023,
	author = {Park, Minjung and Belli, Sirio and Conroy, Charlie and Tacchella, Sandro and Leja, Joel and Cutler, Sam E. and Johnson, Benjamin D. and Nelson, Erica J. and Emami, Razieh},
	bdsk-color = {7},
	doi = {10.3847/1538-4357/acd54a},
	issn = {0004-637X},
	journal = {\apj},
	month = aug,
	note = {Publisher: IOP ADS Bibcode: 2023ApJ...953..119P},
	pages = {119},
	title = {Rapid {Quenching} of {Galaxies} at {Cosmic} {Noon}},
	url = {https://ui.adsabs.harvard.edu/abs/2023ApJ...953..119P},
	urldate = {2024-09-11},
	volume = {953},
	year = {2023}}

@article{Peng2015,
	adsurl = {https://ui.adsabs.harvard.edu/abs/2015Natur.521..192P},
	archiveprefix = {arXiv},
	author = {{Peng}, Y. and {Maiolino}, R. and {Cochrane}, R.},
	doi = {10.1038/nature14439},
	eprint = {1505.03143},
	journal = {\nat},
	month = may,
	number = {7551},
	pages = {192-195},
	primaryclass = {astro-ph.GA},
	title = {{Strangulation as the primary mechanism for shutting down star formation in galaxies}},
	volume = {521},
	year = 2015}

@article{Schawinski2014,
	adsurl = {https://ui.adsabs.harvard.edu/abs/2014MNRAS.440..889S},
	archiveprefix = {arXiv},
	author = {{Schawinski}, Kevin and {Urry}, C. Megan and {Simmons}, Brooke D. and {Fortson}, Lucy and {Kaviraj}, Sugata and {Keel}, William C. and {Lintott}, Chris J. and {Masters}, Karen L. and {Nichol}, Robert C. and {Sarzi}, Marc and {Skibba}, Ramin and {Treister}, Ezequiel and {Willett}, Kyle W. and {Wong}, O. Ivy and {Yi}, Sukyoung K.},
	doi = {10.1093/mnras/stu327},
	eprint = {1402.4814},
	journal = {\mnras},
	month = may,
	number = {1},
	pages = {889-907},
	primaryclass = {astro-ph.GA},
	title = {{The green valley is a red herring: Galaxy Zoo reveals two evolutionary pathways towards quenching of star formation in early- and late-type galaxies}},
	volume = {440},
	year = 2014}

@article{Taylor2024,
	adsurl = {https://ui.adsabs.harvard.edu/abs/2024MNRAS.535.1684T},
	archiveprefix = {arXiv},
	author = {{Taylor}, Elizabeth and {Maltby}, David and {Almaini}, Omar and {Merrifield}, Michael and {Wild}, Vivienne and {Rowlands}, Kate and {Harrold}, Jimi},
	doi = {10.1093/mnras/stae2463},
	eprint = {2411.00102},
	journal = {\mnras},
	month = dec,
	number = {2},
	pages = {1684-1692},
	primaryclass = {astro-ph.GA},
	title = {{High-velocity outflows persist up to 1 Gyr after a starburst in recently quenched galaxies at z > 1}},
	volume = {535},
	year = 2024}

@ARTICLE{Bugiani2024,
       author = {{Bugiani}, Letizia and {Belli}, Sirio and {Park}, Minjung and {Davies}, Rebecca L. and {Mendel}, J. Trevor and {Johnson}, Benjamin D. and {Khoram}, Amir H. and {Benton}, Chlo{\"e} and {Cimatti}, Andrea and {Conroy}, Charlie and {Emami}, Razieh and {Leja}, Joel and {Li}, Yijia and {Maheson}, Gabriel and {Mathews}, Elijah P. and {Naidu}, Rohan P. and {Nelson}, Erica J. and {Tacchella}, Sandro and {Terrazas}, Bryan A. and {Weinberger}, Rainer},
        title = "{Active Galactic Nucleus Feedback in Quiescent Galaxies at Cosmic Noon Traced by Ionized Gas Emission}",
      journal = {\apj},
         year = 2025,
        month = mar,
       volume = {981},
       number = {1},
          eid = {25},
        pages = {25},
          doi = {10.3847/1538-4357/adaeaf},
archivePrefix = {arXiv},
       eprint = {2406.08547},
 primaryClass = {astro-ph.GA},
       adsurl = {https://ui.adsabs.harvard.edu/abs/2025ApJ...981...25B}
}

@article{Fabian2012,
	adsurl = {https://ui.adsabs.harvard.edu/abs/2012ARA&A..50..455F},
	archiveprefix = {arXiv},
	author = {{Fabian}, A.~C.},
	doi = {10.1146/annurev-astro-081811-125521},
	eprint = {1204.4114},
	journal = {\araa},
	month = sep,
	pages = {455-489},
	primaryclass = {astro-ph.CO},
	title = {{Observational Evidence of Active Galactic Nuclei Feedback}},
	volume = {50},
	year = 2012}

@article{Rodriguez2019,
	adsurl = {https://ui.adsabs.harvard.edu/abs/2019MNRAS.490.2139R},
	archiveprefix = {arXiv},
	author = {{Rodr{\'\i}guez Montero}, Francisco and {Dav{\'e}}, Romeel and {Wild}, Vivienne and {Angl{\'e}s-Alc{\'a}zar}, Daniel and {Narayanan}, Desika},
	doi = {10.1093/mnras/stz2580},
	eprint = {1907.12680},
	journal = {\mnras},
	month = dec,
	number = {2},
	pages = {2139-2154},
	primaryclass = {astro-ph.GA},
	title = {{Mergers, starbursts, and quenching in the SIMBA simulation}},
	volume = {490},
	year = 2019}

@article{Ellison2022,
	adsurl = {https://ui.adsabs.harvard.edu/abs/2022MNRAS.517L..92E},
	archiveprefix = {arXiv},
	author = {{Ellison}, Sara L. and {Wilkinson}, Scott and {Woo}, Joanna and {Leung}, Ho-Hin and {Wild}, Vivienne and {Bickley}, Robert W. and {Patton}, David R. and {Quai}, Salvatore and {Gwyn}, Stephen},
	doi = {10.1093/mnrasl/slac109},
	eprint = {2209.07613},
	journal = {\mnras},
	month = nov,
	number = {1},
	pages = {L92-L96},
	primaryclass = {astro-ph.GA},
	title = {{Galaxy mergers can rapidly shut down star formation}},
	volume = {517},
	year = 2022}

@article{Setton2024,
	adsurl = {https://ui.adsabs.harvard.edu/abs/2024ApJ...974..145S},
	archiveprefix = {arXiv},
	author = {{Setton}, David J. and {Khullar}, Gourav and {Miller}, Tim B. and {Bezanson}, Rachel and {Greene}, Jenny E. and {Suess}, Katherine A. and {Whitaker}, Katherine E. and {Antwi-Danso}, Jacqueline and {Atek}, Hakim and {Brammer}, Gabriel and {Cutler}, Sam E. and {Dayal}, Pratika and {Feldmann}, Robert and {Fujimoto}, Seiji and {Furtak}, Lukas J. and {Glazebrook}, Karl and {Goulding}, Andy D. and {Kokorev}, Vasily and {Labbe}, Ivo and {Leja}, Joel and {Ma}, Yilun and {Marchesini}, Danilo and {Nanayakkara}, Themiya and {Pan}, Richard and {Price}, Sedona H. and {Siegel}, Jared C. and {Shipley}, Heath and {Weaver}, John R. and {van Dokkum}, Pieter and {Wang}, Bingjie and {Williams}, Christina C.},
	doi = {10.3847/1538-4357/ad6a18},
	eid = {145},
	eprint = {2402.05664},
	journal = {\apj},
	month = oct,
	number = {1},
	pages = {145},
	primaryclass = {astro-ph.GA},
	title = {{UNCOVER NIRSpec/PRISM Spectroscopy Unveils Evidence of Early Core Formation in a Massive, Centrally Dusty Quiescent Galaxy at z $_{spec}$ = 3.97}},
	volume = {974},
	year = 2024}

@article{Gelli2023,
	adsurl = {https://ui.adsabs.harvard.edu/abs/2023ApJ...954L..11G},
	archiveprefix = {arXiv},
	author = {{Gelli}, Viola and {Salvadori}, Stefania and {Ferrara}, Andrea and {Pallottini}, Andrea and {Carniani}, Stefano},
	doi = {10.3847/2041-8213/acee80},
	eid = {L11},
	eprint = {2303.13574},
	journal = {\apjl},
	month = sep,
	number = {1},
	pages = {L11},
	primaryclass = {astro-ph.GA},
	title = {{Quiescent Low-mass Galaxies Observed by JWST in the Epoch of Reionization}},
	volume = {954},
	year = 2023}

@ARTICLE{Looser2023,
       author = {{Looser}, Tobias J. and {D'Eugenio}, Francesco and {Maiolino}, Roberto and {Tacchella}, Sandro and {Curti}, Mirko and {Arribas}, Santiago and {Baker}, William M. and {Baum}, Stefi and {Bonaventura}, Nina and {Boyett}, Kristan and {Bunker}, Andrew J. and {Carniani}, Stefano and {Charlot}, Stephane and {Chevallard}, Jacopo and {Curtis-Lake}, Emma and {Lola Danhaive}, A. and {Eisenstein}, Daniel J. and {de Graaff}, Anna and {Hainline}, Kevin and {Ji}, Zhiyuan and {Johnson}, Benjamin D. and {Kumari}, Nimisha and {Nelson}, Erica and {Parlanti}, Eleonora and {Rix}, Hans-Walter and {Robertson}, Brant and {Del Pino}, Bruno Rodr{\'\i}guez and {Sandles}, Lester and {Scholtz}, Jan and {Smit}, Renske and {Stark}, Daniel P. and {{\"U}bler}, Hannah and {Williams}, Christina C. and {Willott}, Chris and {Witstok}, Joris},
        title = "{JADES: Differing assembly histories of galaxies: Observational evidence for bursty star formation histories and (mini-)quenching in the first billion years of the Universe}",
      journal = {\aap},
         year = 2025,
        month = may,
       volume = {697},
          eid = {A88},
        pages = {A88},
          doi = {10.1051/0004-6361/202347102},
archivePrefix = {arXiv},
       eprint = {2306.02470},
 primaryClass = {astro-ph.GA},
       adsurl = {https://ui.adsabs.harvard.edu/abs/2025A&A...697A..88L}
}

@ARTICLE{Bertin1996,
       author = {{Bertin}, E. and {Arnouts}, S.},
        title = "{SExtractor: Software for source extraction.}",
      journal = {\aaps},
         year = 1996,
        month = jun,
       volume = {117},
        pages = {393-404},
          doi = {10.1051/aas:1996164},
       adsurl = {https://ui.adsabs.harvard.edu/abs/1996A&AS..117..393B}
}

@article{McLeod2024,
	adsurl = {https://ui.adsabs.harvard.edu/abs/2024MNRAS.527.5004M},
	archiveprefix = {arXiv},
	author = {{McLeod}, D.~J. and {Donnan}, C.~T. and {McLure}, R.~J. and {Dunlop}, J.~S. and {Magee}, D. and {Begley}, R. and {Carnall}, A.~C. and {Cullen}, F. and {Ellis}, R.~S. and {Hamadouche}, M.~L. and {Stanton}, T.~M.},
	doi = {10.1093/mnras/stad3471},
	eprint = {2304.14469},
	journal = {\mnras},
	month = jan,
	number = {3},
	pages = {5004-5022},
	primaryclass = {astro-ph.GA},
	title = {{The galaxy UV luminosity function at z ≃ 11 from a suite of public JWST ERS, ERO, and Cycle-1 programs}},
	volume = {527},
	year = 2024}

@ARTICLE{Calzetti2000,
       author = {{Calzetti}, Daniela and {Armus}, Lee and {Bohlin}, Ralph C. and {Kinney}, Anne L. and {Koornneef}, Jan and {Storchi-Bergmann}, Thaisa},
        title = "{The Dust Content and Opacity of Actively Star-forming Galaxies}",
      journal = {\apj},
         year = 2000,
        month = apr,
       volume = {533},
       number = {2},
        pages = {682-695},
          doi = {10.1086/308692},
archivePrefix = {arXiv},
       eprint = {astro-ph/9911459},
 primaryClass = {astro-ph},
       adsurl = {https://ui.adsabs.harvard.edu/abs/2000ApJ...533..682C}
}

@article{Bruzual2003,
	adsurl = {https://ui.adsabs.harvard.edu/abs/2003MNRAS.344.1000B},
	archiveprefix = {arXiv},
	author = {{Bruzual}, G. and {Charlot}, S.},
	doi = {10.1046/j.1365-8711.2003.06897.x},
	eprint = {astro-ph/0309134},
	journal = {\mnras},
	month = oct,
	number = {4},
	pages = {1000-1028},
	primaryclass = {astro-ph},
	title = {{Stellar population synthesis at the resolution of 2003}},
	volume = {344},
	year = 2003}

@ARTICLE{Chevallard2016,
       author = {{Chevallard}, Jacopo and {Charlot}, St{\'e}phane},
        title = "{Modelling and interpreting spectral energy distributions of galaxies with BEAGLE}",
      journal = {\mnras},
         year = 2016,
        month = oct,
       volume = {462},
       number = {2},
        pages = {1415-1443},
          doi = {10.1093/mnras/stw1756},
archivePrefix = {arXiv},
       eprint = {1603.03037},
 primaryClass = {astro-ph.GA},
       adsurl = {https://ui.adsabs.harvard.edu/abs/2016MNRAS.462.1415C}
}

@article{Pacifici2012,
	author = {Pacifici, Camilla and Charlot, St{\'e}phane and Blaizot, J{\'e}r{\'e}my and Brinchmann, Jarle},
	doi = {10.1111/j.1365-2966.2012.20431.x},
	issn = {0035-8711},
	journal = {\mnras},
	month = apr,
	note = {Publisher: OUP ADS Bibcode: 2012MNRAS.421.2002P},
	pages = {2002--2024},
	title = {Relative merits of different types of rest-frame optical observations to constrain galaxy physical parameters},
	url = {https://ui.adsabs.harvard.edu/abs/2012MNRAS.421.2002P},
	urldate = {2024-09-11},
	volume = {421},
	year = {2012}}

@ARTICLE{Alberts2024a,
       author = {{Alberts}, Stacey and {Williams}, Christina C. and {Helton}, Jakob M. and {Suess}, Katherine A. and {Ji}, Zhiyuan and {Shivaei}, Irene and {Lyu}, Jianwei and {Rieke}, George and {Baker}, William M. and {Bonaventura}, Nina and {Bunker}, Andrew J. and {Carniani}, Stefano and {Charlot}, Stephane and {Curtis-Lake}, Emma and {D'Eugenio}, Francesco and {Eisenstein}, Daniel J. and {de Graaff}, Anna and {Hainline}, Kevin N. and {Hausen}, Ryan and {Johnson}, Benjamin D. and {Maiolino}, Roberto and {Parlanti}, Eleonora and {Rieke}, Marcia J. and {Robertson}, Brant E. and {Sun}, Yang and {Tacchella}, Sandro and {Willmer}, Christopher N.~A. and {Willott}, Chris J.},
        title = "{To High Redshift and Low Mass: Exploring the Emergence of Quenched Galaxies and Their Environments at 3 < z < 6 in the Ultra-deep JADES MIRI F770W Parallel}",
      journal = {\apj},
         year = 2024,
        month = nov,
       volume = {975},
       number = {1},
          eid = {85},
        pages = {85},
          doi = {10.3847/1538-4357/ad66cc},
archivePrefix = {arXiv},
       eprint = {2312.12207},
 primaryClass = {astro-ph.GA},
       adsurl = {https://ui.adsabs.harvard.edu/abs/2024ApJ...975...85A}
}

@ARTICLE{Long2023,
       author = {{Long}, Arianna S. and {Antwi-Danso}, Jacqueline and {Lambrides}, Erini L. and {Lovell}, Christopher C. and {de la Vega}, Alexander and {Valentino}, Francesco and {Zavala}, Jorge A. and {Casey}, Caitlin M. and {Wilkins}, Stephen M. and {Yung}, L.~Y. Aaron and {Arrabal Haro}, Pablo and {Bagley}, Micaela B. and {Bisigello}, Laura and {Chworowsky}, Katherine and {Cooper}, M.~C. and {Cooper}, Olivia R. and {Cooray}, Asantha R. and {Croton}, Darren and {Dickinson}, Mark and {Finkelstein}, Steven L. and {Franco}, Maximilien and {Gould}, Katriona M.~L. and {Hirschmann}, Michaela and {Hutchison}, Taylor A. and {Kartaltepe}, Jeyhan S. and {Kocevski}, Dale D. and {Koekemoer}, Anton M. and {Lucas}, Ray A. and {McKinney}, Jed and {Nere}, Rachel and {Papovich}, Casey and {P{\'e}rez-Gonz{\'a}lez}, Pablo G. and {Pirzkal}, Nor and {Santini}, Paola},
        title = "{Efficient NIRCam Selection of Quiescent Galaxies at 3 < z < 6 in CEERS}",
      journal = {\apj},
         year = 2024,
        month = jul,
       volume = {970},
       number = {1},
          eid = {68},
        pages = {68},
          doi = {10.3847/1538-4357/ad4cea},
archivePrefix = {arXiv},
       eprint = {2305.04662},
 primaryClass = {astro-ph.GA},
       adsurl = {https://ui.adsabs.harvard.edu/abs/2024ApJ...970...68L}
}

@ARTICLE{Castellano2022,
       author = {{Castellano}, Marco and {Fontana}, Adriano and {Treu}, Tommaso and {Santini}, Paola and {Merlin}, Emiliano and {Leethochawalit}, Nicha and {Trenti}, Michele and {Vanzella}, Eros and {Mestric}, Uros and {Bonchi}, Andrea and {Belfiori}, Davide and {Nonino}, Mario and {Paris}, Diego and {Polenta}, Gianluca and {Roberts-Borsani}, Guido and {Boyett}, Kristan and {Brada{\v{c}}}, Maru{\v{s}}a and {Calabr{\`o}}, Antonello and {Glazebrook}, Karl and {Grillo}, Claudio and {Mascia}, Sara and {Mason}, Charlotte and {Mercurio}, Amata and {Morishita}, Takahiro and {Nanayakkara}, Themiya and {Pentericci}, Laura and {Rosati}, Piero and {Vulcani}, Benedetta and {Wang}, Xin and {Yang}, Lilan},
        title = "{Early Results from GLASS-JWST. III. Galaxy Candidates at z  9-15}",
      journal = {\apjl},
         year = 2022,
        month = oct,
       volume = {938},
       number = {2},
          eid = {L15},
        pages = {L15},
          doi = {10.3847/2041-8213/ac94d0},
archivePrefix = {arXiv},
       eprint = {2207.09436},
 primaryClass = {astro-ph.GA},
       adsurl = {https://ui.adsabs.harvard.edu/abs/2022ApJ...938L..15C}
}

@ARTICLE{Shapley2024,
       author = {{Shapley}, Alice E. and {Sanders}, Ryan L. and {Topping}, Michael W. and {Reddy}, Naveen A. and {Pahl}, Anthony J. and {Oesch}, Pascal A. and {Berg}, Danielle A. and {Bouwens}, Rychard J. and {Brammer}, Gabriel and {Carnall}, Adam C. and {Cullen}, Fergus and {Dav{\'e}}, Romeel and {Dunlop}, James S. and {Ellis}, Richard S. and {F{\"o}rster Schreiber}, N.~M. and {Furlanetto}, Steven R. and {Glazebrook}, Karl and {Illingworth}, Garth D. and {Jones}, Tucker and {Kriek}, Mariska and {McLeod}, Derek J. and {McLure}, Ross J. and {Narayanan}, Desika and {Pettini}, Max and {Schaerer}, Daniel and {Stark}, Daniel P. and {Steidel}, Charles C. and {Tang}, Mengtao and {Clarke}, Leonardo and {Donnan}, Callum T. and {Kehoe}, Emily},
        title = "{The AURORA Survey: An Extraordinarily Mature, Star-forming Galaxy at z {\ensuremath{\sim}} 7}",
      journal = {\apj},
         year = 2025,
        month = mar,
       volume = {981},
       number = {2},
          eid = {167},
        pages = {167},
          doi = {10.3847/1538-4357/adaf98},
archivePrefix = {arXiv},
       eprint = {2410.00110},
 primaryClass = {astro-ph.GA},
       adsurl = {https://ui.adsabs.harvard.edu/abs/2025ApJ...981..167S}
}

@ARTICLE{Hartley2023,
       author = {{Hartley}, Abigail I. and {Nelson}, Erica J. and {Suess}, Katherine A. and {Garcia}, Alex M. and {Park}, Minjung and {Hernquist}, Lars and {Bezanson}, Rachel and {Nevin}, Rebecca and {Pillepich}, Annalisa and {Schechter}, Aimee L. and {Terrazas}, Bryan A. and {Torrey}, Paul and {Wellons}, Sarah and {Whitaker}, Katherine E. and {Williams}, Christina C.},
        title = "{The first quiescent galaxies in TNG300}",
      journal = {\mnras},
         year = 2023,
        month = jun,
       volume = {522},
       number = {2},
        pages = {3138-3144},
          doi = {10.1093/mnras/stad1162},
archivePrefix = {arXiv},
       eprint = {2304.09392},
 primaryClass = {astro-ph.GA},
       adsurl = {https://ui.adsabs.harvard.edu/abs/2023MNRAS.522.3138H}
}

@ARTICLE{Kimmig2025,
       author = {{Kimmig}, Lucas C. and {Remus}, Rhea-Silvia and {Seidel}, Benjamin and {Valenzuela}, Lucas M. and {Dolag}, Klaus and {Burkert}, Andreas},
        title = "{Blowing Out the Candle: How to Quench Galaxies at High Redshift{\textemdash}An Ensemble of Rapid Starbursts, AGN Feedback, and Environment}",
      journal = {\apj},
         year = 2025,
        month = jan,
       volume = {979},
       number = {1},
          eid = {15},
        pages = {15},
          doi = {10.3847/1538-4357/ad9472},
archivePrefix = {arXiv},
       eprint = {2310.16085},
 primaryClass = {astro-ph.GA},
       adsurl = {https://ui.adsabs.harvard.edu/abs/2025ApJ...979...15K}
}

@ARTICLE{Gould2023,
       author = {{Gould}, Katriona M.~L. and {Brammer}, Gabriel and {Valentino}, Francesco and {Whitaker}, Katherine E. and {Weaver}, John. R. and {Lagos}, Claudia del P. and {Rizzo}, Francesca and {Franco}, Maximilien and {Hsieh}, Bau-Ching and {Ilbert}, Olivier and {Jin}, Shuowen and {Magdis}, Georgios and {McCracken}, Henry J. and {Mobasher}, Bahram and {Shuntov}, Marko and {Steinhardt}, Charles L. and {Strait}, Victoria and {Toft}, Sune},
        title = "{COSMOS2020: Exploring the Dawn of Quenching for Massive Galaxies at 3 < z < 5 with a New Color-selection Method}",
      journal = {\aj},
         year = 2023,
        month = jun,
       volume = {165},
       number = {6},
          eid = {248},
        pages = {248},
          doi = {10.3847/1538-3881/accadc},
archivePrefix = {arXiv},
       eprint = {2302.10934},
 primaryClass = {astro-ph.GA},
       adsurl = {https://ui.adsabs.harvard.edu/abs/2023AJ....165..248G}
}

@ARTICLE{Dave2017,
       author = {{Dav{\'e}}, Romeel and {Rafieferantsoa}, Mika H. and {Thompson}, Robert J.},
        title = "{mufasa: the assembly of the red sequence}",
      journal = {\mnras},
         year = 2017,
        month = oct,
       volume = {471},
       number = {2},
        pages = {1671-1687},
          doi = {10.1093/mnras/stx1693},
archivePrefix = {arXiv},
       eprint = {1704.01135},
 primaryClass = {astro-ph.GA},
       adsurl = {https://ui.adsabs.harvard.edu/abs/2017MNRAS.471.1671D}
}

@ARTICLE{Rennehan2024,
       author = {{Rennehan}, Douglas},
        title = "{The Manhattan Suite: Accelerated Galaxy Evolution in the Early Universe}",
      journal = {\apj},
         year = 2024,
        month = nov,
       volume = {975},
       number = {1},
          eid = {114},
        pages = {114},
          doi = {10.3847/1538-4357/ad793d},
archivePrefix = {arXiv},
       eprint = {2406.06672},
 primaryClass = {astro-ph.GA},
       adsurl = {https://ui.adsabs.harvard.edu/abs/2024ApJ...975..114R}
}

@article{Beverage2024,
	author = {Beverage, Aliza G. and Kriek, Mariska and Suess, Katherine A. and Conroy, Charlie and Price, Sedona H. and Barro, Guillermo and Bezanson, Rachel and Franx, Marijn and Lorenz, Brian and Ma, Yilun and Mowla, Lamiya A. and Pasha, Imad and van Dokkum, Pieter and Weisz, Daniel R.},
	doi = {10.3847/1538-4357/ad372d},
	issn = {0004-637X},
	journal = {\apj},
	language = {en},
	month = may,
	number = {2},
	pages = {234},
	shorttitle = {The {Heavy} {Metal} {Survey}},
	title = {The {Heavy} {Metal} {Survey}: {The} {Evolution} of {Stellar} {Metallicities}, {Abundance} {Ratios}, and {Ages} of {Massive} {Quiescent} {Galaxies} since z ∼ 2},
	url = {https://ui.adsabs.harvard.edu/abs/2024ApJ...966..234B/abstract},
	urldate = {2024-09-11},
	volume = {966},
	year = {2024}}

@ARTICLE{Maiolino2012,
       author = {{Maiolino}, R. and {Gallerani}, S. and {Neri}, R. and {Cicone}, C. and {Ferrara}, A. and {Genzel}, R. and {Lutz}, D. and {Sturm}, E. and {Tacconi}, L.~J. and {Walter}, F. and {Feruglio}, C. and {Fiore}, F. and {Piconcelli}, E.},
        title = "{Evidence of strong quasar feedback in the early Universe}",
      journal = {\mnras},
         year = 2012,
        month = sep,
       volume = {425},
       number = {1},
        pages = {L66-L70},
          doi = {10.1111/j.1745-3933.2012.01303.x},
archivePrefix = {arXiv},
       eprint = {1204.2904},
 primaryClass = {astro-ph.CO},
       adsurl = {https://ui.adsabs.harvard.edu/abs/2012MNRAS.425L..66M}
}

@ARTICLE{Strait2023,
       author = {{Strait}, Victoria and {Brammer}, Gabriel and {Muzzin}, Adam and {Desprez}, Guillaume and {Asada}, Yoshihisa and {Abraham}, Roberto and {Brada{\v{c}}}, Maru{\v{s}}a and {Iyer}, Kartheik G. and {Martis}, Nicholas and {Mowla}, Lamiya and {Noirot}, Ga{\"e}l and {Sarrouh}, Ghassan T.~E. and {Sawicki}, Marcin and {Willott}, Chris and {Gould}, Katriona and {Grindlay}, Tess and {Matharu}, Jasleen and {Rihtar{\v{s}}i{\v{c}}}, Gregor},
        title = "{An Extremely Compact, Low-mass Galaxy on its Way to Quiescence at z = 5.2}",
      journal = {\apjl},
         year = 2023,
        month = jun,
       volume = {949},
       number = {2},
          eid = {L23},
        pages = {L23},
          doi = {10.3847/2041-8213/acd457},
archivePrefix = {arXiv},
       eprint = {2303.11349},
 primaryClass = {astro-ph.GA},
       adsurl = {https://ui.adsabs.harvard.edu/abs/2023ApJ...949L..23S}
}

@ARTICLE{Sanchez2006,
       author = {{S{\'a}nchez-Bl{\'a}zquez}, P. and {Peletier}, R.~F. and {Jim{\'e}nez-Vicente}, J. and {Cardiel}, N. and {Cenarro}, A.~J. and {Falc{\'o}n-Barroso}, J. and {Gorgas}, J. and {Selam}, S. and {Vazdekis}, A.},
        title = "{Medium-resolution Isaac Newton Telescope library of empirical spectra}",
      journal = {\mnras},
         year = 2006,
        month = sep,
       volume = {371},
       number = {2},
        pages = {703-718},
          doi = {10.1111/j.1365-2966.2006.10699.x},
archivePrefix = {arXiv},
       eprint = {astro-ph/0607009},
 primaryClass = {astro-ph},
       adsurl = {https://ui.adsabs.harvard.edu/abs/2006MNRAS.371..703S}
}

@ARTICLE{Falcon2011,
       author = {{Falc{\'o}n-Barroso}, J. and {S{\'a}nchez-Bl{\'a}zquez}, P. and {Vazdekis}, A. and {Ricciardelli}, E. and {Cardiel}, N. and {Cenarro}, A.~J. and {Gorgas}, J. and {Peletier}, R.~F.},
        title = "{An updated MILES stellar library and stellar population models}",
      journal = {\aap},
         year = 2011,
        month = aug,
       volume = {532},
          eid = {A95},
        pages = {A95},
          doi = {10.1051/0004-6361/201116842},
archivePrefix = {arXiv},
       eprint = {1107.2303},
 primaryClass = {astro-ph.CO},
       adsurl = {https://ui.adsabs.harvard.edu/abs/2011A&A...532A..95F}
}

@ARTICLE{Bressen2012,
       author = {{Bressan}, Alessandro and {Marigo}, Paola and {Girardi}, L{\'e}o. and {Salasnich}, Bernardo and {Dal Cero}, Claudia and {Rubele}, Stefano and {Nanni}, Ambra},
        title = "{PARSEC: stellar tracks and isochrones with the PAdova and TRieste Stellar Evolution Code}",
      journal = {\mnras},
         year = 2012,
        month = nov,
       volume = {427},
       number = {1},
        pages = {127-145},
          doi = {10.1111/j.1365-2966.2012.21948.x},
archivePrefix = {arXiv},
       eprint = {1208.4498},
 primaryClass = {astro-ph.SR},
       adsurl = {https://ui.adsabs.harvard.edu/abs/2012MNRAS.427..127B}
}

@ARTICLE{Marigo2013,
       author = {{Marigo}, Paola and {Bressan}, Alessandro and {Nanni}, Ambra and {Girardi}, L{\'e}o and {Pumo}, Maria Letizia},
        title = "{Evolution of thermally pulsing asymptotic giant branch stars - I. The COLIBRI code}",
      journal = {\mnras},
         year = 2013,
        month = sep,
       volume = {434},
       number = {1},
        pages = {488-526},
          doi = {10.1093/mnras/stt1034},
archivePrefix = {arXiv},
       eprint = {1305.4485},
 primaryClass = {astro-ph.SR},
       adsurl = {https://ui.adsabs.harvard.edu/abs/2013MNRAS.434..488M}
}

@ARTICLE{Salim2018,
       author = {{Salim}, Samir and {Boquien}, M{\'e}d{\'e}ric and {Lee}, Janice C.},
        title = "{Dust Attenuation Curves in the Local Universe: Demographics and New Laws for Star-forming Galaxies and High-redshift Analogs}",
      journal = {\apj},
         year = 2018,
        month = may,
       volume = {859},
       number = {1},
          eid = {11},
        pages = {11},
          doi = {10.3847/1538-4357/aabf3c},
archivePrefix = {arXiv},
       eprint = {1804.05850},
 primaryClass = {astro-ph.GA},
       adsurl = {https://ui.adsabs.harvard.edu/abs/2018ApJ...859...11S}
}

@ARTICLE{Charlot2000,
       author = {{Charlot}, St{\'e}phane and {Fall}, S. Michael},
        title = "{A Simple Model for the Absorption of Starlight by Dust in Galaxies}",
      journal = {\apj},
         year = 2000,
        month = aug,
       volume = {539},
       number = {2},
        pages = {718-731},
          doi = {10.1086/309250},
archivePrefix = {arXiv},
       eprint = {astro-ph/0003128},
 primaryClass = {astro-ph},
       adsurl = {https://ui.adsabs.harvard.edu/abs/2000ApJ...539..718C}
}

@ARTICLE{Inoue2014,
       author = {{Inoue}, Akio K. and {Shimizu}, Ikkoh and {Iwata}, Ikuru and {Tanaka}, Masayuki},
        title = "{An updated analytic model for attenuation by the intergalactic medium}",
      journal = {\mnras},
         year = 2014,
        month = aug,
       volume = {442},
       number = {2},
        pages = {1805-1820},
          doi = {10.1093/mnras/stu936},
archivePrefix = {arXiv},
       eprint = {1402.0677},
 primaryClass = {astro-ph.CO},
       adsurl = {https://ui.adsabs.harvard.edu/abs/2014MNRAS.442.1805I}
}

@article{Skilling2004,
    author = {Skilling, John},
    title = {Nested Sampling},
    journal = {AIP Conference Proceedings},
    volume = {735},
    number = {1},
    pages = {395-405},
    year = {2004},
    month = {11},
    issn = {0094-243X},
    doi = {10.1063/1.1835238},
    url = {https://doi.org/10.1063/1.1835238},
    eprint = {https://pubs.aip.org/aip/acp/article-pdf/735/1/395/11702789/395_1_online.pdf},
}

@ARTICLE{Feroz2019,
       author = {{Feroz}, Farhan and {Hobson}, Michael P. and {Cameron}, Ewan and {Pettitt}, Anthony N.},
        title = "{Importance Nested Sampling and the MultiNest Algorithm}",
      journal = {The Open Journal of Astrophysics},
         year = 2019,
        month = nov,
       volume = {2},
       number = {1},
          eid = {10},
        pages = {10},
          doi = {10.21105/astro.1306.2144},
archivePrefix = {arXiv},
       eprint = {1306.2144},
 primaryClass = {astro-ph.IM},
       adsurl = {https://ui.adsabs.harvard.edu/abs/2019OJAp....2E..10F}
}

@ARTICLE{Buchner2014,
       author = {{Buchner}, J. and {Georgakakis}, A. and {Nandra}, K. and {Hsu}, L. and {Rangel}, C. and {Brightman}, M. and {Merloni}, A. and {Salvato}, M. and {Donley}, J. and {Kocevski}, D.},
        title = "{X-ray spectral modelling of the AGN obscuring region in the CDFS: Bayesian model selection and catalogue}",
      journal = {\aap},
         year = 2014,
        month = apr,
       volume = {564},
          eid = {A125},
        pages = {A125},
          doi = {10.1051/0004-6361/201322971},
archivePrefix = {arXiv},
       eprint = {1402.0004},
 primaryClass = {astro-ph.HE},
       adsurl = {https://ui.adsabs.harvard.edu/abs/2014A&A...564A.125B}
}

@ARTICLE{Pacifici2016,
       author = {{Pacifici}, Camilla and {Kassin}, Susan A. and {Weiner}, Benjamin J. and {Holden}, Bradford and {Gardner}, Jonathan P. and {Faber}, Sandra M. and {Ferguson}, Henry C. and {Koo}, David C. and {Primack}, Joel R. and {Bell}, Eric F. and {Dekel}, Avishai and {Gawiser}, Eric and {Giavalisco}, Mauro and {Rafelski}, Marc and {Simons}, Raymond C. and {Barro}, Guillermo and {Croton}, Darren J. and {Dav{\'e}}, Romeel and {Fontana}, Adriano and {Grogin}, Norman A. and {Koekemoer}, Anton M. and {Lee}, Seong-Kook and {Salmon}, Brett and {Somerville}, Rachel and {Behroozi}, Peter},
        title = "{The Evolution of Star Formation Histories of Quiescent Galaxies}",
      journal = {\apj},
         year = 2016,
        month = nov,
       volume = {832},
       number = {1},
          eid = {79},
        pages = {79},
          doi = {10.3847/0004-637X/832/1/79},
archivePrefix = {arXiv},
       eprint = {1609.03572},
 primaryClass = {astro-ph.GA},
       adsurl = {https://ui.adsabs.harvard.edu/abs/2016ApJ...832...79P}
}

@ARTICLE{Trussler2020,
       author = {{Trussler}, James and {Maiolino}, Roberto and {Maraston}, Claudia and {Peng}, Yingjie and {Thomas}, Daniel and {Goddard}, Daniel and {Lian}, Jianhui},
        title = "{Both starvation and outflows drive galaxy quenching}",
      journal = {\mnras},
         year = 2020,
        month = feb,
       volume = {491},
       number = {4},
        pages = {5406-5434},
          doi = {10.1093/mnras/stz3286},
archivePrefix = {arXiv},
       eprint = {1811.09283},
 primaryClass = {astro-ph.GA},
       adsurl = {https://ui.adsabs.harvard.edu/abs/2020MNRAS.491.5406T}
}

@ARTICLE{Eisenstein2023c,
       author = {{Eisenstein}, Daniel J. and {Johnson}, Benjamin D. and {Robertson}, Brant and {Tacchella}, Sandro and {Hainline}, Kevin and {Jakobsen}, Peter and {Maiolino}, Roberto and {Bonaventura}, Nina and {Bunker}, Andrew J. and {Cameron}, Alex J. and {Cargile}, Phillip A. and {Curtis-Lake}, Emma and {Hausen}, Ryan and {Pusk{\'a}s}, D{\'a}vid and {Rieke}, Marcia and {Sun}, Fengwu and {Willmer}, Christopher N.~A. and {Willott}, Chris and {Alberts}, Stacey and {Arribas}, Santiago and {Baker}, William M. and {Baum}, Stefi and {Bhatawdekar}, Rachana and {Carniani}, Stefano and {Charlot}, Stephane and {Chen}, Zuyi and {Chevallard}, Jacopo and {Curti}, Mirko and {DeCoursey}, Christa and {D'Eugenio}, Francesco and {de Graaff}, Anna and {Egami}, Eiichi and {Helton}, Jakob M. and {Ji}, Zhiyuan and {Jones}, Gareth C. and {Kumari}, Nimisha and {L{\"u}tzgendorf}, Nora and {Laseter}, Isaac and {Looser}, Tobias J. and {Lyu}, Jianwei and {Maseda}, Michael V. and {Nelson}, Erica and {Parlanti}, Eleonora and {Rauscher}, Bernard J. and {Rawle}, Tim and {Rieke}, George and {Rix}, Hans-Walter and {Rujopakarn}, Wiphu and {Sandles}, Lester and {Saxena}, Aayush and {Scholtz}, Jan and {Sharpe}, Katherine and {Shivaei}, Irene and {Simmonds}, Charlotte and {Smit}, Renske and {Topping}, Michael W. and {{\"U}bler}, Hannah and {Venturi}, Giacomo and {Williams}, Christina C. and {Witstok}, Joris and {Woodrum}, Charity},
        title = "{The JADES Origins Field: A New JWST Deep Field in the JADES Second NIRCam Data Release}",
      journal = {arXiv e-prints},
         year = 2023,
        month = oct,
          eid = {arXiv:2310.12340},
        pages = {arXiv:2310.12340},
          doi = {10.48550/arXiv.2310.12340},
archivePrefix = {arXiv},
       eprint = {2310.12340},
 primaryClass = {astro-ph.GA},
       adsurl = {https://ui.adsabs.harvard.edu/abs/2023arXiv231012340E}
}

@ARTICLE{deGraaf2024,
       author = {{de Graaff}, Anna and {Brammer}, Gabriel and {Weibel}, Andrea and {Lewis}, Zach and {Maseda}, Michael V. and {Oesch}, Pascal A. and {Bezanson}, Rachel and {Boogaard}, Leindert A. and {Cleri}, Nikko J. and {Cooper}, Olivia R. and {Gottumukkala}, Rashmi and {Greene}, Jenny E. and {Hirschmann}, Michaela and {Hviding}, Raphael E. and {Katz}, Harley and {Labb{\'e}}, Ivo and {Leja}, Joel and {Matthee}, Jorryt and {McConachie}, Ian and {Miller}, Tim B. and {Naidu}, Rohan P. and {Price}, Sedona H. and {Rix}, Hans-Walter and {Setton}, David J. and {Suess}, Katherine A. and {Wang}, Bingjie and {Whitaker}, Katherine E. and {Williams}, Christina C.},
        title = "{RUBIES: A complete census of the bright and red distant Universe with JWST/NIRSpec}",
      journal = {\aap},
         year = 2025,
        month = may,
       volume = {697},
          eid = {A189},
        pages = {A189},
          doi = {10.1051/0004-6361/202452186},
archivePrefix = {arXiv},
       eprint = {2409.05948},
 primaryClass = {astro-ph.GA},
       adsurl = {https://ui.adsabs.harvard.edu/abs/2025A&A...697A.189D}
}

@ARTICLE{Heintz2024,
       author = {{Heintz}, Kasper E. and {Watson}, Darach and {Brammer}, Gabriel and {Vejlgaard}, Simone and {Hutter}, Anne and {Strait}, Victoria B. and {Matthee}, Jorryt and {Oesch}, Pascal A. and {Jakobsson}, P{\'a}ll and {Tanvir}, Nial R. and {Laursen}, Peter and {Naidu}, Rohan P. and {Mason}, Charlotte A. and {Killi}, Meghana and {Jung}, Intae and {Hsiao}, Tiger Yu-Yang and {Abdurro'uf} and {Coe}, Dan and {Arrabal Haro}, Pablo and {Finkelstein}, Steven L. and {Toft}, Sune},
        title = "{Strong damped Lyman-{\ensuremath{\alpha}} absorption in young star-forming galaxies at redshifts 9 to 11}",
      journal = {Science},
         year = 2024,
        month = may,
       volume = {384},
       number = {6698},
        pages = {890-894},
          doi = {10.1126/science.adj0343},
archivePrefix = {arXiv},
       eprint = {2306.00647},
 primaryClass = {astro-ph.GA},
       adsurl = {https://ui.adsabs.harvard.edu/abs/2024Sci...384..890H}
}

@article{Horne1986,
	author = {Horne, K.},
	doi = {10.1086/131801},
	issn = {0004-6280},
	journal = {Publications of the Astronomical Society of the Pacific},
	month = jun,
	note = {Publisher: IOP ADS Bibcode: 1986PASP...98..609H},
	pages = {609--617},
	title = {An optimal extraction algorithm for {CCD} spectroscopy.},
	url = {https://ui.adsabs.harvard.edu/abs/1986PASP...98..609H},
	urldate = {2024-09-19},
	volume = {98},
	year = {1986}}

@article{Liu2025,
    author = {Liu, Feng-Yuan and Dunlop, James S and McLure, Ross J and McLeod, Derek J and Barrufet, Laia and Carnall, Adam C and Begley, Ryan and Pérez-González, Pablo G and Donnan, Callum T and Ellis, Richard S and Grogin, Norman A and Magee, Dan and Illingworth, Garth D and Cullen, Fergus and Stevenson, Struan D and Koekemoer, Anton M and Fontana, Adriano and Bowler, Rebecca A A},
    title = {JWST PRIMER: A deep JWST study of all ALMA-detected galaxies in PRIMER COSMOS – dust-obscured star-formation history back to z ≃ 7},
    journal = {\mnras},
    pages = {staf1961},
    year = {2025},
    month = {11},
    issn = {0035-8711},
    doi = {10.1093/mnras/staf1961},
    url = {https://doi.org/10.1093/mnras/staf1961},
    eprint = {https://academic.oup.com/mnras/advance-article-pdf/doi/10.1093/mnras/staf1961/65252987/staf1961.pdf},
}

@ARTICLE{Baker2025a,
       author = {{Baker}, William M. and {Lim}, Seunghwan and {D'Eugenio}, Francesco and {Maiolino}, Roberto and {Ji}, Zhiyuan and {Arribas}, Santiago and {Bunker}, Andrew J. and {Carniani}, Stefano and {Charlot}, Stephane and {de Graaff}, Anna and {Hainline}, Kevin and {Looser}, Tobias J. and {Lyu}, Jianwei and {Rinaldi}, Pierluigi and {Robertson}, Brant and {Schaller}, Matthieu and {Schaye}, Joop and {Scholtz}, Jan and {{\"U}bler}, Hannah and {Williams}, Christina C. and {Willmer}, Christopher N.~A. and {Willott}, Chris and {Zhu}, Yongda},
        title = "{The abundance and nature of high-redshift quiescent galaxies from JADES spectroscopy and the FLAMINGO simulations}",
      journal = {\mnras},
         year = 2025,
        month = may,
       volume = {539},
       number = {1},
        pages = {557-589},
          doi = {10.1093/mnras/staf475},
archivePrefix = {arXiv},
       eprint = {2410.14773},
 primaryClass = {astro-ph.GA},
       adsurl = {https://ui.adsabs.harvard.edu/abs/2025MNRAS.539..557B}
}

@ARTICLE{McLeod2021,
       author = {{McLeod}, D.~J. and {McLure}, R.~J. and {Dunlop}, J.~S. and {Cullen}, F. and {Carnall}, A.~C. and {Duncan}, K.},
        title = "{The evolution of the galaxy stellar-mass function over the last 12 billion years from a combination of ground-based and HST surveys}",
      journal = {\mnras},
         year = 2021,
        month = may,
       volume = {503},
       number = {3},
        pages = {4413-4435},
          doi = {10.1093/mnras/stab731},
archivePrefix = {arXiv},
       eprint = {2009.03176},
 primaryClass = {astro-ph.GA},
       adsurl = {https://ui.adsabs.harvard.edu/abs/2021MNRAS.503.4413M}
}

@ARTICLE{Baldwin1981,
       author = {{Baldwin}, J.~A. and {Phillips}, M.~M. and {Terlevich}, R.},
        title = "{Classification parameters for the emission-line spectra of extragalactic objects.}",
      journal = {\pasp},
         year = 1981,
        month = feb,
       volume = {93},
        pages = {5-19},
          doi = {10.1086/130766},
       adsurl = {https://ui.adsabs.harvard.edu/abs/1981PASP...93....5B}
}

@MISC{Eisenstein2017,
       author = {{Eisenstein}, Daniel J. and {Ferruit}, Pierre and {Rieke}, Marcia J. and {Willmer}, Christopher Nicholas Andrew and {Willott}, Chris J.},
        title = "{NIRCam-NIRSpec galaxy assembly survey - GOODS-S - part \#1a}",
 howpublished = {JWST Proposal. Cycle 1, ID. \#1180},
         year = 2017,
        month = jun,
        pages = {1180},
       adsurl = {https://ui.adsabs.harvard.edu/abs/2017jwst.prop.1180E}
}

@MISC{Rieke2017,
       author = {{Rieke}, George and {Alberts}, Stacey and {Lyu}, Jianwei and {Morrison}, Jane and {Shivaei}, Irene},
        title = "{MIRI in the Hubble Ultra-Deep Field}",
 howpublished = {JWST Proposal. Cycle 1, ID. \#1207},
         year = 2017,
        month = jul,
        pages = {1207},
       adsurl = {https://ui.adsabs.harvard.edu/abs/2017jwst.prop.1207R}
}

@MISC{Ferruit2017a,
       author = {{Ferruit}, Pierre},
        title = "{NIRSpec WIDE MOS Survey - GOODS-S}",
 howpublished = {JWST Proposal. Cycle 1, ID. \#1212},
         year = 2017,
        month = jul,
        pages = {1212},
       adsurl = {https://ui.adsabs.harvard.edu/abs/2017jwst.prop.1212F}
}

@MISC{Ferruit2017b,
       author = {{Ferruit}, Pierre},
        title = "{NIRSpec WIDE MOS Survey - COSMOS}",
 howpublished = {JWST Proposal. Cycle 1, ID. \#1214},
         year = 2017,
        month = jul,
        pages = {1214},
       adsurl = {https://ui.adsabs.harvard.edu/abs/2017jwst.prop.1214F}
}

@MISC{Ferruit2017c,
       author = {{Ferruit}, Pierre},
        title = "{NIRSpec WIDE MOS Survey - UDS}",
 howpublished = {JWST Proposal. Cycle 1, ID. \#1215},
         year = 2017,
        month = jul,
        pages = {1215},
       adsurl = {https://ui.adsabs.harvard.edu/abs/2017jwst.prop.1215F}
}

@MISC{Ferruit2017d,
       author = {{Ferruit}, Pierre and {Eisenstein}, Daniel J. and {Rieke}, Marcia J. and {Willott}, Chris J.},
        title = "{NIRCam-NIRSpec galaxy assembly survey - GOODS-S - part \#2}",
 howpublished = {JWST Proposal. Cycle 1, ID. \#1286},
         year = 2017,
        month = jul,
        pages = {1286},
       adsurl = {https://ui.adsabs.harvard.edu/abs/2017jwst.prop.1286F}
}

@ARTICLE{Belli2019,
       author = {{Belli}, Sirio and {Newman}, Andrew B. and {Ellis}, Richard S.},
        title = "{MOSFIRE Spectroscopy of Quiescent Galaxies at 1.5 < z < 2.5. II. Star Formation Histories and Galaxy Quenching}",
      journal = {\apj},
         year = 2019,
        month = mar,
       volume = {874},
       number = {1},
          eid = {17},
        pages = {17},
          doi = {10.3847/1538-4357/ab07af},
archivePrefix = {arXiv},
       eprint = {1810.00008},
 primaryClass = {astro-ph.GA},
       adsurl = {https://ui.adsabs.harvard.edu/abs/2019ApJ...874...17B}
}

@MISC{Belli2021,
       author = {{Belli}, Sirio and {Conroy}, Charlie and {Davies}, Rebecca and {Emami}, Razieh and {Johnson}, Benjamin D. and {Leja}, Joel and {Mendel}, Jon Trevor and {Naidu}, Rohan and {Nelson}, Erica and {Tacchella}, Sandro and {Terrazas}, Bryan and {Weinberger}, Rainer},
        title = "{The Stellar and Gas Content of Galaxies at Cosmic Noon}",
 howpublished = {JWST Proposal. Cycle 1, ID. \#1810},
         year = 2021,
        month = mar,
        pages = {1810},
       adsurl = {https://ui.adsabs.harvard.edu/abs/2021jwst.prop.1810B}
}

@MISC{Barrufet2021,
       author = {{Barrufet}, Laia and {Oesch}, Pascal and {Fudamoto}, Yoshinobu and {Illingworth}, Garth D. and {Labbe}, Ivo and {Magee}, Daniel K. and {Stefanon}, Mauro and {van Dokkum}, Pieter},
        title = "{Quiescent or dusty? Unveiling the nature of extremely red galaxies at z>3}",
 howpublished = {JWST Proposal. Cycle 1, ID. \#2198},
         year = 2021,
        month = mar,
        pages = {2198},
       adsurl = {https://ui.adsabs.harvard.edu/abs/2021jwst.prop.2198B}
}

@MISC{Glazebrook2021,
       author = {{Glazebrook}, Karl and {Nanayakkara}, Themiya and {Esdaile}, James and {Espejo}, Juan Manuel and {Jacobs}, Colin and {Kacprzak}, Glenn G. and {Labbe}, Ivo and {Marchesini}, Danilo and {Marsan}, Cemile and {Oesch}, Pascal and {Papovich}, Casey and {Remus}, Rhea-Silvia and {Schreiber}, Corentin and {Straatman}, Caroline and {Tran}, Kim-Vy and {Urbina}, Claudia Lagos},
        title = "{How Many Quiescent Galaxies are There at 3}",
 howpublished = {JWST Proposal. Cycle 1, ID. \#2565},
         year = 2021,
        month = mar,
        pages = {2565},
       adsurl = {https://ui.adsabs.harvard.edu/abs/2021jwst.prop.2565G}
}

@MISC{Eisenstein2023a,
       author = {{Eisenstein}, Daniel J. and {Maiolino}, Roberto and {Alberts}, Stacey and {Arribas}, Santiago and {Baum}, Stefi A. and {Bhatawdekar}, Rachana and {Bonaventura}, Nina and {Boyett}, Kristan and {Bunker}, Andrew and {Cameron}, Alex James and {Carniani}, Stefano and {Charlot}, Stephane and {Chen}, Zuyi and {Chevallard}, Jacopo and {Curti}, Mirko and {Curtis-Lake}, Emma and {D'Eugenio}, Francesco and {Egami}, Eiichi and {Hainline}, Kevin and {Helton}, Jakob and {Jakobsen}, Peter and {Ji}, Zhiyuan and {Johnson}, Benjamin D. and {Jones}, Gareth C. and {Kumari}, Nimisha and {Looser}, Tobias Jakob and {Luetzgendorf}, Nora and {Lyu}, Jianwei and {Maseda}, Michael and {Nelson}, Erica and {Puskas}, David and {Rawle}, Tim and {Rieke}, Marcia J. and {Robertson}, Brant and {Sandles}, Lester and {Saxena}, Aayush and {Scholtz}, Jan and {Smit}, Renske and {Sun}, Fengwu and {Tacchella}, Sandro and {Topping}, Michael and {Uebler}, Hannah and {Whitler}, Lily and {Williams}, Christina C. and {Willmer}, Christopher Nicholas Andrew and {Willott}, Chris J. and {Witstok}, Joris and {de Graaff}, Anna G.},
        title = "{Unveiling the Redshift Frontier with JWST}",
 howpublished = {JWST Proposal. Cycle 2, ID. \#3215},
         year = 2023,
        month = may,
        pages = {3215},
       adsurl = {https://ui.adsabs.harvard.edu/abs/2023jwst.prop.3215E}
}

@MISC{deGraaff2023,
       author = {{de Graaff}, Anna G. and {Brammer}, Gabriel and {Bezanson}, Rachel and {Gould}, Katriona Mai Landau and {Hirschmann}, Michaela and {Katz}, Harley and {Labbe}, Ivo and {Leja}, Joel and {Lewis}, Zach and {Maseda}, Michael and {Matthee}, Jorryt and {Naidu}, Rohan and {Oesch}, Pascal and {Rix}, Hans-Walter R. and {Setton}, David and {Shivaei}, Irene and {Suess}, Katherine and {Whitaker}, Katherine E. and {Williams}, Christina C.},
        title = "{A complete census of the rare, extreme and red: a NIRCam-selected extragalactic community survey with JWST/NIRSpec}",
 howpublished = {JWST Proposal. Cycle 2, ID. \#4233},
         year = 2023,
        month = may,
        pages = {4233},
       adsurl = {https://ui.adsabs.harvard.edu/abs/2023jwst.prop.4233D}
}

@MISC{Dickinson2024,
       author = {{Dickinson}, Mark and {Amorin}, Ricardo and {Arrabal Haro}, Pablo and {Bagley}, Micaela and {Barro}, Guillermo and {Buat}, Veronique and {Burgarella}, Denis and {Calabro'}, Antonello and {Carnall}, Adam and {Casey}, Caitlin M. and {Chworowsky}, Katherine and {Cleri}, Nikko J. and {Cole}, Justin and {Cooper}, Michael and {Cullen}, Fergus and {Daddi}, Emanuele and {Donnan}, Callum and {Dunlop}, James S. and {Elbaz}, David and {Ferguson}, Henry C. and {Fernandez}, Vital and {Finkelstein}, Steven L. and {Fontana}, Adriano and {Fujimoto}, Seiji and {Giavalisco}, Mauro and {Hamilton}, Timothy S. and {Hathi}, Nimish P. and {Hirschmann}, Michaela and {Hutchison}, Taylor Alexandra and {Juneau}, Stephanie and {Jung}, Intae and {Kartaltepe}, Jeyhan and {Kocevski}, Dale D. and {Koekemoer}, Anton M. and {Larson}, Rebecca L. and {Long}, Arianna and {Lucas}, Ray A. and {Mascia}, Sara and {McGrath}, Elizabeth and {McLeod}, Derek and {McLure}, Ross and {Napolitano}, Lorenzo and {Papovich}, Casey and {Pentericci}, Laura and {Perez Gonzalez}, Pablo G. and {Simons}, Raymond and {Somerville}, Rachel S. and {Trump}, Jonathan R. and {Wang}, Xin and {Weiner}, Benjamin and {Wilkins}, Stephen Matthew and {Yung}, L.~Y. Aaron and {Zavala}, Jorge},
        title = "{The CANDELS-Area Prism Epoch of Reionization Survey (CAPERS)}",
 howpublished = {JWST Proposal. Cycle 3, ID. \#6368},
         year = 2024,
        month = mar,
        pages = {6368},
       adsurl = {https://ui.adsabs.harvard.edu/abs/2024jwst.prop.6368D}
}

@MISC{Coulter2024,
       author = {{Coulter}, David and {Engesser}, Mike and {Pierel}, Justin and {Akins}, Hollis and {Casey}, Caitlin M. and {DeCoursey}, Christa Noel and {Egami}, Eiichi and {Fox}, Ori Dosovitz and {Franco}, Maximilien and {Fujimoto}, Seiji and {Gezari}, Suvi and {Gomez}, Sebastian and {Guolo}, Muryel and {Harish}, Santosh and {Jencson}, Jacob and {Karmen}, Mitchell and {Kartaltepe}, Jeyhan and {Koekemoer}, Anton M. and {McCracken}, Henry Joy and {Moriya}, Takashi and {Quimby}, Robert M. and {Rest}, Armin and {Rhodes}, Jason D. and {Robertson}, Brant and {Shahbandeh}, Melissa and {Shuntov}, Marko and {Siebert}, Matthew Ryan and {Strolger}, Louis-Gregory and {Suzuki}, Nao and {Wang}, Qinan and {Yang}, Lilan and {Zenati}, Yossef},
        title = "{The High-z Menagerie: A Rare Chance to Study the Early and Exotic Transient Universe}",
 howpublished = {JWST Proposal. Cycle 2, ID. \#6585},
         year = 2024,
        month = mar,
        pages = {6585},
       adsurl = {https://ui.adsabs.harvard.edu/abs/2024jwst.prop.6585C}
}

@ARTICLE{Alberts2024b,
       author = {{Alberts}, Stacey and {Lyu}, Jianwei and {Shivaei}, Irene and {Rieke}, George H. and {P{\'e}rez-Gonz{\'a}lez}, Pablo G. and {Bonaventura}, Nina and {Zhu}, Yongda and {Helton}, Jakob M. and {Ji}, Zhiyuan and {Morrison}, Jane and {Robertson}, Brant E. and {Stone}, Meredith A. and {Sun}, Yang and {Williams}, Christina C. and {Willmer}, Christopher N.~A.},
        title = "{SMILES Initial Data Release: Unveiling the Obscured Universe with MIRI Multiband Imaging}",
      journal = {\apj},
         year = 2024,
        month = dec,
       volume = {976},
       number = {2},
          eid = {224},
        pages = {224},
          doi = {10.3847/1538-4357/ad7396},
archivePrefix = {arXiv},
       eprint = {2405.15972},
 primaryClass = {astro-ph.GA},
       adsurl = {https://ui.adsabs.harvard.edu/abs/2024ApJ...976..224A}
}

@article{Lange2023,
    author = {Lange, Johannes U},
    title = "{nautilus: boosting Bayesian importance nested sampling with deep learning}",
    journal = {\mnras},
    volume = {525},
    number = {2},
    pages = {3181-3194},
    year = {2023},
    month = {08},
    doi = {10.1093/mnras/stad2441},
    url = {https://doi.org/10.1093/mnras/stad2441},
    eprint = {https://academic.oup.com/mnras/article-pdf/525/2/3181/51331635/stad2441.pdf},
}

@ARTICLE{Civano2016,
       author = {{Civano}, F. and {Marchesi}, S. and {Comastri}, A. and {Urry}, M.~C. and {Elvis}, M. and {Cappelluti}, N. and {Puccetti}, S. and {Brusa}, M. and {Zamorani}, G. and {Hasinger}, G. and {Aldcroft}, T. and {Alexander}, D.~M. and {Allevato}, V. and {Brunner}, H. and {Capak}, P. and {Finoguenov}, A. and {Fiore}, F. and {Fruscione}, A. and {Gilli}, R. and {Glotfelty}, K. and {Griffiths}, R.~E. and {Hao}, H. and {Harrison}, F.~A. and {Jahnke}, K. and {Kartaltepe}, J. and {Karim}, A. and {LaMassa}, S.~M. and {Lanzuisi}, G. and {Miyaji}, T. and {Ranalli}, P. and {Salvato}, M. and {Sargent}, M. and {Scoville}, N.~J. and {Schawinski}, K. and {Schinnerer}, E. and {Silverman}, J. and {Smolcic}, V. and {Stern}, D. and {Toft}, S. and {Trakhtenbrot}, B. and {Treister}, E. and {Vignali}, C.},
        title = "{The Chandra Cosmos Legacy Survey: Overview and Point Source Catalog}",
      journal = {\apj},
         year = 2016,
        month = mar,
       volume = {819},
       number = {1},
          eid = {62},
        pages = {62},
          doi = {10.3847/0004-637X/819/1/62},
archivePrefix = {arXiv},
       eprint = {1601.00941},
 primaryClass = {astro-ph.GA},
       adsurl = {https://ui.adsabs.harvard.edu/abs/2016ApJ...819...62C}
}

@article{Kocevski2018,
	author = {Kocevski, Dale D. and Hasinger, Guenther and Brightman, Murray and Nandra, Kirpal and Georgakakis, Antonis and Cappelluti, Nico and Civano, Francesca and Li, Yuxuan and Li, Yanxia and Aird, James and Alexander, David M. and Almaini, Omar and Brusa, Marcella and Buchner, Johannes and Comastri, Andrea and Conselice, Christopher J. and Dickinson, Mark A. and Finoguenov, Alexis and Gilli, Roberto and Koekemoer, Anton M. and Miyaji, Takamitsu and Mullaney, James R. and Papovich, Casey and Rosario, David and Salvato, Mara and Silverman, John D. and Somerville, Rachel S. and Ueda, Yoshihiro},
	doi = {10.3847/1538-4365/aab9b4},
	journal = {\apjs},
	month = {jun},
	number = {2},
	pages = {48},
	publisher = {The American Astronomical Society},
	title = {X-UDS: The Chandra Legacy Survey of the UKIDSS Ultra Deep Survey Field},
	url = {https://dx.doi.org/10.3847/1538-4365/aab9b4},
	volume = {236},
	year = {2018}}

@ARTICLE{Luo2017,
       author = {{Luo}, B. and {Brandt}, W.~N. and {Xue}, Y.~Q. and {Lehmer}, B. and {Alexander}, D.~M. and {Bauer}, F.~E. and {Vito}, F. and {Yang}, G. and {Basu-Zych}, A.~R. and {Comastri}, A. and {Gilli}, R. and {Gu}, Q. -S. and {Hornschemeier}, A.~E. and {Koekemoer}, A. and {Liu}, T. and {Mainieri}, V. and {Paolillo}, M. and {Ranalli}, P. and {Rosati}, P. and {Schneider}, D.~P. and {Shemmer}, O. and {Smail}, I. and {Sun}, M. and {Tozzi}, P. and {Vignali}, C. and {Wang}, J. -X.},
        title = "{The Chandra Deep Field-South Survey: 7 Ms Source Catalogs}",
      journal = {\apjs},
         year = 2017,
        month = jan,
       volume = {228},
       number = {1},
          eid = {2},
        pages = {2},
          doi = {10.3847/1538-4365/228/1/2},
archivePrefix = {arXiv},
       eprint = {1611.03501},
 primaryClass = {astro-ph.GA},
       adsurl = {https://ui.adsabs.harvard.edu/abs/2017ApJS..228....2L}
}

@ARTICLE{Almaini2025,
       author = {{Almaini}, Omar and {Wild}, Vivienne and {Maltby}, David and {Taylor}, Elizabeth and {Rowlands}, Kate and {de Lisle}, Thomas and {Alatalo}, Katherine and {Harrold}, Jimi and {Hewitt}, Guillaume and {Patil}, Pallavi and {Skarbinski}, Maya},
        title = "{No evidence for excess AGN activity in recently quenched massive galaxies at cosmic noon}",
      journal = {\mnras},
         year = 2025,
        month = jun,
       volume = {539},
       number = {4},
        pages = {3568-3581},
          doi = {10.1093/mnras/staf659},
archivePrefix = {arXiv},
       eprint = {2504.15342},
 primaryClass = {astro-ph.GA},
       adsurl = {https://ui.adsabs.harvard.edu/abs/2025MNRAS.539.3568A}
}

@ARTICLE{Baker2025b,
       author = {{Baker}, William M. and {Valentino}, Francesco and {Lagos}, Claudia del P. and {Ito}, Kei and {Jespersen}, Christian Kragh and {Gottumukkala}, Rashmi and {Hjorth}, Jens and {Langeroodi}, Danial and {Sedgewick}, Aidan},
        title = "{Exploring over 700 massive quiescent galaxies at z = 2─7: Demographics and stellar mass functions}",
      journal = {\aap},
         year = 2025,
        month = oct,
       volume = {702},
          eid = {A270},
        pages = {A270},
          doi = {10.1051/0004-6361/202555829},
archivePrefix = {arXiv},
       eprint = {2506.04119},
 primaryClass = {astro-ph.GA},
       adsurl = {https://ui.adsabs.harvard.edu/abs/2025A&A...702A.270B}
}

@ARTICLE{Lagos2025,
       author = {{Lagos}, Claudia del P. and {Valentino}, Francesco and {Wright}, Ruby J. and {de Graaff}, Anna and {Glazebrook}, Karl and {De Lucia}, Gabriella and {Robotham}, Aaron S.~G. and {Nanayakkara}, Themiya and {Chandro-Gomez}, Angel and {Bravo}, Mat{\'\i}as and {Baugh}, Carlton M. and {Harborne}, Katherine E. and {Hirschmann}, Michaela and {Fontanot}, Fabio and {Xie}, Lizhi and {Chittenden}, Harry},
        title = "{The diverse star formation histories of early massive, quenched galaxies in modern galaxy formation simulations}",
      journal = {\mnras},
         year = 2025,
        month = jan,
       volume = {536},
       number = {3},
        pages = {2324-2354},
          doi = {10.1093/mnras/stae2626},
archivePrefix = {arXiv},
       eprint = {2409.16916},
 primaryClass = {astro-ph.GA},
       adsurl = {https://ui.adsabs.harvard.edu/abs/2025MNRAS.536.2324L}
}

@ARTICLE{deLucia2024,
       author = {{De Lucia}, Gabriella and {Fontanot}, Fabio and {Xie}, Lizhi and {Hirschmann}, Michaela},
        title = "{Tracing the quenching journey across cosmic time}",
      journal = {\aap},
         year = 2024,
        month = jul,
       volume = {687},
          eid = {A68},
        pages = {A68},
          doi = {10.1051/0004-6361/202349045},
archivePrefix = {arXiv},
       eprint = {2401.06211},
 primaryClass = {astro-ph.GA},
       adsurl = {https://ui.adsabs.harvard.edu/abs/2024A&A...687A..68D}
}

@ARTICLE{Lacey2016,
       author = {{Lacey}, Cedric G. and {Baugh}, Carlton M. and {Frenk}, Carlos S. and {Benson}, Andrew J. and {Bower}, Richard G. and {Cole}, Shaun and {Gonzalez-Perez}, Violeta and {Helly}, John C. and {Lagos}, Claudia D.~P. and {Mitchell}, Peter D.},
        title = "{A unified multiwavelength model of galaxy formation}",
      journal = {\mnras},
         year = 2016,
        month = nov,
       volume = {462},
       number = {4},
        pages = {3854-3911},
          doi = {10.1093/mnras/stw1888},
archivePrefix = {arXiv},
       eprint = {1509.08473},
 primaryClass = {astro-ph.GA},
       adsurl = {https://ui.adsabs.harvard.edu/abs/2016MNRAS.462.3854L}
}

@ARTICLE{Crain2015,
       author = {{Crain}, Robert A. and {Schaye}, Joop and {Bower}, Richard G. and {Furlong}, Michelle and {Schaller}, Matthieu and {Theuns}, Tom and {Dalla Vecchia}, Claudio and {Frenk}, Carlos S. and {McCarthy}, Ian G. and {Helly}, John C. and {Jenkins}, Adrian and {Rosas-Guevara}, Yetli M. and {White}, Simon D.~M. and {Trayford}, James W.},
        title = "{The EAGLE simulations of galaxy formation: calibration of subgrid physics and model variations}",
      journal = {\mnras},
         year = 2015,
        month = jun,
       volume = {450},
       number = {2},
        pages = {1937-1961},
          doi = {10.1093/mnras/stv725},
archivePrefix = {arXiv},
       eprint = {1501.01311},
 primaryClass = {astro-ph.GA},
       adsurl = {https://ui.adsabs.harvard.edu/abs/2015MNRAS.450.1937C}
}

@ARTICLE{Schaye2015,
       author = {{Schaye}, Joop and {Crain}, Robert A. and {Bower}, Richard G. and {Furlong}, Michelle and {Schaller}, Matthieu and {Theuns}, Tom and {Dalla Vecchia}, Claudio and {Frenk}, Carlos S. and {McCarthy}, I.~G. and {Helly}, John C. and {Jenkins}, Adrian and {Rosas-Guevara}, Y.~M. and {White}, Simon D.~M. and {Baes}, Maarten and {Booth}, C.~M. and {Camps}, Peter and {Navarro}, Julio F. and {Qu}, Yan and {Rahmati}, Alireza and {Sawala}, Till and {Thomas}, Peter A. and {Trayford}, James},
        title = "{The EAGLE project: simulating the evolution and assembly of galaxies and their environments}",
      journal = {\mnras},
         year = 2015,
        month = jan,
       volume = {446},
       number = {1},
        pages = {521-554},
          doi = {10.1093/mnras/stu2058},
archivePrefix = {arXiv},
       eprint = {1407.7040},
 primaryClass = {astro-ph.GA},
       adsurl = {https://ui.adsabs.harvard.edu/abs/2015MNRAS.446..521S}
}

@ARTICLE{Springel2018,
       author = {{Springel}, Volker and {Pakmor}, R{\"u}diger and {Pillepich}, Annalisa and {Weinberger}, Rainer and {Nelson}, Dylan and {Hernquist}, Lars and {Vogelsberger}, Mark and {Genel}, Shy and {Torrey}, Paul and {Marinacci}, Federico and {Naiman}, Jill},
        title = "{First results from the IllustrisTNG simulations: matter and galaxy clustering}",
      journal = {\mnras},
         year = 2018,
        month = mar,
       volume = {475},
       number = {1},
        pages = {676-698},
          doi = {10.1093/mnras/stx3304},
archivePrefix = {arXiv},
       eprint = {1707.03397},
 primaryClass = {astro-ph.GA},
       adsurl = {https://ui.adsabs.harvard.edu/abs/2018MNRAS.475..676S}
}

@ARTICLE{Pillepich2018,
       author = {{Pillepich}, Annalisa and {Springel}, Volker and {Nelson}, Dylan and {Genel}, Shy and {Naiman}, Jill and {Pakmor}, R{\"u}diger and {Hernquist}, Lars and {Torrey}, Paul and {Vogelsberger}, Mark and {Weinberger}, Rainer and {Marinacci}, Federico},
        title = "{Simulating galaxy formation with the IllustrisTNG model}",
      journal = {\mnras},
         year = 2018,
        month = jan,
       volume = {473},
       number = {3},
        pages = {4077-4106},
          doi = {10.1093/mnras/stx2656},
archivePrefix = {arXiv},
       eprint = {1703.02970},
 primaryClass = {astro-ph.GA},
       adsurl = {https://ui.adsabs.harvard.edu/abs/2018MNRAS.473.4077P}
}

@ARTICLE{Dave2019,
       author = {{Dav{\'e}}, Romeel and {Angl{\'e}s-Alc{\'a}zar}, Daniel and {Narayanan}, Desika and {Li}, Qi and {Rafieferantsoa}, Mika H. and {Appleby}, Sarah},
        title = "{SIMBA: Cosmological simulations with black hole growth and feedback}",
      journal = {\mnras},
         year = 2019,
        month = jun,
       volume = {486},
       number = {2},
        pages = {2827-2849},
          doi = {10.1093/mnras/stz937},
archivePrefix = {arXiv},
       eprint = {1901.10203},
 primaryClass = {astro-ph.GA},
       adsurl = {https://ui.adsabs.harvard.edu/abs/2019MNRAS.486.2827D}
}

@ARTICLE{Hough2023,
       author = {{Hough}, Renier T. and {Rennehan}, Douglas and {Kobayashi}, Chiaki and {Loubser}, S. Ilani and {Dav{\'e}}, Romeel and {Babul}, Arif and {Cui}, Weiguang},
        title = "{SIMBA-C: an updated chemical enrichment model for galactic chemical evolution in the SIMBA simulation}",
      journal = {\mnras},
         year = 2023,
        month = oct,
       volume = {525},
       number = {1},
        pages = {1061-1076},
          doi = {10.1093/mnras/stad2394},
archivePrefix = {arXiv},
       eprint = {2308.03436},
 primaryClass = {astro-ph.GA},
       adsurl = {https://ui.adsabs.harvard.edu/abs/2023MNRAS.525.1061H}
}

@ARTICLE{Szpila2025,
       author = {{Szpila}, Jakub and {Dav{\'e}}, Romeel and {Rennehan}, Douglas and {Cui}, Weiguang and {Hough}, Renier T.},
        title = "{The nature and evolution of early massive quenched galaxies in the SIMBA-C simulation}",
      journal = {\mnras},
         year = 2025,
        month = feb,
       volume = {537},
       number = {2},
        pages = {1849-1868},
          doi = {10.1093/mnras/staf132},
archivePrefix = {arXiv},
       eprint = {2402.08729},
 primaryClass = {astro-ph.GA},
       adsurl = {https://ui.adsabs.harvard.edu/abs/2025MNRAS.537.1849S}
}

@ARTICLE{CidFernandes2011,
       author = {{Cid Fernandes}, R. and {Stasi{\'n}ska}, G. and {Mateus}, A. and {Vale Asari}, N.},
        title = "{A comprehensive classification of galaxies in the Sloan Digital Sky Survey: how to tell true from fake AGN?}",
      journal = {\mnras},
         year = 2011,
        month = may,
       volume = {413},
       number = {3},
        pages = {1687-1699},
          doi = {10.1111/j.1365-2966.2011.18244.x},
archivePrefix = {arXiv},
       eprint = {1012.4426},
 primaryClass = {astro-ph.CO},
       adsurl = {https://ui.adsabs.harvard.edu/abs/2011MNRAS.413.1687C}
}

@ARTICLE{Kauffmann2003,
       author = {{Kauffmann}, Guinevere and {Heckman}, Timothy M. and {Tremonti}, Christy and {Brinchmann}, Jarle and {Charlot}, St{\'e}phane and {White}, Simon D.~M. and {Ridgway}, Susan E. and {Brinkmann}, Jon and {Fukugita}, Masataka and {Hall}, Patrick B. and {Ivezi{\'c}}, {\v{Z}}eljko and {Richards}, Gordon T. and {Schneider}, Donald P.},
        title = "{The host galaxies of active galactic nuclei}",
      journal = {\mnras},
         year = 2003,
        month = dec,
       volume = {346},
       number = {4},
        pages = {1055-1077},
          doi = {10.1111/j.1365-2966.2003.07154.x},
archivePrefix = {arXiv},
       eprint = {astro-ph/0304239},
 primaryClass = {astro-ph},
       adsurl = {https://ui.adsabs.harvard.edu/abs/2003MNRAS.346.1055K}
}

@ARTICLE{Kewley2001,
       author = {{Kewley}, L.~J. and {Dopita}, M.~A. and {Sutherland}, R.~S. and {Heisler}, C.~A. and {Trevena}, J.},
        title = "{Theoretical Modeling of Starburst Galaxies}",
      journal = {\apj},
         year = 2001,
        month = jul,
       volume = {556},
       number = {1},
        pages = {121-140},
          doi = {10.1086/321545},
archivePrefix = {arXiv},
       eprint = {astro-ph/0106324},
 primaryClass = {astro-ph},
       adsurl = {https://ui.adsabs.harvard.edu/abs/2001ApJ...556..121K}
}

@ARTICLE{Yan2012,
       author = {{Yan}, Renbin and {Blanton}, Michael R.},
        title = "{The Nature of LINER-like Emission in Red Galaxies}",
      journal = {\apj},
         year = 2012,
        month = mar,
       volume = {747},
       number = {1},
          eid = {61},
        pages = {61},
          doi = {10.1088/0004-637X/747/1/61},
archivePrefix = {arXiv},
       eprint = {1109.1280},
 primaryClass = {astro-ph.CO},
       adsurl = {https://ui.adsabs.harvard.edu/abs/2012ApJ...747...61Y}
}

\appendix

\section{The impact of \textit{JWST} MIRI data on \lowercase{$\mathbf{z>2}$} quiescent galaxy selection}\label{appendix:miri}

\begin{figure*}
    \includegraphics[width=\textwidth]{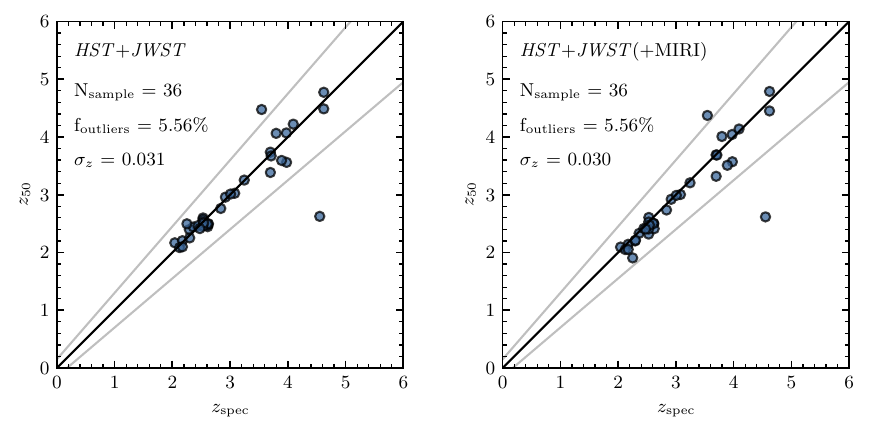}\\\vspace{-0.1cm}
    \includegraphics[width=\textwidth]{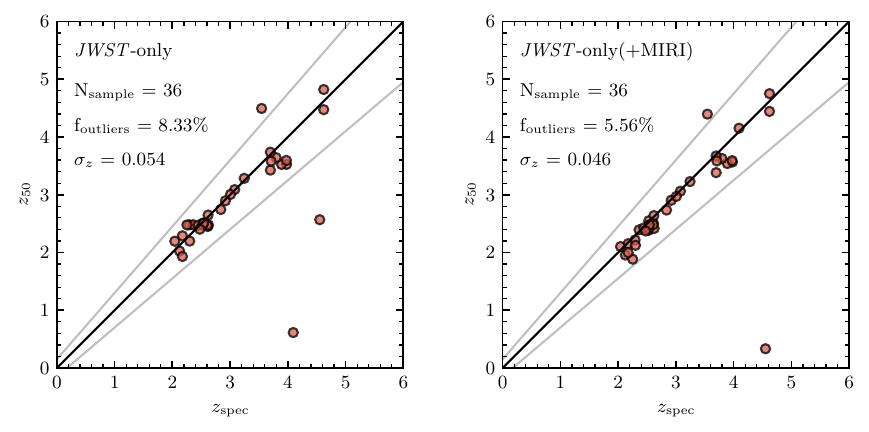}
    \caption{A comparison between the photometric redshifts derived in our `\textit{HST}+\textit{JWST}', `\textit{HST}+\textit{JWST}+(MIRI)', `\textit{JWST}-only' and `\textit{JWST}-only(+MIRI)' runs, for the same spectroscopic objects shown in Fig. \ref{figure:zzplot}. In each plot we use the same sample as described in Appendix \ref{appendix:miri}, those selected as high-redshift massive quiescent galaxy candidates in our main \textit{HST}+\textit{JWST} run that are in either the UDS or COSMOS field, have clean MIRI F770W coverage, and corresponding spectroscopic data.}
    \label{figure:zzplot_miri}
\end{figure*}

To investigate the impact of \textit{JWST} MIRI data on the selection of high-redshift massive quiescent galaxies, we refit our primary sample, described in Section \ref{section:methods:selection}, with additional available F770W MIRI photometry. Specifically, we use imaging from the PRIMER survey, which covers sections of the UDS and COSMOS fields, the latter of which also benefits from F770W MIRI coverage from the COSMOS-Web survey (GO 1727, \citealt{Casey2023}).

From the 183 massive quiescent galaxy candidates from our primary sample in UDS and COSMOS, we were able to extract clean F770W MIRI imaging for 101, representing $\simeq 55$ per cent coverage. In the following analysis, we only consider the 101 massive quiescent galaxy candidates benefitting from clean MIRI coverage (33 in COSMOS and 68 in UDS). Due to the extremely wide tails present in the F770W point-spread function (PSF), we extract photometry in 1.5 arcsec diameter apertures. The calculated F770W flux is then first corrected to a total flux using enclosed energy estimates from the PSF growth curve ($+17.6$ per cent)\footnote{\href{https://jwst-docs.stsci.edu/jwst-mid-infrared-instrument/miri-performance/miri-point-spread-functions}{https://jwst-docs.stsci.edu/jwst-mid-infrared-instrument/miri-performance/miri-point-spread-functions}}, then by a zero-point correction derived from the median ratio of model F770W flux to observed F770W flux ($+15.3$ per cent).

The final F770W photometry is then used in two refitting runs of our primary sample with {\sc Bagpipes}: (i) a fit with 9 bands of only NIRCam and MIRI photometry, \textit{JWST}-only(+MIRI), and (ii) a fit with 12 bands of ACS, NIRCam and MIRI photometry, \textit{HST}+\textit{JWST}(+MIRI). We use the same priors and settings for our {\sc Bagpipes} fits as described in Section \ref{section:methods:sedfit}.

\subsection*{Objects in \textit{HST}+\textit{JWST} but not \textit{JWST}-only(+MIRI)}

In this section we argue that, in the selection of massive quiescent galaxies at $z>2$, the addition of mid-IR MIRI F770W data does not constitute a substitute for optical \textit{HST} ACS data, by comparing the sample selected in our flagship \textit{HST}+\textit{JWST} run with the sample selected in the \textit{JWST}-only(+MIRI) run. In total, we lose $14/101$ massive quiescent galaxy candidates, representing a $\simeq 14$ per cent loss of candidates selected in our primary \textit{HST}+\textit{JWST} run. Of these, seven are pushed slightly below $z=2$ but remain at quiescent levels of SFR (only one of which has a spectrum, and is at $z_{\mathrm{spec}}=2.25$). Six others are fitted with excessively high SFRs, far above the quiescent sSFR criterion, hence now characterised as significantly star-forming. However, two of these are spectroscopically confirmed to be massive quiescent galaxies in the EXCELS survey (see Section \ref{section:data:spectroscopy:excels}, IDs 34495 and 113667). The remaining lost candidate is fitted at low redshift and high star-formation rate, however this is a well known spectral contaminant AGN at $z_{\mathrm{spec}}=4.56$ (PRIMER-UDS-116064, see Fig. \ref{figure:xrayspectra}). Regarding the difference in robust candidate identification, 52 robust candidates are similarly selected in both runs, 15 are lost and only eight gained. This suggests a general deficiency in the identification of robust candidates in the absence of \textit{HST} data, despite the presence of \textit{JWST} MIRI data.

Therefore, our results suggest that the inclusion of mid-IR MIRI data does not approach the level of constraining power displayed by optical \textit{HST} ACS data when selecting samples of high-redshift massive quiescent galaxies. Echoing the result from our \textit{JWST}-only run (see Section \ref{section:sample:hst:candidateslost}), in which 18 per cent of candidates selected in the presence of \textit{HST} data were lost, we lose 14 per cent of \textit{HST}+\textit{JWST} selected candidates in the \textit{JWST}-only(+MIRI) run. Whilst this is a marginal improvement, it nevertheless still represents a significant decrease in the accurate selection of both general and robust massive quiescent galaxy candidates compared with \textit{HST}+\textit{JWST} data.

\subsection*{\textit{HST}+\textit{JWST} versus \textit{HST}+\textit{JWST}(+MIRI)}

In the following analysis we show that the addition of F770W MIRI data makes relatively little change to our primary sample of massive quiescent galaxies selected with ACS and NIRCam data, by comparing the samples selected in our flagship \textit{HST}+\textit{JWST} run with our supplementary \textit{HST}+\textit{JWST}(+MIRI) run. Both samples are largely identical, with only a $<10$ per cent loss of massive quiescent galaxy candidates between them, where each loss is invariably due to minute random scatter about the $z=2$ and sSFR selection criteria boundaries. There are no catastrophic property changes for any massive quiescent galaxy candidate selected in \textit{HST}+\textit{JWST} when MIRI F770W data is subsequently included. Regarding the difference in robust candidate identification, 59 robust candidates are similarly selected in both runs, eight are lost and six gained. This shows a general agreement in the robust candidate samples of \textit{HST}+\textit{JWST} and \textit{HST}+\textit{JWST}(+MIRI), suggesting that the inclusion of MIRI F770W data (when \textit{HST} data is already present) does not greatly aid in their identification.

To further illustrate the agreement between the fitting runs \textit{HST}+\textit{JWST} and \textit{HST}+\textit{JWST}(+MIRI), we include a comparison of their respective photometric redshifts in the top panels of Fig. \ref{figure:zzplot_miri}. These are derived by crossmatching the subset of massive quiescent galaxy candidates with clean MIRI F770W coverage, with the spectroscopic sample described in Section \ref{section:sample:contamination:spectra}. Comparing only the runs with \textit{HST} data included, the number and exact nature of redshift outliers is identical, and there is virtually no change in the accuracy of photometric redshifts.

Considering now only the runs with \textit{HST} data excluded (lower panels of Fig. \ref{figure:zzplot_miri}), there is a notably positive effect when MIRI is included, with a decrease in redshift outliers and slight increase in general redshift accuracy. While not as drastic an effect as when \textit{HST} data is included, this result nonetheless motivates the inclusion of MIRI data where no \textit{HST} data is available.

It is also worth noting that MIRI does not systematically remove contaminants in samples selected in the presence of \textit{HST} data. From the two spectroscopically identified contaminants with good MIRI coverage, PRIMER-UDS-116064 and PRIMER-UDS-125909 (which are also the two redshift outliers in the top two panels of Fig. \ref{figure:zzplot_miri}), neither are rejected by the addition of MIRI F770W data.

To summarise, in the selection of massive quiescent galaxies at high redshift, mid-IR MIRI F770W data has a notably positive effect when combined with near-IR NIRCam data alone. However, the benefit is markedly less than the benefit of \textit{HST} ACS optical data, and MIRI F770W data is effectively irrelevant when optical \textit{HST} is already included.\\~\\

\section{Object list}

\begin{figure*}
    \includegraphics[width=\textwidth]{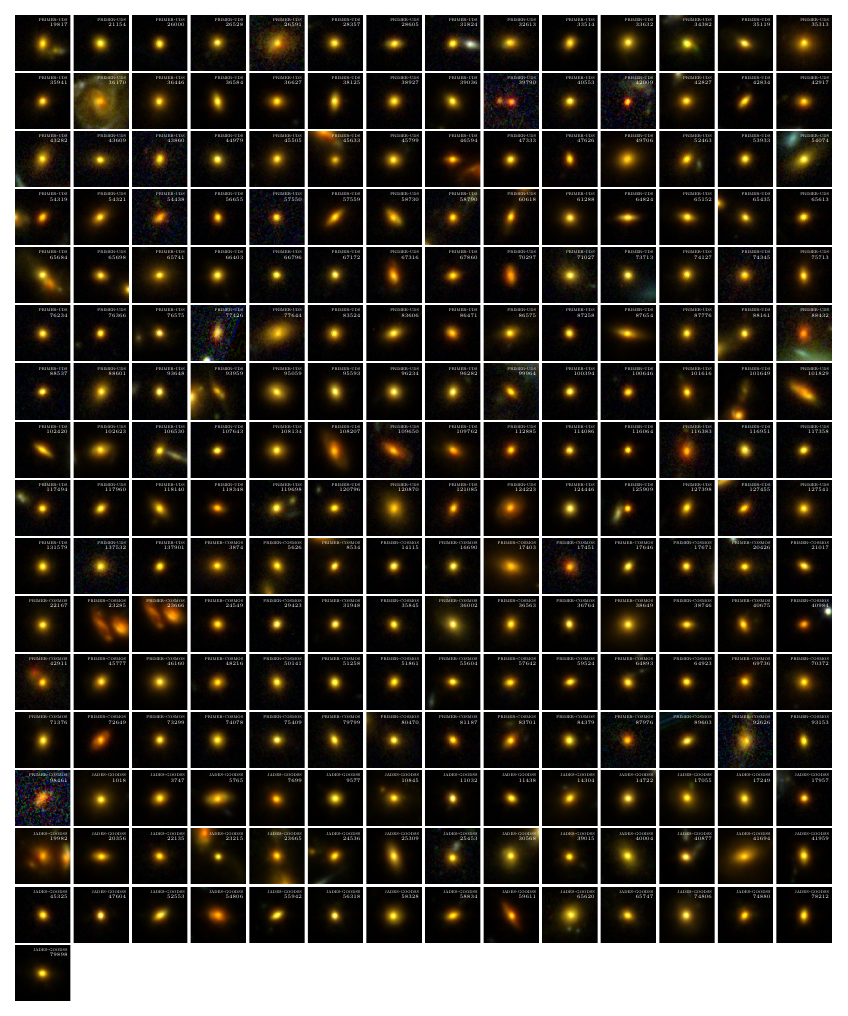}
    \caption{Imaging cutouts ($2\times2$ arcsec) of each of the galaxies in our sSFR selected massive quiescent galaxy sample. In the creation of RGB images we use the F356W+F444W filters for the red image, F200W+F277W filters for the green image and F090W+F115W filters for the blue image. Each galaxy is labelled by its survey and photometric catalogue ID.}
    \label{figure:rgb}
\end{figure*}

Imaging cutouts ($2\times2$ arcsec) for our full sample of massive quiescent galaxy candidates are shown in Fig. \ref{figure:rgb}. 

\bsp	
\label{lastpage}
\end{document}